# PEBP1/RKIP: from multiple functions to a common role in cellular processes.

Françoise Schoentgen, Slavica Jonic

Institut de Minéralogie, de Physique des Matériaux, et de Cosmochimie (IMPMC), Sorbonne Université, UPMC Univ Paris 06, UMR CNRS 7590, Muséum National d'Histoire Naturelle, IRD UMR 206, IUC, 4 Place Jussieu, F-75005 Paris, France.

## Abstract

PEBPs (PhosphatidylEthanolamine Binding Proteins) form a protein family widely present in the living world since they are encountered in microorganisms, plants and animals. In all organisms PEBPs appear to regulate important mechanisms that govern cell cycle, proliferation, differentiation and motility. In humans, three PEBPs have been identified, namely PEBP1, PEBP2 and PEBP4. PEBP1 and PEBP4 are the most studied as they are implicated in the development of various cancers. PEBP2 is specific of testes in mammals and was essentially studied in rats and mice where it is very abundant. A lot of information has been gained on PEBP1 also named RKIP (Raf Kinase Inhibitory protein) due to its role as a metastasis suppressor in cancer. In contrast, PEBP4 is known to favor tumor growth in various cancers, but was less studied than PEBP1. The many studies carried out on PEBP1 has demonstrated its implication in cancer but also in Alzheimer's disease, diabetes and nephropathies. PEBP1 was also described to be involved in many cellular processes, among them are signal transduction, inflammation, cell cycle, proliferation, adhesion, differentiation, apoptosis, autophagy, circadian rhythm and mitotic spindle checkpoint. On the molecular level, PEBP1 was demonstrated to regulate several signaling pathways such as Raf/MEK/ERK, NF-κB, PI3K/Akt/mTOR, p38, Notch and Wnt. PEBP1 acts by inhibiting most of the kinases of these signaling cascades. Moreover, PEBP1 is able to bind to a variety of small ligands such as ATP, phospholipids, nucleotides, flavonoids or drugs. Considering PEBP1 is a small cytoplasmic protein (21kDa), its molecular features and its involvement in so many diseases and cellular mechanisms are amazing. The aim of this review is to highlight the molecular systems that are common to all these cellular mechanisms in order to decipher the specific role of PEBP1. Recent discoveries in the organization of cellular systems enable us to propose that PEBP1 is a modulator of molecular interactions that control signal transduction during membrane and cytoskeleton reorganization.

## 1- Introduction

PhosphatidylEthanolamine Binding Proteins (PEBPs) are a family of proteins implicated in numerous cellular processes in a large number of living organisms such as bacteria, yeasts, parasites, plants, insects and animals. Several PEBP isoforms were encountered in each specific species. In mammals, the main expressed isoform was firstly discovered and characterized in the brain from bovine (**1, Bernier 1984; 2, Schoentgen**), human (**3, Seddiqi**) and rat (**4, Grandy**), where it was associated with membranes and phospholipids. Studies measuring the affinity of this isoform toward different phospholipids showed a preference for phosphatidylethanolamine (PE) (**5, Besnier 1986**), then designated PEBP1. Later on, it was described to also bind other small ligands such as nucleotides (**6, Bucquoy**), flavonoids (**7, Kim**) or drugs (**8, Dadvar**). The PEBP1 isoform was described to be present in almost all the tissues in mammals. Furthermore, it was described to be downregulated in numerous types of cancer, the disappearance of PEBP1 in tumor cells being concomitant with metastases development (**9, Lamiman,** for a review), thus PEBP1 is considered a suppressor of metastases (**10, Keller**). In mammals, another PEBP isoform was identified, it appeared to be specific of testes with a role in mammal spermatogenesis and post-testicular sperm maturation (**11, Hickox**) and it was called PEBP2. Lastly, in humans, a third isoform was identified to be overexpressed in cells of several cancer tumors, it was called PEBP4 (**12, Wang 2004**). It should be noted that PEBP3 was never described to be expressed in any human tissue, probably corresponding to a pseudogene.

Given the presence of PEBP1 in almost all human tissues and the observation that PEBP1 downregulation leads to metastases, numerous studies were conducted to understand its mechanism of action. One major discovery was the inhibitory effect of PEBP1 toward Raf1, a kinase of the signaling pathway Raf/MEK/ERK (**13, yeung 1999**). This discovery prompted the authors to redefine PEBP1 as a Raf Kinase inhibitory protein (RKIP). Since then, PEBP1 was described to regulate many other protein kinases implicated in various signaling pathways such as NF-κB, GPCRs, STAT3, GSK3 (**14, Shin; 15, Raikumar**) PI3K/Akt/mTOR (**16, Lin**), Wnt (**14, shin**), p38 (**17, Lai**) and Notch1 (**18, Noh**). Among the various properties of PEBP1, it was identified as a precursor of a hippocampal cholinergic neurostimulating peptide (HCNP) (**19, Ojika 1992**), so it is sometimes called HCNPpp (HCNP precursor protein) (**20, Bassi**). HCNP corresponds to the first 11 amino acids in N-terminal of PEBP1. This peptide was encountered in brain tissue where it enhances acetylcholine synthesis in the hippocampal medial septal nuclei (**21, Ojika 1994**). It was also observed to display a cardiotropic action by enhancing the cholinergic-mediated negative inotropy of rat heart through direct interaction with the G-protein-coupled muscarinic receptor pathway (**22, Angelone**). The crystallographic structure of bovine (**23, Serre**) and human (**24, Banfield**) PEBP1 was solved, revealing one unique domain having a beta-fold and formed by two antiparallel β sheets in a Greek-key topology. A small cavity close to the protein surface has a high affinity for anions, such as phosphate, acetate, phosphorylethanolamine or cacodilate molecules and was identified as the binding site of PEBP1.

Thus, presently, a lot of information is available about the molecular features of PEBP1, its binding properties toward small ligands, its implication in main cellular processes through its interaction with numerous protein partners, the regulation of its expression and its role in several diseases. However, there is no general explanation for the apparent diversity of PEBP1 biological functions. Clearly, it is not just a phosphatidylethanolamine binding protein, nor only a Raf kinase inhibitory protein, neither just an hippocampal cholinergic neurostimulating peptide precursor (**25, Skinner**). Then, one can wonder: how does PEBP1, a small 21kDa protein, perform such diverse functions? What is its mode of action? Are there common molecular mechanisms in the main cellular processes that could explain the implication of PEBP1 in several of them?

In this review article, we attempt to highlight the facts and the convergences that could explain the general and original behavior of PEBP1 and its astonishing and unique role in cellular regulation and homeostasis. PEBP1 appears to be a fascinating protein controlling the intimate and subtle interactions between important molecular players in the cells.

The first sections of this review are dedicated to an overview of the main properties and functions of PEBP1; then the following sections are devoted to the relationships and common features between the cellular processes implying PEBP1. These comparisons aim to decipher the special role of PEBP1 in cells.

# 2- Generalities on PEBP1

## 2.1- Role of PEBP1 in physiological processes and diseases

Surprisingly, PEBP1 was described to influence a considerable number of cell physiological behaviors, among them we can enumerate cell growth and differentiation (**26, Hellmann; 27, Sagisaka 2010; 28, Toyoda**), proliferation (**29, Li 2014; 30, Zou; 31, Liu; 32, Zang**), migration (**33, Lei; 34, Xinzhou ; 35, Han**), motility (**36, Al-Mulla 2012; 33, Lei; 37, Dai**), cell cycle (**38, Al-Mulla 2011**), genomic stability (**39, Eves; 40, Al-Mulla 2008**), apoptosis (**31, Liu 2014; 41, Zuo; 42, Lin; 43, Wei**), autophagy (**44, Noh; 45 Wen**), drug resistance (**46, Li 2014; 47, Chen; 48, Bonavida**), spermatogenesis (**49, Moffit; 50, Klysik** ), mechanical and oxidative stress (**51, Glombik; 52, Li 2010; 53, Al-Mulla 2012**), adhesion (**54, Mc Henry; 55, Das**), neuronal synapse (**56, Kato; 57, Mizuno**) and immunity (**58, Gu; 59, Wright; 60, Wang 2016**). However, the growing interest of the researchers in investigating PEBP1 properties is due mainly to its implication in diseases. Some of the pathologies modulated by PEBP1, such as cancer, Alzheimer's disease diabetes and ciliopathies, are major challenges in public health.

### 2.1.1- PEBP1 in cancer

A great number of papers have indicated that the downregulation of PEBP1 expression in tumor cells is crucial for development of metastases and patient survival. Loss of PEBP1 expression has been clearly demonstrated in a large variety of cancers, affecting breast, prostate, brain, lung, liver, colorectal, endometrial, esophageal, gastric, renal, pancreatic, nasopharyngeal, melanoma (**9, Lamiman for a review**). For the most part, the effect of PEBP1 decrease in cancer appeared to be due to promotion of metastases and loss of PEBP1 pro-apoptotic activity. Finally, PEBP1 appeared to be a clinically relevant suppressor of metastases that may function by decreasing vascular invasion (**61, Fu**).

### 2.1.2- PEBP1 in brain functioning

In brain, apart its role in cancer, PEBP1 was also described to be involved in circadian rhythm, synaptic long term depression (LTD), neural development and differentiation, stress and depression, and Alzheimer's disease (**62, Ling**).

**Circadian clock**

Activation of the MAPK/ERK signaling cascade in the suprachiasmatic nucleus (SCN), a region of the anterior hypothalamus, is a key event that couples light to **circadian clock** entrainment. An important role of PEBP1 in the regulation of MAPK/ERK signaling in the SCN and in photic entrainment of the SCN clock was demonstrated (**63, Antoun**).

**Long-term depression**

Long-term depression (LTD) is a mechanism that is used in different brain regions to influence synaptic plasticity which is the basis for many forms of learning and memory. It was suggested that a positive feedback loop, including protein kinase C (PKC), mitogen-activated protein kinase (MAPK) and PEBP1, is required for the gradual expression of cerebellar LTD (**64, Yamamoto 2012**).

**Neural development**

First studies showed that the peptide HCNP stimulated the development and differentiation of cholinergic neurons in cultures of rat medial septal nuclei, by enhancing the activity of choline acetyltransferase and thus promoting acetylcholine synthesis (**19, Ojika 1992; 21, Ojika 1994**). Similar findings were observed in rat hippocampal progenitor cells, where endogenous PEBP1 expression closely correlated with differentiation into neurons and oligodendrocytes, but not astrocytes (**27, Sagisaka**). Consistent with this, overexpression of PEBP1 promoted neuronal and oligodendrocyte differentiation, whereas PEBP1 silencing promoted differentiation into astrocytes (**27, Sagisaka**).

**Stress and depression**

Chronic restraint-induced stress downregulated the expression of PEBP1 in the rat brain; a parallel reduction in PEBP1 was observed in rat hippocampal cell lines stimulated with the synthetic glucocorticoid (stress hormone) dexamethasone (**65, Kim 2007**). PEBP1 transgenic mice exhibited enhanced depressive-like behavior that was apparent in aged subjects (30 weeks old) but not a younger cohort (12 weeks old) (**66, Matsukawa**). The cognitive impact of PEBP1 overexpression was restricted to depression, as behaviors associated with anxiety and spatial learning and memory were not affected in these animals (**66, Matsukawa**).

**Alzheimer's disease**

Several hypotheses have been proposed to provide a causative basis for Alzheimer's disease (AD) (**67, de la Torre**). One of these, the cholinergic hypothesis, proposed that the cognitive impairments in AD patients are the result of a deficiency in acetylcholine synthesis, which, is promoted by HCNP, the peptide corresponding to the N-terminal part of PEBP1 (**68, Bartus; 69, Maki**). Additionally, HCNP/PEBP1 has been shown to accumulate in Hirano bodies, which are intracellular aggregates found in nerve cells of the elderly and patients with certain forms of neurodegenerative diseases (**70, Mitake**).

## 2.1.3- PEBP1 in diabetic nephropathy and retinopathy

In rats with diabetic nephropathy, the expression level of PEBP1 in glomerular tissue was downregulated and the expression level of NF-κB was upregulated. With rituximab treatment, expression levels of PEBP1 and NF-κB were, respectively, upregulated and downregulated, and the concentrations of 24-h urine protein were also decreased. These results suggested that expression levels of PEBP1 might be regulated by rituximab via NF-κB (**71, Li 2013**). Otherwise, PEBP1 was upregulated in the insulin-targeting tissues (liver, muscle and fat) in type II diabetic Otsuka Long-Evans Tokushima Fatty (OLETF) rats fed with fenofibrate, suggesting a role of PEBP1 in the improvement of insulin sensitivity and control of lipid metabolism (**72, Hahm**). It is worth noting that lower levels of PEBP1 were detected in the vitreous humor from type 2 patients with proliferative diabetic retinopathy, this result was related to the possible involvement of PEBP1 in neovascularization and cell cycle progression (**72, Hahm**).

## 2.1.4- PEBP1 in olfaction deficit and ciliopathies

The PEBP1−/− mice are viable and PEBP1 function appears to be largely dispensable during embryonic development. However, the PEBP1−/− mice display an olfaction deficit that is apparent by four months of age but increases in severity by 5–8 months. Moreover, the results also suggested that the mutation is associated with a general learning deficit (**73, Theroux**). This data are to be compared with the observation that in *Drosophila*, PEBP orthologues reside in antennae and olfactory hairs and play a role in odorant binding (**74, Pikielny**).

In another study, mutations in CEP290 (a centrosomal protein of 290 kDa) resulted in relatively early onset severe retinal degeneration and dysfunction in mice and humans. PEBP1 prevents cilia formation and is associated with Cep290-mediated photoreceptor degeneration in retina. Cep290-mediated photoreceptor degeneration is associated with aberrant accumulation of PEBP1 (**75, Murga-Zamalloa**). This data implicates CEP290-PEBP1 pathway in CEP290-retinal degeneration and suggests that targeting PEBP1 levels can delay photoreceptor degeneration (**76, Subramanian**).

## 2.2- PEBP1 is a modulator of signaling pathways

PEBP1 was described to regulate the Raf/MEK/ERK by competitively disrupting the interaction between Raf-1 and MEK. By inhibiting the phosphorylation and activation of MEK by Raf-1, PEBP1 appeared as the only physiological endogenous inhibitor of the Raf/MEK/ERK cascade, a highly conserved signaling pathway that regulates cell growth, differentiation, migration, and apoptosis (**13, Yeung 1999**). Today, PEBP1 is admitted to interplay with many pivotal intracellular signaling cascades that control cellular growth, proliferation, division, differentiation, motility, apoptosis, genomic integrity, and therapeutic resistance (**77, Al-Mulla 2013**). Here, we discuss briefly the main cascades and downstream effectors known to be regulated by PEBP1 and especially, we put emphasis on the crosstalk between most of these pathways (**see Figure 1**).

### 2.2.1- Raf/MEK/ERK pathway

The discovery that PEBP1 was an endogenous inhibitor of the Raf/MEK/ERK signaling pathway was the first study elucidating the role for PEBP1 in a main cellular signaling cascade. It was demonstrated that PEBP1 binds directly either to Raf1 and MEK. Furthermore, PEBP1 was also found to bind MEK and ERK, which subsequently caused Raf-1 to dissociate from MEK (**78, Yeung 2000**). In cancer, the downregulation of PEBP1 in metastases and invasive cells has been described for different types of cancer such as breast (**79, Hagan**), prostate (**80, Fu 2006**), colorectal (**81, Koelzer**), cervical cancers (**82, Martinho**) and melanoma (**83, Schuierer**). The effect of PEBP1 downregulation on ERK activation was reported in prostate cancer (**61, Fu 2003**), hepatocellular cancer cells (**84, Lee**), and gastric cancer (**85, Guo**). However, whether cancer metastasis is linked to the ERK pathway is not always clear and the effect of PEBP1 loss on the Raf/MEK/ERK cascade seems to be cell-type dependent (**86, Vandamme**). In addition to Raf-1, PEBP1 could also bind to the B-Raf isoform and antagonized its kinase activity in melanoma cancer cells. (**87, Park**).

### 2.2.2- G protein-coupled receptors (GPCRs) and G receptor kinase (GRK2)

PEBP1 has been implicated in the control of G-protein coupled receptors (GPCR) signaling cascade (**88, Lorenz**). GPCRs belong to a large family of membrane receptors implicated in numerous cellular processes. G protein-coupled receptor kinase 2 (GRK2) phosphorylates activated receptors which causes them to dissociate from G proteins and leads to GPCR internalization. It was demonstrated that the PKC isoforms -$\alpha$,$-\beta$I, $-\beta$II, $-\gamma$, and atypical PKC$\zeta$ phosphorylate PEBP1 at Ser153. The phosphorylated PEBP1 releases from Raf1 and then binds to GRK2. The inhibition of GRK2 by the phosphorylated form of PEBP1 then blocks the GPCRs internalization (**89, Corbit**). More recently, it was seen that, in non-inflamed tissue, Delta Opioid Receptors (DOR) are maintained in an analgesically incompetent state by association with GRK2. This interaction prevents G$\beta$ subunit association with the receptor DOR, reducing its activity. In inflamed tissue, PKC phosphorylates PEBP1 which, in turn, sequesters GRK2 in the cytosol, restoring DOR functionality in sensory neurons (**90, Brackley**).

### 2.2.3- NF-$\kappa$B pathway

In response to stimulation with necrosis factor alpha (TNF$\alpha$) or interleukin 1 beta, PEBP1 was demonstrated to act upstream of the kinase complex that mediates the phosphorylation and inactivation of the inhibitor of NF$\kappa$B (I$\kappa$B). Similarly to its action upon the Raf/MEK/ERK pathway, PEBP1 physically interacted with four kinases of the NF-kappaB activation pathway:

NF-kappaB-inducing kinase (NIK), transforming growth factor beta-activated kinase 1 (TAK1), and the inhibitor of kappaB (IκB) kinases IKKalpha, and IKKbeta, inhibiting them (**91, Yeung 2001**). Since then, other studies gave evidence of the NFκB pathway modulation by PEBP1. Indeed, it was demonstrated that PEBP1 plays a key role in neural cell apoptosis caused by oxygen-glucose deprivation (OGD) partly via regulating NFκB and ERK. PEBP1 overexpression significantly increased the cell viability of OGD cells, and attenuated apoptosis, cell cycle arrest, and reactive oxygen species generation. It was found that PEBP1 interacts with TAK1, NIK, IKK, and Raf-1 and negatively regulates the NFκB and ERK pathways (**92, Su**). In cultured cancer cells, the presence of PEBP1 favored the optimal ubiquitination of interleukine-1 receptor-associated kinase (IRAK), TNF receptor-associated factor 6 (TRAF6), and TAK1 as well as the assembly of the IKK complex. PEBP1 seemed to function as a scaffold protein facilitating the phosphorylation of IκB by upstream kinases IKKs. Interestingly, contrary to what one would expect of a scaffold protein, the results showed that PEBP1 has an overall inhibitory effect on the NFκB transcriptional activities (**93, Tang**). In the intestine, the PEBP1 expression is positively correlated with the severity of inflammatory bowel disease. Mechanistically, PEBP1 enhances the induction of p53-upregulated modulator of apoptosis by interacting with TAK1 and promoting TAK1-mediated NFκB activation (**42, Lin W**).

### 2.2.4- PI3K/Akt pathway

In nasopharyngeal carcinoma (NPC), PEBP1 was frequently downregulated in the radioresistant tissues compared with radiosensitive tissues. PEBP1 reduction promoted NPC cell radioresistance by increasing ERK and Akt activity, and Akt may be a downstream transducer of ERK signaling in clinical samples. These data highlighted a PEBP1-ERK-Akt signaling axis in NPC radiosensitization (**94, Yuan**). The expression of PEBP1 was markedly decreased in mouse retina following optic nerve crush (ONC). Furthermore, overexpression of PEBP1 inhibits retinal ganglion cells apoptosis and promotes axonal regeneration after ONC. It was shown that overexpression of PEBP1 increased the phosphorylation in ERK1/2 and Akt in the injured retina (**95, Wei J**). Finally, it was shown that the regulation of PEBP1 by didymin in hepatic fibrosis leads to inhibit ERK and PI3K/Akt pathways (**16, Lin X**).

### 2.2.5- GSK3β, p38 MAPK and Wnt pathways

PEBP1 levels in human colorectal cancer positively correlate with the expression of glycogen synthase kinase-3β (GSK3β). GSK3β suppresses tumor progression by downregulating multiple oncogenic pathways including Wnt signaling and cyclin D1 activation. PEBP1 binds GSK3 proteins and maintains GSK3β protein levels. Depletion of PEBP1 augments oxidative stress–mediated activation of the p38 mitogen activated protein kinase, which, in turn, inactivates GSK3β (**96, Al-Mulla 2011**). On HEK293 cells, PEBP1 silencing induced an intense oxidative stress response that includes the activation of p38 MAPK. Activated p38 MAPK triggers the phosphorylation of numerous downstream proteins and kinases. Particularly, activated p38 MAPK leads to phosphorylation and stabilization of the tumor suppressor protein p53, resulting in the induction of a p53-dependent G1/S and G2/M arrest, cellular growth inhibition, and apoptosis (**86, Vandamme**). Furthermore, the activation of p38 MAPK inhibits GSK3β by phosphorylating its inhibitory T390 residue (**97, Thornton**). PEBP1 prevents the inhibitory phosphorylation of GSK3β by p38 MAPK and also stabilizes GSK3β. It was found that PEBP1 regulates GSK3β in two ways : first, by binding and maintaining GSK3β protein levels, and second, by preventing the p38 MAPK mediated inhibitory phosphorylation of GSK3β (**96, Al-**

**Mulla 2011).** It should be mentioned that PEBP1 was also observed to stabilize Kelch-like ECH-associated protein 1 (KEAP1) in colorectal cancer. KEAP1 is known to interact with nuclear factor 2 (NRF2) playing a central role in protecting cells against oxidative and xenobiotic stress. The Keap1–Nrf2 pathway counteracts ROS-mediated damage in cancers and neurodegenerative diseases. In colorectal cancer, PEBP1 loss causes a rapid reduction in the basal level of KEAP 1 protein by accelerating its rate of degradation, inducing NRF2 nuclear accumulation and the transcriptional upregulation of NRF2-dependant drug resistant genes (**53, Al-Mulla 2012**).

### 2.2.6- STAT3

The signal transducer and activator of transcription 3 (STAT3) was described to play a role in correlation with PEBP1 in various types of cancer. In breast and prostate cancer cells, PEBP1 overexpression inhibited c-Src auto-phosphorylation and activation, as well as IL-6-, JAK1 and 2-, and Raf-mediated STAT3 activation. PEBP1 overexpression resulted in constitutive physical interaction with STAT3 and blocked c-Src and STAT3 association. Thus, PEBP1 was demonstrated to physically interact with STAT3. PEBP1 synergizes with microtubule inhibitors such as taxols to induce apoptosis and inhibit STAT3 activation of breast and prostate cancer cells (**98, Yousuf**). In gastric cancer, data suggested that PEBP1 inhibits gastric cancer metastasis via the downregulation of its downstream target genes STAT3 and cyclin D1 (**99, Zhang**). Consistently, in nasopharyngeal carcinoma (NPC) PEBP1 downregulation promoted NPC invasion, metastasis and epithelial–mesenchymal transition by activating STAT3 signaling. A direct interaction between PEBP1 and STAT3 was shown, leading to a reduced phosphorylation of STAT3 (**100, He**).

### 2.2.7- LIN28 and Let-7 miRNA

PEBP1 was described to suppress a metastasis signaling cascade involving LIN28 and let-7 miRNA (**101, Dangi-Garimella**). The role of PEBP1 and let-7 was particularly supported in metastatic breast cancer. In this case, PEBP1 inhibits the let-7 targets HMGA2 and BACH1 that in turn upregulate bone metastasis genes encoding matrix metalloproteinase 1 (MMP1), osteopontin (OPN) and Chemokine (C-X-C motif) receptor 4 (CXCR4) (**102, Yun**). The implication of PEBP1 and let-7 was also highlight in neuroblastoma, where didymin overcomes drug-resistance in p53-mutant neuroblastoma through PEBP1-mediated inhibition of MYCN (**103, Singhal**). In esophageal cancer, other studies clearly demonstrated that PEBP1 inhibits cell invasion by downregulating the expression of GRK-2, LIN28 and MMP-14 (**104, Zhao**). It is known that the let-7 microRNAs and their antagonists, the Lin28 RNA-binding proteins, control the timing of embryonic development (**105, Moss**). This pathway was also shown to regulate glucose metabolism in adult mice and to reprogram metabolism during tissue injury and repair (**106, Shyh-Chang**). In cancer, malignant cells exhibit major metabolic alterations. When expressed in multiple cancer cell lines, Lin28 actively promotes aerobic glycolysis while inhibiting mitochondrial OxPhos to facilitate cancer proliferation. Moreover, conditionally overexpressing human LIN28B in the liver resulted in the development of liver cancer, with histological features of both hepatoblastoma and hepatocellular carcinoma (**107, Nguyen**).

### 2.2.8- Crosstalk between pathways

It is to note here that PEBP1 interacts with various kinases implicated in several cellular signaling cascades. Another important note is that crosstalk was described between most of these pathways, either in normal or in pathological conditions (**Figure 1**). In particular, crosstalk between two main pathways regulated by PEBP1, namely Raf/MEK/ERK and PI3K/Akt/mTOR was found during brain ischemia/reperfusion. The findings suggested that Raf/MAPK/ERK1/2 signal pathway was inhibited by Akt via direct phosphorylation and inhibition at Raf-1 node during ischemia. During reperfusion, the ROS-dependent increase in phosphatase and tensin homolog activity in relieved ERK1/2 from inhibition of Akt (**108, Zhou 2015**). Crosstalk between PI3 Kinase/PDK1/Akt/ Rac1 and Ras/Raf/MEK/ERK pathways was described downstream of platelet-derived growth factor (PDGF) receptor in human lung cells. It was proved that PI3K, PDK1, Akt, and Rac1 could activate ERK in MSTO-211H cells and that MEK and ERK could otherwise activate Akt and Rac1 (**109, Niba**). A functional and complex interplay between Akt/mTOR and Ras/MAKP pathways was described to promote rapid hepatocarcinogenesis in a well-defined mouse model, *in vivo* (**110, Wang 2013**). It appeared that a relationship may exist between Raf/MEK/ERK and Wnt pathways. Thus, crosstalk was described between Raf/MEK/ERK and Wnt during cartilage regeneration (**111, Zhang 2014**). In a theoretical study, a mathematical model suggested that multiple feedback loops implying PEBP1, Snail and GSK3beta are involved in crosstalk between Raf/MEK/ERK and Wnt pathways to regulate Epithelial-Mesenchymal Transition in the generation of invasive tumor cells (**14, Shin 2010**). In a series of various cancer types, a crosstalk between MAPK and Hedgehog-GLI signaling pathways was also described (**112, Rovida**). Finally, there is compelling evidence that GSK3-Wnt-Lin28-let-7miRNA constitutes a central signaling axis in regulating proliferation and neurogenic potential of Müller glial cells in adult mammalian retina (**113, Yao**).

## 2.3- Regulation of PEBP1 expression

The downregulation of PEBP1 expression in cancer cells was correlated with development of metastases and resistance against chemical and immunological treatment. Thus, delineating the molecular mechanisms controlling the expression level of PEBP1was the focus a several studies. First, Snail was determined to be the repressor of PEBP1 transcription in metastatic prostate cancer cells (**114, Beach, 2008**). Then, a NFκB-snail-RKIP circuitry was described to regulate both the metastatic cascade and resistance to apoptosis by cytotoxic drugs (**115, Wu, 2009**) and finally, a NFκB/Snail/YY1/RKIP loop was proposed to regulate the resistance of cancer cells to chemo-immunological treatment (**48, Bonavida 2014**). More recently, the inverse correlation of expression between the metastasis suppressor PEBP1 and the metastasis inducer Yin Yang 1 (YY1) was studied (**116, Wottrich 2017**). Analyses of the molecular regulation of the expression patterns of PEBP1 and YY1 as well as epigenetic, post-transcriptional, and post-translational regulation revealed the existence of several effector mechanisms and crosstalk pathways, of which five pathways of relevance have been identified and analyzed. The five examined crosstalk pathways include the following loops: PEBP1/NF-κB/Snail/YY1, p38/MAPK/PEBP1/GSK3β/Snail/YY1, PEBP1/Smurf2/YY1/Snail, PEBP1/MAPK/Myc/Let-7/HMGA2/Snail/YY1, as well as PEBP1/GPCR/STAT3/miR-34/YY1. Each loop is comprised of multiple interactions and cascades that provide evidence for the negative regulation of PEBP1 expression by YY1 and *vice versa* (**116, Wottrich 2017).**

Contrary to the five loops that regulate PEBP1 expression in an inverse correlation of YY1, TWEAK (necrosis factor-like weak inducer of apoptosis) was described to regulate positively PEBP1 in colorectal cancer (**117, Lin 2012**). Tumor TWEAK is a member of the TNF (tumor necrosis factor) superfamily of cytokines that together are involved in many critical biological processes, including embryo development, organogenesis, tissue repair, and innate and adaptive immune responses. TWEAK has been also implicated in tumor development and progression. A cDNA microarray analysis identified a signature of 46 genes that were upregulated and downregulated after recombinant TWEAK treatment for 3 and 6 h. One of the most strongly upregulated genes was PEBP1. Additionally, functional assays indicated that TWEAK protein reduced the invasive ability of colon cancer cells (**117, Lin 2012**).

In human gastric cell line SGC-7901, the downregulation of PEBP1 expression was attributed to miR-224. It was concluded that miR-224 could negatively regulate the expression and biological characteristics of PEBP1, contributing to suppress the proliferation and invasion of gastric cells (**31, Liu 2014**).

Aberrant DNA methylation in the promoter regions of genes, which leads to inactivation of tumor suppressor and other cancer-related genes, is the most well-defined epigenetic hallmark in gastric cancer (**118, Qu**). The frequency of methylation in the promotor region of PEBP1 encoding gene in gastric cardia adenocarcinoma (GCA) tumor tissues was 62.1 %, i.e. significantly higher than that in corresponding normal tissues (4.1 %). The high promotor methylation level was associated with tumor stage, histological differentiation, depth of invasion, lymph node metastasis, distant metastasis and upper gastrointestinal cancers (UGIC) family history. These results suggested that epigenetic silencing of the PEBP1 promoter via hypermethylation may be one of the mechanisms for inactivation of PEBP1 in GCA, especially in patients with UGIC family history of North China (**85, Guo 2013**). The promoter hypermethylation of the gene encoding PEBP1 was also found to occur in Esophageal Squamous Cell Carcinoma (ESCC) and a close correlation was noted between PEBP1 gene methylation and the loss of mRNA and protein expression in ESCC specimens (**119, Guo 2012**).

Independently of the regulation of the gene encoding PEBP1, the degradation of the expressed PEBP1 itself was also reported. In neurons, it was demonstrated that PEBP1 is a substrate of cyclin-dependent kinase 5 (CDK5) and that the phosphorylation of PEBP1 at T42 causes the release of Raf-1. Moreover, T42 phosphorylation promoted the exposure and recognition of the target motif "KLYEQ" in the C-terminus of PEBP1 by chaperone Hsc70 and the subsequent degradation of PEBP1 via chaperone-mediated autophagy. Furthermore, in the brain sample of Parkinson's disease patients, CDK5-mediated phosphorylation and autophagy of PEBP1 were involved in the overactivation of the ERK/MAPK cascade, leading to S-phase reentry and neuronal loss (**45, Wen**).

## 2.4- PEBP1 is a target for treatment and prevention in diseases

The fight against cancer have resulted in two main research axes concerning PEBP1, namely 1) countering the resistance of cancer cells to drugs and 2) increasing the levels of PEBP1 expression in order to prevent metastases development.

**2.4.1- The resistance of cancer cells to drugs**

As early as 2004, studies suggested that PEBP1 may represent a novel effector of signal transduction pathways leading to apoptosis, and a prognostic marker of the pathogenesis of human cancer cells and tumors after treatment with clinically relevant chemotherapeutic drugs (**120, Chatterjee 2004**). Several papers indicated an effect of some anti-cancer drugs on PEBP1 expression level. Indeed, during treatment with rituximab, a chemosensitization of non-hodgkin lymphoma B cells was observed (**121, Jazirehi**). Rituximab was shown to sensitize non-Hodgkin's lymphoma (NHL) cell lines to chemotherapeutic drug-induced apoptosis. Rituximab treatment of NHL B cells significantly up-regulated PEBP1 expression, thus interrupting the ERK1/2 signaling pathway through the physical association between Raf-1 and PEBP1, which was concomitant with Bcl-xL downregulation (**121, Jazirehi**). Nitrite oxide was also found to sensitize prostate carcinoma cell lines to TRAIL-mediated apoptosis (**122, Huerta-Yepez**).

Later on, studies have implicated the role of Nitric Oxide (NO) in the regulation of tumor cell behavior and have shown that NO either promotes or inhibits tumorigenesis. Treatment of metastatic cancer cell lines with DETANONOate inhibited the epithelial mesenchymal transition (EMT) phenotype, and thus tumor metastasis, through the NF-κB/Snail/YY1/PEBP1 loop (**123, Bonavida 2011**). Nitric oxide have the dual function of both inhibiting metastasis and sensitizing tumor cells to chemotherapy and immunotherapy (**124, Bonavida 2008**). A gene regulatory mechanism that inhibits, in large part, the cell death apoptotic pathways is the result of several inter-related gene products that form a loop and consist of the NF-κB/Snail/YY1/PEBP1/PTEN. The expressions and activities of NFκB, Snail and YY1 are upregulated whereas the expressions and the activities of PEBP1 and PTEN are downregulated. The upregulated gene products are involved in the cell survival and growth and the expression of anti-apoptotic gene products. The downregulated gene products are involved in the inhibition of cell survival and anti-apoptotic gene products. Treatment of tumor cells with high levels of NO donors resulted in the downregulation of the expression of NFκB, Snail and YY1 while upregulation of PEBP1 and PTEN. Hence, treatment of resistant tumor cells with NO resulted in both the chemo and immunosensitization of tumor cells to apoptosis, *in vitro* and *in vivo* (**125, Bonavida 2015**).

More specifically, PhotoDynamic Therapy (PDT) against cancer has gained attention due to the successful outcome in some cancers, particularly those on the skin. NO revealed to be a modulator of PDT. By acting on the NFκB/Snail/PEBP1 survival/anti-apoptotic loop, NO can either stimulate or inhibit apoptosis. Low-dose PDT induces low NO levels by stimulating the anti-apoptotic nature of the above loop, whereas high-dose PDT stimulates high NO levels inhibiting the loop and activating apoptosis (**126, Rapozzi 2013**). The anti-cytoprotective effect of PDT-induced NO was observed at high NO levels, which inhibit the anti-apoptotic (prosurvival) NFκB and YY1 while inducing the pro-apoptotic (antisurvival) and metastasis suppressor PEBP1, resulting in significant anti-tumor activity. Thus, a successful application of NO in anticancer therapy requires control of its concentration in the target tissue. To address this issue Rapozzi et al. (**127, Rapozzi 2017**) proposed a bimolecular conjugate composed of a photosensitizer (Pheophorbide a) and a non-steroidal anti-androgen molecule capable of releasing NO under the exclusive control of light.

## 2.4.2- Physical and chemical factors influencing levels of PEBP1 expression

Independently of the role of NO on PEBP1 expression level, other physical and chemical factors were described to influence the levels of PEBP1 expression. Indeed, the expression level of PEBP1 is variable according to the tissue and to the cells investigated. Particularly, PEBP1 expression differs between healthy and cancer tumor cells, a low level of PEBP1 promoting metastasis formation and being considered as a factor of bad prognostic in cancer. Apart from these observations other variations in PEBP1 expression were described depending on physical or chemical treatment of various cell types.

The mechanical effect of flow was demonstrated on renal epithelial cells. In these cells, primary cilia act as mechanosensors in response to changes in luminal fluid flow. Primary cilia are nonmotile projections from polarized, highly differentiated epithelial cells (**128, Praetorius**). In the kidney, primary cilia are present on nearly all epithelial cells and extend from the apical surface into the tubular lumen, where they respond to luminal fluid flow. Recent studies suggest that primary cilia also serve an important role in the maintenance of cell differentiation. To determine the role of cilia bending in the mechanosensory function of cilia, a proteomic analysis of collecting duct cell lines with or without cilia was performed (**129, Sas**). A comparison between cells kept stationary or rotated to stimulate cilia bending revealed that expression of PEBP1 was significantly elevated in rotated cilia (+) cells. Downstream of PEBP1, expression of phosphorylated ERK was decreased only in cells that had cilia and were subjected to constant cilia bending. A PKC inhibitor significantly reduced the levels of PEBP1 in rotated cilia cells suggesting that PKC has a role in PEBP1 regulation in response to cilia bending (**129, Sas**) These findings indicated that it may not be just the presence of cilia but rather ciliary movement that is essential for the maintenance of cell differentiation and suppression of cell proliferation and that this occurs, at least in part, through regulation of PEBP1. In terms of polycystic kidney disease, loss of cilia and therefore sensitivity to flow may lead to reduced PEBP1 levels, activation of the MAPK pathway, and contribute to the formation of cysts (**129, Sas**).

Several drugs and natural products were demonstrated to act on the level of PEBP1 expression. In prostate cancer cells, a significant increase in PEBP1 expression was observed in androgen-independent PC3 cells after treatment with the anti-cancer drug 9-nitrocamptothecin. Similarly, exposure to other genotoxic stimuli such as etoposide or cisplatin resulted in the induction of PEBP1 expression in the prostate carcinoma cell line DU145. In addition, similar results were also found in drug-sensitive and drug-resistant breast carcinoma cell lines (**130, Odabaei**).

Proteomic analysis of rat hippocampus after chronic treatment with antidepressants, such as fluoxetine and venlafaxine, revealed up-regulation of PEBP1 (**131, Khawaja**). This was in accordance with the observation that PEBP1 decreased by stress exposure in rat brain could be restored by antidepressant treatment (**65, Kim 2007**). Chronic methamphetamine (a strong central nervous system stimulant) intake has been shown to induce a neuroinflammatory state leading to significant changes in brain functioning including behavioral changes. These changes can persist for years after drug use is discontinued and likely contribute to the risk of relapse. While methamphetamine intake was associated with reduced PEBP1 protein levels, treatment with ibudilast (an anti-inflammatory phosphodiesterase inhibitor) reversed this effect. Furthermore, Raf-1, MEK, and ERK expression levels were also attenuated by ibudilast treatment. It was suggested that PEBP1, given its synaptic localization and its role as a modulator of the ERK/MAPK pathway, could be a potential therapeutic target mediating drug-seeking behaviors associated with neuroinflammation (**132, Charntikov**).

Very interestingly, several natural products were reported to upregulate PEBP1. Naturally occurring polyphenols found in food sources provide health benefits. Chemically, all polyphenols have one or more hydroxylated aromatic rings that account for both structural and physicochemical properties, and allow their classification into several chemical classes including lignans, flavonoids, stilbenes, isoflavones and phenolic acid derivatives. All polyphenols have reducing properties; they can donate hydrogen to oxidized cellular constituent, and play a significant role against oxidative stress-related pathologies, like cardiovascular diseases, cancer and variety of neurodegenerative disorders. Flavonoids are the most abundant polyphenols consumed by peoples worldwide; therefore, many researchers focused on the effects of many flavonoids like resveratrol, quercetin, epigallocathechin-3-gallate (EGCG), rutin and curcumin, as a health promoting compounds in treatments of several diseases (**133, Hussain**). As PEBP1 appears to be a target for cancer therapy, modulation of PEBP1 expression using different natural agents was studied in cancer cells (**134, Farooqi**). American ginseng extracts tested on MCF-7 human breast cancer cells produced a decrease in phospho-MEK1/2 and -ERK1/2 and an increase in phospho-Raf-1. Furthermore, in these tests, American ginseng appeared to inhibit breast cancer cell proliferation by increasing the expression of PEBP1, resulting in inhibition of the MAPK pathway (**135, King**).

PEBP1 expression was considerably enhanced in pancreatic adenocarcinoma cells treated with epigallocatechin gallate (EGCG). EGCG induced PEBP1 upregulation *via* the inhibition of histone deacetylase (HDAC) activity which increased histone H3 expression and inhibited Snail expression, NFκB nuclear translocation, MMP-2 and -9 activity and Matrigel invasion in AsPC-1 cells. These results strongly suggest that EGCG regulates RKIP/ERK/NFκB and/or RKIP/NFκB/Snail and inhibits invasive metastasis (**7, Kim 2013**).

Dihydroartemisinin (DHA) inhibited the proliferation of Hela and Caski cervical cancer cells in a dose-dependent and time-dependent manner. Apoptosis was significantly upregulated in DHA-treated cells and this upregulation of apoptosis was associated with an increase in the levels of PEBP1 and a decrease in the levels of bcl-2. In addition, DHA inhibited tumor growth in nude mice bearing Hela or Caski tumors (**136, Hu**).

In human hepatocellular carcinoma cell line HepG-2, silibinin inhibited cell proliferation, matrix metalloproteinase 2 enzymatic activity, NO production and ERK 1/2 phosphorylation in a dose-dependent manner without exerting any cytotoxicity effect. An expressive increase in mRNA levels of PEBP1, Spred-1 and Spred-2 was coupled with a significant reduction in transcriptional levels of Hec1 and MMP-2. These data suggested that silibinin treatment could inhibit cell proliferation and invasive potential of HepG2 cells through inhibition of ERK 1/2 cascade both directly (through suppression of ERK 1/2 phosphorylation) and indirectly (through up-regulation of PEBP1, Spred-1 and Spred-2) (**137, Momeny**).

Today, didymin is a flavonoid that is studied for its action in cancer. The orally administered didymin, an active citrus flavonoid, causes regression of neuroblastoma xenografts in mouse models, without toxicity to non-malignant cells, neural tissues, or neural stem cells. In these models, PEBP1 that regulates the activation of the transcription factor MYCN, is transcriptionally upregulated by didymin, and appears to play a key role in the anti-neuroblastoma actions of didymin (**103, Singhal**). Didymin from Origanum vulgare (OVD) was assessed on human HepG2 liver carcinoma cells. The results showed that OVD induces apoptosis against of HepG2 cells through mitochondrial dysfunction and inactivation of the ERK/MAPK and PI3K/Akt pathways by up-regulating PEBP1. OVD significantly induced apoptosis and induced cell cycle arrest at G2/M phase by regulating cyclin B1, cyclin D1 and CDK4. The anti-proliferative and pro-apoptotic effects were associated with changes in the Bcl-2/Bax ratio and induction of caspase-mediated apoptosis. Moreover, OVD attenuated the mitochondrial membrane potential, accompanied by the release of cytochrome C (**43, Wei 2017**). Other studies showed that OVD treatment significantly reduced CYP2E1 activity, lipid peroxidation level, ROS generation, NO production and pro-inflammatory cytokines (such as TNF-α, IL-6 and IL-1β) in liver tissues and RAW 264.7 cells, but enhanced the hepatic antioxidative enzymes activities. OVD significantly inhibited the NFκB and MAPK pathways by up-regulating PEBP1 expression (**138, Huang**).

# 3- Molecular partners of PEBP1

## 3.1- Proteins interacting with PEBP1

Above, we have described the physical interaction between PEBP1 and kinases belonging to several signaling pathways. However, PEBP1 was suggested to interact with many other proteins including kinases, small GTPases, centrosomal proteins, cytoskeletal proteins, glycolytic enzymes, molecular chaperones, mitochondrial proteins as well as proteins implicated in protein biosynthesis and transport of proteins and vesicles. By describing these partners of PEBP1, we take the opportunity to underline the properties of these proteins as much as their functions correlate with those of PEBP1.

### 3.1.1- Phosphorylation of PEBP1 by several kinases

Apart from the phosphorylation of PEBP1 at S153 by PKC, several other phosphorylated sites were described on PEBP1, particularly residues S6, T42, S52, S54, S98, S99 and S109. As described below, different kinases are responsible for these phosphorylations, particularly CDK5, TBK1 and ERK. The phosphorylated sites have the effect of modifying and specifying the interactions between PEBP1 and its partners. The phosphorylation of residues S6 and S98 was observed in liver phosphoproteome (**139, Bian**). A quantitative atlas of mitotic phosphorylation indicated that the PEBP1 T42 was unphosphorylated in the G1-phase but was phosphorylated in M-phase while S52 and S54 were phosphorylated in G1 and M (**140, Dephoure**). In neurons, it was demonstrated that PEBP1 is a substrate of CDK5 leading to the phosphorylation of PEBP1 at T42 that promotes the recognition of PEBP1 by Hsc70 and the subsequent degradation of PEBP1 *via* chaperone-mediated autophagy (**45, Wen**). S99 was mentioned as a residue subjected to phosphorylation by ERK activation (**77, Al-Mulla 2013**). Finally, upon viral infection, PEBP1 was observed to be phosphorylated at S109 by TANK-binding kinase 1 (TBK1) in bone marrow-derived macrophages. Phosphorylation at S109 was observed only in PEBP1 purified from vesicular stomatitis virus-treated cells while that at S52 was observed in PEBP1 from both vesicular stomatitis-treated or untreated cells (**58, Gu 2016**).

### 3.1.2- Interaction of PEBP1 with LC3, syntenin, CEP290, Rab8A and relationships with Aurora B

Recently, PEBP1 was described to be implicated in important cellular mechanisms such as autophagy, primary cilium formation, mitotic spindle checkpoint, chromosome segregation and vesicular traffic by directly interacting with key proteins known to control these mechanisms.

### 3.1.2.1- LC3

Autophagy plays a critical role in maintaining cell homeostasis in response to various stressors, especially starvation or hypoxia. During autophagy, autophagosomes engulf cytoplasmic components, including cytosolic proteins and organelles. Concomitantly, a cytosolic form of microtubule-associated protein 1A/1B-light chain 3 (LC3) is conjugated to phosphatidylethanolamine to form LC3-phosphatidylethanolamine conjugate (LC3-II), which is recruited to autophagosomal membranes. Then, autophagosomes fuse with lysosomes to form autolysosomes, in such a way that intra-autophagosomal components are degraded by lysosomal hydrolases (**141, Tanida**). LC3 is conjugated with phosphatidylethanolamine in the membranes and regulates initiation of autophagy through interaction with many autophagy-related proteins possessing an LC3-interacting region (LIR) motif constituted by the amino acids WXXL (where X represents any of the 20 amino acids). PEBP1 was demonstrated to bind with LC3 by direct interaction through the LIR motif [55]WDGL[58] that is situated in an exposed external loop of PEBP1. PEBP1 was observed specifically bound to PE-unconjugated LC3 in cells, and mutation of the LIR motif disrupted its interaction with LC3 (**44, Noh 2016**). Interestingly, under starvation conditions, ablation of PEBP1 expression dramatically promoted the autophagic process. In contrast, overexpression of PEBP1 significantly inhibited starvation-induced autophagy by activating the Akt/mTORC1 signaling pathway. Thus, PEBP1-dependent suppression of autophagy was not associated with the MAPK pathway but through stimulation of the Akt-mTORC1 pathway and direct interaction with LC3. Furthermore, PEBP1 phosphorylation at S153 caused dissociation of LC3 from the PEBP1-LC3 complex for autophagy induction (**44, Noh**).

LC3 is considered to be involved in microtubule assembly/disassembly. To elucidate the mechanism by which LC3 facilitates epithelial ovarian cancer (EOC) cell migration and invasion under conditions of hypoxia, the effects of LC3B inhibition under hypoxic conditions on migration, invasion, and adhesion in HO8910PM and HO8910 EOC cell lines were investigated. The conclusion was that LC3B may promote the migration and invasion of EOC cells by affecting the cytoskeleton *via* the RhoA pathway (**142, Tang**).

### 3.1.2.2- Syntenin

Also called MDA-9, syntenin functions as a positive regulator of melanoma progression and metastasis. Tissue microarray and cell lines analyses showed that syntenin transcriptionally downregulated PEBP1. Furthermore, syntenin and PEBP1 physically interacted in a manner that correlated with a suppression of focal adhesion kinase (FAK) and c-Src phosphorylation, crucial steps necessary for syntenin to promote FAK/c-Src complex formation and initiate signaling cascades that drive the metastatic phenotype. Ectopic PEBP1 expression in melanoma cells overrode syntenin-mediated signaling, inhibiting cell invasion, anchorage-independent growth, and *in vivo* dissemination of tumor cells (**55, Das**). It should be pointed out here some functional properties of syntenin. Syntenin was initially identified as a molecule linking syndecan-mediated signaling to the cytoskeleton. Syntenin contains tandemly repeated PDZ domains that bind the cytoplasmic C-terminal domains of a variety of transmembrane proteins. Syntenin may also affect cytoskeletal-membrane organization (**143, Grootjans**), cell adhesion, cell migration, protein trafficking, and cell cycle progression in cancer cells (**144, Kashyap**). The protein was primarily localized to membrane-associated adherens junctions and focal adhesions but was also found at the endoplasmic reticulum and nucleus. Syntenin is also known to be implicated in the biogenesis of exosomes (**145, Friand; 146, Fares**).

### 3.1.2.3- CEP290

It was shown that PEBP1 interacts with the centrosomal protein of 290 kDa (CEP290) at the Connecting Cilium (Transition Zone) of Photoreceptors. PEBP1 prevented cilia formation and was associated with CEP290-mediated photoreceptor degeneration. In-frame deletion of 299 amino acids in CEP290 is associated with retinal degeneration in the *rd16* mouse, this domain more specifically interacted with PEBP1. The CEP290 mutant mouse rd16, revealed that CEP290-mediated photoreceptor degeneration is associated with aberrant accumulation of PEBP1. In *rd16* retina cells, there was 2.5-fold increase in the levels of endogenous PEBP1, as compared with the wild type (**75, Murga-Zamalloa**).

CEP290 is essential for the formation of the primary cilium, a small antenna-like projections of the cell membrane that plays a main role in the photoreceptors at the back of the retina, in brain, kidney, and many other organs. CEP290 is a microtubule and membrane binding protein that might serve as a structural link between the microtubule core of the cilium and the overlying ciliary membrane. Knocking down the level of CEP290 resulted in dramatic suppression of ciliogenesis in retinal pigment epithelial cells in culture (**147, Drivas).** CEP290 is also one of the centrosomal protein. The centrosome is an important cellular organelle which nucleates microtubules to form the cytoskeleton during interphase and the mitotic spindle during mitosis. Depletion of CEP290 caused a reduction of the astral spindle, leading to misorientation of the mitotic spindle. Microtubules polymerization also decreased in CEP290-depleted cells, indicating that CEP290 is involved in spindle nucleation (**148, Song**).

### 3.1.2.4- Rab8A

In ciliopathies, the study of photoreceptor degeneration mediated by Cep290 revealed the direct interaction of PEBP1 with Cep290 and an aberrant accumulation of PEBP1. Moreover, in zebrafish and cultured cells, ectopic accumulation of PEBP1 led to defective cilia formation, an effect mediated by its interaction with the ciliary GTPase Rab8A. Overexpression of PEBP1 revealed mislocalization of Rab8A in about 40% of the cells and it was shown that PEBP1 directly interacted preferentially with the GDP-locked mutant of Rab8A while little interaction was detected with the GTP-locked mutant of Rab8A (**75, Murga-Zamalloa**). Moreover, the GTPase Rab8A is a critical component of the protein trafficking machinery of photoreceptors and studies showed that overexpression of the GDP-locked variant of Rab8A resulted in mistrafficking of rhodopsin and photoreceptor degeneration (**149, Deretic**). Whereas the GDP-locked variant Rab8A disrupted cilia formation, expression of the GTP-locked variant Rab8A promoted ciliogenesis. As PEBP1 interacts preferably with the GDP-bound form of Rab8A in the retina and accumulation of PEBP1 due to Cep290 mutation is associated with retinopathy, it was proposed that the PEBP1-Rab8A (GDP) complex may need to dissociate for the release of Rab8A-GDP and subsequent conversion to Rab8A-GTP for appropriate ciliary transport (**75, Murga-Zamalloa**). Finally**,** Rab8 was recognized as a shared regulator of ciliogenesis and immune synapse assembly (**150, Patrussi**).

Rab8 was also described to be implicated in autophagy. Numerous Rab proteins have been shown to be involved in various stages of autophagy. Rab1, Rab5, Rab7, Rab9A, Rab11, Rab23, Rab32, and Rab33B participate in autophagosome formation, whereas Rab9 is required in non-canonical autophagy. Rab7, Rab8B, and Rab24 have a key role in autophagosome maturation. Rab8A and Rab8B are also known to be involved in autophagy (**151, Ao**).

Rab8 activation increased Rac1 activity, whereas its depletion activated RhoA, which led to reorganization of the actin cytoskeleton. Rab8 was also associated with focal adhesions, promoting their disassembly in a microtubule-dependent manner. This Rab8 effect involved calpain, MMP14 and Rho GTPases. Moreover, data revealed that Rab8 drives cell motility by mechanisms both dependent and independent of Rho GTPases, thereby regulating the establishment of cell polarity, turnover of focal adhesions and actin cytoskeleton rearrangements, thus determining the directionality of cell migration (**152, Bravo-Cordero**).

### 3.1.2.5- Aurora B

Chromosome segregation in mitosis is orchestrated by the dynamic interactions between the kinetochore and spindle microtubules. The microtubule depolymerase mitotic centromere-associated kinesin (MCAK) is a key regulator for an accurate kinetochore-microtubule attachment. It was reported that the phosphorylation of shugoshin (hSgo2) by Aurora B, promotes its association with MCAK, the recruitment of MCAK to inner-centromeres, and chromosome alignment (**153, Tanno**). Therefore, although Aurora B blocks MCAK activity at early mitosis for initial spindle assembly, it also recruits MCAK to the centromeres and activates polo-like kinase 1 (PLK1) to ensure subsequent activation of MCAK to timely correct the improper attachments of kinetochores to microtubules (**154, Shao 2015**).

Interestingly enough, PEBP1 was shown to associate with centrosomes and kinetochores, and to regulate Aurora B and the spindle checkpoint in mammalian cells. Kinetochores form the interface between microtubules of the mitotic spindle and chromosomes and regulate chromosome movements during mitosis. PEBP1 depletion altered localization and kinase activity of Aurora B at kinetochores. In addition, PEBP1 depletion led to a decrease in mitotic index, an acceleration in timing of the metaphase/anaphase transition and a defect in the spindle checkpoint in HeLa and H19-7 cells. However, a direct interaction between PEBP1 and Aurora B was not proved and it was suggested that Aurora B kinase activity could be inhibited directly by ERK1/2 phosphorylation of Aurora B or indirectly by mechanisms such as phosphorylation of other proteins required for recruitment of the kinase to the kinetochores (**155, Eves 2006**). In another note, by monitoring mitotic index and transit time from nuclear envelope breakdown to anaphase, it was demonstrated that PEBP1 depletion led to a defective spindle checkpoint and genomic instability, particularly in response to drugs that disrupt microtubule function (**39, Eves 2010**). In contrast to elevated MAP kinase signaling during the G1, S or G2 phases of the cell cycle that activates checkpoints and induces arrest or senescence, loss of PEBP1 during M phase leads to bypass of the spindle assembly checkpoint and the generation of chromosomal abnormalities. These results revealed a role for PEBP1 in ensuring the fidelity of chromosome segregation prior to cell division and raised the possibility that PEBP1 status in tumors could influence the efficacy of treatments, such as Taxol, that stimulate the Aurora B-dependent spindle assembly checkpoint (**156, Rosner**).

### 3.1.3- Interactome of PEBP1 in gastric cancer cell line SGC7901

Mapping the interactome of overexpressed PEBP1 in a gastric cancer cell line (SGC7901 cells) was of particular interest as a total of 72 PEBP1-interacting proteins were identified by MS/MS (**157, Gu 2013**). These proteins play roles in enzyme metabolism, molecular chaperoning, biological oxidation, cytoskeleton organization, signal transduction, and enzymolysis (**Table 1**). For the first time, interactions of PEBP1 with HSP90, 14-3-3epsilon and keratin 8 were found and were confirmed by Western blot analysis and co-immunoprecipitation. For data analysis, three databases, including Michigan Molecular Interactions (MiMI), Functional Linkage Network, and Predictome, were chosen to acquire the interaction network diagrams. Among the 72 proteins, 7 were identified as first level neighbors of PEBP1 with the three protein network diagrams and only 35 proteins were found through the MiMI analysis to have existing interactions with PEBP1 (first and second level neighbors). These 35 proteins were also consistently found to be closely related to PEBP1 with Functional Linkage Network and Predictome analysis. **Table 1** presents the 35 proteins classified according to their main cellular function, i.e. proteins related to cytoskeleton, protein biosynthesis, glycolytic enzymes, molecular chaperones, mitochondrial proteins and proteins/vesicles transport. We discuss here the main properties of these 35 close partners of PEBP1, starting with the 7 proteins considered as the closer partners of PEBP1 by the three diagram analysis. These 7 proteins are Ras GTPase-activating-like protein IQGAP, pyruvate kinase isozymes M1/M2, elongation factor 1-alpha 1, ATP synthase subunit alpha mitochondrial, isoleucyl-tRNA synthase cytoplasmic, voltage-dependent anion-selective channel protein1 (VDAC) and Clathrin heavy chain.

### 3.1.3.1- The seven proteins identified as first level neighbors of PEBP1

**The Ras GTPase-activating-like protein IQGAP1** is a scaffold protein involved in the regulation of various cellular processes ranging from organization of the actin cytoskeleton, transcription and cellular adhesion to cell cycle control. Over-expression of IQGAP1 was associated with increased migration and invasion in the human breast epithelial cancer cell line MCF-7 (**158, White; 159, Jadeski**). IQGAP1 may also be involved in the deregulation of proliferation and differentiation through its modulation of the MEK/ERK pathway (**160, Brown**). **Elongation factor 1-alpha 1** (eEF1A1) is an isoform of the alpha subunit of the elongation factor-1 complex, which is responsible for the enzymatic delivery of aminoacyl tRNAs to the ribosome. eEF1A1 is expressed in most cells and induces HSP70 during heat shock. The eEF1A1 is a GTPase that couples the hydrolysis of GTP to GDP with the delivery of aminoacyl tRNAs to the ribosome during protein translation. eEF1A also has translation-independent roles in embryogenesis, senescence, oncogenic transformation, cell proliferation, apoptosis, cytoskeletal organization and protein degradation (**161, Blanch**). Furthermore, it participates in several processes including mitotic apparatus formation and signal transduction (**162, Sasikumar**).

eEF1a1 colocalizes with filamentous actin and putatively binds actin and microtubules at synapses to modulate the cytoskeleton. In breast and prostate cancer samples, the upregulation of eEF1A was reported (**163, Pecorari; 164, Liu 2010**).

**Isoleucyl-tRNA synthase cytoplasmic** catalyzes the aminoacylation of the tRNA by isoleucine. The three branched amino acids (BCAAs) Leucine, Isoleucine and Valine, are among the nine essential amino acids for humans and provide several metabolic and physiologic roles. Metabolically, BCAAs promote protein synthesis and turnover, signaling pathways and metabolism of glucose. BCAAs are a marker of insulin resistance (**165, Lynch**), as insulin resistance increases the rate of appearance of BCAAs and is linked to reduced expression of mitochondrial BCAA catabolic enzymes. Circulating levels of BCAAs tend to be increased in individuals with obesity and are associated with worse metabolic health and future insulin resistance or type 2 diabetes mellitus (**166, Newgard**).

**Pyruvate kinase isozymes M1/M2** are glycolytic enzymes. They catalyze the last step within glycolysis, the dephosphorylation of phosphoenolpyruvate to pyruvate, and are responsible for net ATP production within the glycolytic sequence. The final step of glycolysis is highly regulated and irreversible because pyruvate is a crucial intermediate building block for further metabolic pathways. Once pyruvate kinase synthesizes pyruvate, pyruvate either enters the tricarboxylic acid (TCA) cycle for further production of ATP under aerobic conditions, or is reduced to lactate under anaerobic conditions (**167, Gupta**). In contrast to mitochondrial respiration, energy regeneration by pyruvate kinase is independent from oxygen supply and allows survival of the organs under hypoxic conditions often found in solid tumors (**168, Vaupel**). PKM2 is a cytosolic enzyme that is associated with other glycolytic enzymes within a so-called glycolytic enzyme complex. Particularly, *in vitro* studies on breast cancer cell line have revealed that glycolytic enzymes may form a multienzymatic complex consisting of glyceraldehyde 3-phosphate dehydrogenase, phosphoglycerate kinase, enolase, pyruvate kinase, adenylate kinase, nucleoside diphosphate kinase type A (NDPK A), hexokinase, phosphoglycerate mutase (PGAM), lactate dehydrogenase, and A-Raf kinase (**169, Kowalski**). Pyruvate kinase inhibits taxol-induced tubulin polymerization into microtubules and partially disassembles taxol-stabilized microtubules into less sedimentable oligomers leading to the appearance of tubulin in the supernatant fractions. Thus, these data suggest that pyruvate kinase may display multiple regulatory functions as a glycolytic control enzyme and as a modulator of microtubule dynamics (**170, Vertessy**).

**ATP synthase subunit alpha mitochondrial** is a subunit of mitochondrial $F_1F_o$ ATP synthase which catalyzes ATP synthesis from ADP and inorganic phosphate. The formation of ATP from ADP and $P_i$ is energetically unfavorable and would normally proceed in the reverse direction. In order to drive this reaction forward, ATPase couples ATP synthesis during cellular respiration to an electrochemical gradient created by the difference in proton ($H^+$) concentration across the mitochondrial inner membrane (**171, Bernardi; 172, Bernardi**). Subunit alpha does not bear the catalytic high-affinity ATP-binding sites, however, mutations affecting the ATP synthase subunit alpha gene cause combined oxidative phosphorylation deficiency, a mitochondrial disorder characterized by intrauterine growth retardation, microcephaly, hypotonia, pulmonary hypertension, encephalopathy, and heart failure (**173, Janer**).

Microtubule cytoskeleton is reformed during apoptosis, forming a cortical structure beneath plasma membrane, which plays an important role in preserving cell morphology and plasma membrane integrity. Apoptotic microtubule network (AMN) was organized in apoptotic cells with high ATP levels and hyperpolarized mitochondria and, on the contrary, was dismantled in apoptotic cells with low ATP levels and mitochondrial depolarization. Furthermore, high ATP levels and mitochondria polarization collapse after oligomycin treatment in apoptotic cells suggested that ATP synthase works in "reverse" mode during apoptosis (**174, Oropesa**).
**The voltage-dependent anion-selective channel protein 1 (VDAC1)** forms an ion channel in the outer mitochondrial membrane (OMM) and allows ATP and other small metabolites to diffuse out of the mitochondria into the cytoplasm. VDAC1 is involved in cell metabolism, allowing regulation of the tricarboxylic acid (TCA) cycle and, by extension, reactive oxygen species (ROS) production (**175, Huang 2014**). Besides metabolic permeation, VDAC1 also acts as a scaffold for proteins such as hexokinase that can in turn regulate metabolism (**176, Reina**). There are three VDAC isoforms (VDAC1, 2 and 3), of the three isoforms, VDAC1 is the main calcium ion transport channel and the most abundantly transcribed, its dysfunction is implicated in cancer (**177, Gao**), Parkinson's disease (**178, Chu**) and Alzheimer disease (**179, Smilansky**). In Parkinson disease, VDAC1 increases calcium ion levels within the mitochondria, resulting in increased mitochondrial permeability, disrupted mitochondrial membrane potential, elevated ROS production, cell death, and neuronal degeneration. (**178, Chu**). VDAC1 has been implicated in cancer through its interactions with the anti-apoptotic family of proteins, Bcl-2 proteins, particularly Bcl-xl, and Mcl-1, which are overexpressed during cancer. These two Bcl-2 proteins interact with VDAC1 to regulate calcium ion transport across the OMM and, ultimately, ROS production (**182, Shoshan-Barmatz**). VDAC1 has also been described in association with apoptotic processes and as interacting with cytoskeletal proteins (**176, Reina**).
**Clathrin heavy chain** is a subunit of the clathrin coat that is composed of three heavy chains and three light chains forming a triskelion (**181, Halebian**). Clathrin is a major protein component of the cytoplasmic face of intracellular organelles, called coated vesicles and coated pits. These specialized organelles are involved in the intracellular trafficking of receptors and endocytosis of a variety of macromolecules (**182, McMahon**). In a cell, clathrin triskelion binds to an adaptor protein that has bound membrane, linking one of its three feet to the membrane at a time; many proteins involved in clathrin-mediated endocytosis have been described (**183, Pucadyil**). Formation of clathrin-coated vesicles is shut down in cells undergoing mitosis. During mitosis, clathrin binds to the spindle apparatus, in complex with TACC3 and CKAP5. Clathrin aids in the congression of chromosomes by stabilizing kinetochore fibers of the mitotic spindle (**184, Royle**). Otherwise, clathrin-mediated endocytosis (CME) involves the recruitment of numerous proteins to sites on the plasma membrane with prescribed timing to mediate specific stages of the process. Actin assembly generally preceded and then promoted dynamin-2 recruitment during the late phases of CME. Thus, precise temporal and quantitative regulation of the dynamin-2 recruitment is influenced by actin polymerization (**185, Grassart**).

### 3.1.3.2- The 28 other proteins neighbors of PEBP1

Among the 28 remaining proteins presented in Table1, nine are components of cytoskeleton or are associated to filaments and microtubules formation. Eukaryotic cells contain three main kinds of cytoskeletal filaments: microtubules, actin microfilaments and intermediate filaments. (**186, Fletcher**). The cytoskeleton is involved in a multitude of functions. It gives the cell its shape and mechanical resistance to deformation and allows cells to migrate (**186, Fletcher**). It is implicated in many cell signaling pathways, in endocytosis, in chromosomes segregation during cellular division, and intracellular transport. Furthermore, it forms specialized structures, such as cilia (**187, Pedersen**), lamellepodia (**188, Guo**) and podosomes (**189, Veillat**).

#### 3.1.3.2.1- Components of cytoskeleton

**Tubulin-beta** is a major constituent of microtubules that are made up of polymerised α- and β-tubulin dimers (**190, Nogales**). Microtubules are typically nucleated and organized by dedicated organelles called microtubule-organizing centers (MTOCs). However, microtubules can be nucleated at cilia at their base termed basal bodies and the microtubule can dynamically switch between growing and shrinking phases in this region (**191, Pearson**). Tubulin dimers can bind two molecules of GTP, one of which can be hydrolyzed subsequent to assembly. During polymerization, the tubulin dimers are in the GTP-bound state. The GTP bound to α-tubulin is stable and it plays a structural function in this bound state. However, the GTP bound to β-tubulin may be hydrolyzed to GDP shortly after assembly and GDP-tubulin is more prone to depolymerization (**192, Piedra**).

**Actin cytoplasmic 1**, is also known as beta-actin, it is one of six different actin isoforms which have been identified in humans and one of the two nonmuscle cytoskeletal actins. The most abundant isoactin in many nonmuscle cells including myeloid and neuronal cells is β-actin (**193, Cheever**). Cytoplasmic β- actin is essential for cell migration, cell shape maintenance, mitosis and intracellular transport processes. Cellular studies showed that the β-isoform is preferentially recruited into cellular protrusions (**194, Svitkina**), stress fibers, circular bundles and at cell–cell contacts (**195, Collinet**). The rapid cytoskeletal rearrangements observed for these structures appear to be linked to the highly dynamic turnover of actin filaments made from β-actin (**196, Hundt**). It is noted that actin is an ATPase. Each molecule of actin is bound to a molecule of ATP or ADP that is associated with an $Mg^{2+}$ cation. The most commonly found forms of actin, compared to all the possible combinations, are ATP-G-Actin (globular actin, free monomer) and ADP-F-actin (filamentous, part of a linear polymer microfilament) (**197, Graceffa**).

**Plectin, F-actin-capping protein and vinculin** are actin-binding proteins. Plectin acts as a link between the three main components of the cytoskeleton: actin microfilaments, microtubules and intermediate filaments (**198, Svitkina**). Plectin links the cytoskeleton to junctions found in the plasma membrane that structurally connect different cells. By holding these different networks together, plectin plays an important role in maintaining the mechanical integrity and viscoelastic properties of tissues (**199, Wiche**).

The F-actin capping protein binds in a calcium-independent manner to the fast-growing ends of actin filaments (barbed end), thereby blocking the exchange of subunits at these ends. In its absence, cells progressively accumulate actin filaments and eventually die (**200, Delalle**).

Vinculin is a membrane-cytoskeletal protein in focal adhesion plaques that is involved in linkage of integrin adhesion molecules to the actin cytoskeleton. Vinculin is a cytoskeletal protein associated with cell-cell and cell-matrix junctions, where it is thought to function as one of several interacting proteins involved in anchoring F-actin to the membrane. Thus it plays a role in cellular adhesion and migration (**201, Dumbauld**).

Finally, **Myosin-9** (Myosin, heavy chain 9, non-muscle) is also a multifunctional protein, it binds PKC epsilon, affects actin cytoskeleton (**202, Yuan**) and plays a role in cytokinesis and secretion (**203, Atanasova**).

Three proteins are involved in ***intermediate filaments***: keratin type1 cytoskeletal 18, keratin type 2 cytoskeletal 8 and vimentin. Intermediate filaments are an important component of the cellular cytoskeleton, they play a role in the organization of microtubules and microfilaments, the regulation of nuclear structure and activity, the control of cell cycle and the regulation of signal transduction pathways. It was shown that intermediate filaments associate with key proteins of the vesicular membrane transport machinery. In particular, the contribution of intermediate filaments to the endocytic pathway was strongly suggested (**204, Margiotta**). It is to note that unlike the other cytoskeleton components, intermediate filament (IF) monomers do not have any enzymatic activity. Their assembly in mature insoluble filaments and their disassembly in soluble components (from tetramers to monomers) are regulated by phosphorylation/dephosphorylation cycles (**204, Margiotta).**

More particularly, when phosphorylated, **keratin type 1 cytoskeletal 18** plays a role in filament reorganization during apoptosis (**205, Weerasinghe**). **Keratin type 2 cytoskeletal 8** is present in certain forms of cancer and can be used to differentiate lobular carcinoma of the breast from ductal carcinoma of the breast (**206, Kongara**). It hampers the biogenesis of mutated CFTR (Cystic fibrosis transmembrane conductance regulator) and its insertion at the plasma membrane (**207, Premchandar**) and at the interleukin-6 (IL-6)-mediated barrier protection (**208, Wang 2007). Vimentin** is the major cytoskeletal component of mesenchymal cells, it is often used as a marker of mesenchymally-derived cells or cells undergoing an epithelial-to-mesenchymal transition (EMT) during both normal development and metastatic progression. Vimentin plays a significant role in supporting and anchoring the position of the organelles in the cytosol. Vimentin is attached to the nucleus, endoplasmic reticulum, and mitochondria, either laterally or terminally (**209, Katsumoto**). Moreover, it is accepted that vimentin is the cytoskeletal component responsible for maintaining cell integrity (**210, Goldman**).

### 3.1.3.2.2- protein biosynthesis

Apart from Elongation factor-1-alpha1 and Isoleucyl-tRNA synthase described above as being among the seven first level partners of PEBP1, the interactome of overexpressed PEBP1 in gastric cancer cells, revealed 6 proteins associated with ribosomes and implicated in protein biosynthesis. These are valyl-tRNA synthase, elongation factor 1-gamma, elongation factor 2, initiation factor 4A, guanine nucleotide-binding protein subunit beta-2-like 1 and 40S ribosomal protein S3.

Among them **valyl-tRNA synthase** catalyzes the aminoacylation of the tRNA by valine. Like Isoleucine and Leucine, valine is a branched chain amino-acid (BCAA)**.** As Isoleucyl-tRNA synthase**,** valyl-tRNA synthase promotes protein synthesis and turnover, but is also implicated in metabolism of glucose (**211, Higuchi),** insulin resistance (**212, Mahendran**) and signaling pathways (**213, Mattick).**

**Elongation factor 1-gamma** is a subunit of the elongation factor-1 complex, which is responsible for the enzymatic delivery of aminoacyl tRNAs to the ribosome. This subunit contains an N-terminal glutathione transferase domain, which may be involved in regulating the assembly of multisubunit complexes containing this elongation factor and aminoacyl-tRNA synthetases (**214, Achilonu**).

**The eukaryotic Elongation factor 2 (eEF2)** is a member of the GTP-binding translation elongation factor family. This protein is an essential factor for protein synthesis. It promotes the GTP-dependent translocation of the ribosome. eEF2 is completely inactivated by EF-2 kinase phosphorylation which is regulated by p90$^{RSK1}$ and p70 S6 kinase (**215, Hizli).** Findings suggest that eEF2K likely contributes to neuronal function by regulating the synthesis of microtubule-related proteins. Modulation of the elongation phase of protein synthesis is important for numerous physiological processes in neurons. The synthesis of microtubule-related proteins is up-regulated by inhibition of elongation suggesting that translation elongation is a key regulator of cytoskeletal dynamics in neurons (**216, Kenney**).

**The eukaryotic initiation factor-4A (eIF4A)** family consists of 3 closely related proteins EIF4A1, EIF4A2, and EIF4A3. These factors are required for the binding of mRNA to 40S ribosomal subunits. In addition these proteins are helicases that function to unwind double-stranded RNA (**217, Marintchev**). These proteins stabilize the formation of the functional ribosome around the start codon and also provide regulatory mechanisms in translation initiation.

**Guanine nucleotide-binding protein subunit beta-2-like 1** was alternatively named Receptor of activated protein kinase C 1 (RACK1). RACK1 is a major component of translating ribosomes, which harbor significant amounts of PKC and which recruits and binds a variety of signaling molecules such as integrins, proto-oncogene tyrosine-protein kinase Src (pp60c-src) and focal adhesion kinase (**218, Nielsen**).

**40S ribosomal protein S3** is a ribosomal protein that is a component of the 40S subunit, in which it forms part of the domain where translation is initiated. Studies of the mouse and rat proteins have demonstrated that 40S ribosomal protein S3 has an extraribosomal role as an endonuclease involved in the repair of mitochondrial DNA damage (**219, Kim 2013**). Successive structural re-arrangements in ribosomal protein S3 promote maturation of the 40S ribosomal subunit.

### 3.1.3.2.3- *Glycolytic enzymes*

In addition to pyruvate kinase described above, five glycolytic enzymes were suggested to interact with PEBP1 according to the interactome analysis. These are glyceraldehyde-3-phosphate dehydrogenase, L-lactate dehydrogenase chain A and chain B, alpha enolase and fructose-bisphosphate aldolase A.

**Glyceraldehyde-3-phosphate dehydrogenase (GAPDH)** catalyzes the sixth step of glycolysis corresponding to the conversion of glyceraldehyde-3- phosphate in D-glycerate 1,3 bisphosphate. In addition to this long established metabolic function, it was implicated in several non-metabolic processes, including transcription activation, initiation of apoptosis (**220, Tarze**), endoplasmic reticulum to Golgi vesicle shuttling, and fast axonal or axoplasmic transport (**221, Zala**). In sperm, a testis-specific isoenzyme named GAPDHS is expressed (**222, Margaryan**).

**Lactate dehydrogenase (LDH) chain A and chain B** catalyze the interconversion of pyruvate and lactate with concomitant interconversion of NADH and NAD$^+$. Pyruvate, the final product of glycolysis is converted to lactate when oxygen is absent or in short supply, and the reverse reaction is performed during the Cori cycle in the liver. It has also been shown that LDHA plays an important role in the development, invasion and metastasis of malignancies (**223, Miao**). Using the cancer cell line MDA-MB-435, it was shown that knockdown of LDHA results in elevated mitochondrial ROS production and a concomitant decrease in cell proliferation and motility. Thus, in cancer, aerobic glycolysis and reduced mitochondrial ROS production create an environment conducive to cytoskeletal remodeling and high cell motility (**224, Arseneault**).

**Alpha enolase (ENO1)** is expressed in most tissues, it catalyzes the conversion of 2-phosphoglycerate to phosphoenolpyruvate. In addition, it functions as a structural lens protein (tau-crystallin) in the monomeric form. Alternative splicing of the gene encoding alpha enolase results in a shorter isoform that was shown to bind to the *c-myc* promoter and functions as a tumor suppressor. ENO1 also plays a role in other functions, including a cell surface receptor for plasminogen on pathogens, an oxidative stress protein, a heat shock protein (HSP48), and a binding partner of cytoskeletal (**225, Díaz-Ramos**). It was demonstrated that enolase isoforms interact with microtubules during muscle satellite cell differentiation suggesting that binding of enolase to microtubules could contribute to the regulation of the dynamism of the cytoskeletal filaments known to occur during the transition from myoblasts to myotubes (**226, Keller**). ENO1 overexpression was associated with multiple tumors (**227, Capello**). In many of these tumors, ENO1 promoted cell proliferation by regulating the PI3K/Akt signaling pathway and induced tumorigenesis by activating plasminogen (**228, Fu**). Moreover, ENO1 is expressed on the tumor cell surface during pathological conditions such as inflammation, autoimmunity, and malignancy. Its role as a plasminogen receptor leads to extracellular matrix degradation and cancer invasion (**229, Hsiao**).

**Fructose-bisphosphate aldolase A** is an enzyme catalyzing a reversible reaction that splits fructose 1,6-bisphosphate, into the triose phosphates dihydroxyacetone phosphate (DHAP) and glyceraldehyde 3-phosphate (G3P). Aldolase A is found in the developing embryo and is produced in even greater amounts in adult muscle. Signal transduction *via* PI3K allows for the physical dissociation of aldolase from F-actin into the cytoplasm where it is active. This simple, biophysical mechanism of activating aldolase through recruitment from the cytoskeleton is a rapid and efficient way for cells to increase metabolic flux. Redistribution of aldolase in response to PI3K signaling achieves coordination of cytoskeletal dynamics and glycolysis, while avoiding the time- and energy-consuming path of transcriptional activation and biosynthesis of new enzyme molecules (**230, Hu**).

In the end, cancer cells enhance their glycolysis, producing lactate, even in the presence of oxygen. Glycolysis is a series of ten metabolic reactions catalyzed by enzymes whose expression is most often increased in tumor cells. HKII and phosphoglucose isomerase (PGI) have mainly an anti-apoptotic effect; PGI and glyceraldehyde-3-phosphate dehydrogenase activate survival pathways; phosphofructokinase 1 and triose phosphate isomerase participate in cell cycle activation; aldolase promotes epithelial mesenchymal transition; PKM2 enhances various nuclear effects such as transcription and stabilization. Thus, the multiple non-glycolytic roles of glycolytic enzymes appeared essential for promoting survival, proliferation, chemoresistance and dissemination of cancer cells (**231, Lincet**). Interestingly, the phosphoinositide 3-kinase (PI3K) pathway regulates multiple steps in glucose metabolism and also cytoskeletal functions, such as cell movement and attachment. PI3K directly coordinates glycolysis with cytoskeletal dynamics in an Akt-independent manner. Growth factors or insulin stimulate the PI3K-dependent activation of Rac, leading to disruption of the actin cytoskeleton, release of filamentous actin-bound aldolase A, and an increase in aldolase activity. These results point toward a master regulatory function of PI3K that integrates an epithelial cell metabolism and its shape and function, coordinating glycolysis with the energy-intensive dynamics of actin remodeling (**230, Hu 2016**).

### 3.1.3.2.3- Molecular chaperones

Five molecular chaperones were identified by the interactome analysis of gastric cancer cells overexpressing PEBP1: HSP-90 beta, HScognate 71 kDa, HSP-90 alpha, Stress-70 protein mitochondrial, 60kDa HSP mitochondrial.

**Hsp90 (heat shock protein 90) alpha and beta** are cytoplasmic homologues, human HSP90 alpha sharing 85% sequence identity with HSP90 beta. A proteomic differential display analysis between the regressive cancer cell line QR-32 and the inflammatory cell-promoting progressive cancer cell line QRsP-11 of murine fibrosarcoma, indicated that calreticulin precursor, tropomyosin 1 alpha chain, annexin A5, Prx II, heat shock protein (HSP)90-alpha, HSP90-beta and PEBP1, were overexpressed (**232, Hayashi**). The two HSP90 isoforms are chaperone proteins that assist other proteins to fold properly, stabilize proteins against heat stress, and aid in protein degradation. Hsp90 alpha and beta contain three functional domains, the ATP-binding, protein-binding, and dimerizing domain, each of which playing a crucial role in the function of the protein. The region of the protein near the N-terminus has a high-affinity ATP-binding site formed by a sizable cleft. When given a suitable protein substrate, Hsp90 cleaves the ATP into ADP and $P_i$ (**233, Didenko**). The protein-binding region of Hsp90 is located toward the C-terminus of the amino sequence. The Hsp90 conformation is an open ATP-bound state or a closed ADP-bound state. Thus, when a bound substrate is in place, the energy-releasing ATP hydrolysis by the ATPase near the N-terminal domain forces conformational changes that clamp the Hsp90 down onto the substrate (**234, Pearl**). The ability of Hsp90 to clamp onto proteins allows it to perform several functions including assisting folding, preventing aggregation, and facilitating transport. In particular, HSP90 stabilizes the 26S proteasome which enables the cell to degrade unwanted proteins and modulates tumor cell apoptosis through effects on Akt, TNFR and NFκB (**235, Graner**).

**HScognate 71 kDa** also known as **Hsc70** is a member of the heat shock protein 70 family. It is a chaperone protein which facilitates the proper folding of newly translated and misfolded proteins, as well as stabilizes or degrades mutant proteins (**236, Mayer**). HScognate 71 kDa contributes to biological processes including signal transduction, apoptosis, protein homeostasis, cell growth and differentiation and it was associated with an extensive number of cancers, neurodegenerative diseases, cell senescence and aging (**237, Liu 2012**). As HSP90, it has a C-terminal protein substrate-binding domain and an N-terminal ATP-binding domain. HScognate 71 kDa is known to localize to the cytoplasm and lysosome, where it participates in chaperone-mediated autophagy by aiding the unfolding and translocation of substrate proteins across the membrane into the lysosomal lumen (**238, Xie**). HScognate 71 kDa additionally serves as a positive regulator of cell cycle transition and carcinogenesis. For example, HScognate 71 kDa regulates the nuclear accumulation of cyclin D1, which is a key player in G1 to S phase cell cycle transition (**239, Hatakeyama**). Interestingly, another function of HScognate 71 kDa is as an ATPase in the disassembly of clathrin-coated vesicles during transport of membrane components through the cell (**240, Cho**). HScognate 71 kDa plays a protective role in neurodegenerative diseases, such as Alzheimer's disease, Parkinson's disease, Huntington's disease, spinocerebellar ataxias, aging and cell senescence, as observed in centenarians subjected to heat shock challenge (**241, Jinwal**). In neurons, cyclin-dependent kinase 5 (CDK5) phosphorylates PEBP1 at T42 promoting the exposure and recognition of the target motif "KLYEQ" in the C-terminus of PEBP1 by HScognate 71 kDa and the subsequent degradation of PEBP1 *via* chaperone-mediated autophagy. In the brain sample of Parkinson's disease patients, CDK5-mediated autophagy of PEBP1 is involved in the over-activation of the ERK cascade, leading to S-phase reentry and neuronal loss (**45, Wen**).

**Stress-70 protein mitochondrial** also named mitochondrial 70kDa HSP and HSPA9, is also known as mortalin. 70kDa HSP is a heat-shock cognate protein; it plays a role in the control of cell proliferation and may also act as a chaperone. 70kDa HSP is primarily localized to the mitochondria but is also found in the endoplasmic reticulum, plasma membrane and cytoplasmic vesicles. 70kDa HSP has been attributed many cellular functions, including energy generation, proliferation, stress response and carcinogenesis (**242, Wu**). 70kDa HSP regulates the functions of the tumor suppressor protein p53 and plays important roles in stress response and maintenance of the mitochondria and endoplasmic reticulum. Furthermore, 70kDa HSP appears to have roles in membrane trafficking and viral release regulation. More recently, mutations in 70kDa HSP gene have been found in Parkinson´s disease patients. It was shown that 70kDa HSP expression differs in hippocampus of patients with Alzheimer´s disease and could regulate the β-amyloid toxicity pathway (**243, Flachbartova**).

**60kDa HSP mitochondrial** (HSP60) is a mitochondrial chaperonin that is typically held responsible for the transportation and refolding of proteins from the cytoplasm into the mitochondrial matrix. HSP60 functions as a chaperonin to assist in folding linear amino acid chains into their respective three-dimensional structure. Further studies have linked HSP60 to diabetes, stress response, cancer and certain types of immunological disorders (**244, Capello**). Under normal physiological conditions, HSP60 is a 60 kilodalton oligomer composed of monomers that form a complex arranged as two stacked heptameric rings (**245, Cheng**). This double ring structure forms a large central cavity in which the unfolded protein binds *via* hydrophobic interactions. Each subunit of HSP60 has three domains: the apical domain, the equatorial domain, and the intermediate domain. The equatorial domain contains the binding site for ATP. The intermediate domain induces a conformational change when ATP is bound allowing for an alternation between the hydrophilic and hydrophobic substrate binding sites (**246, Okamoto**). HSP60 possesses two main responsibilities with respect to mitochondrial protein transport. It functions to catalyze the folding of proteins destined for the mitochondrial matrix and maintains protein in an unfolded state for transport across the inner membrane of the mitochondria. Subsequent changes in ATP concentrations hydrolyze the bonds between the protein and HSP60 which signals the protein to exit the mitochondria (**247, Koll**).

In sum, PEBP1 interacts with cytoplasmic and mitochondrial HSPs implicated in proteins and vesicular transport. All these HSPs are implicated in numerous processes and several diseases such as cancer and neurodegenerescence. They all display an ATP binding site and an important ATPase activity.

### 3.1.3.2.4- Transport of proteins

In addition to Clathrin heavy chain, importin subunit beta 1 is a partner of PEBP1 that is implicated in transport of proteins and vesicles.

**Importin subunit beta 1** is a type of karyopherin that transports proteins into the nucleus by binding to specific recognition sequences, called nuclear localization sequences (NLS) (**248, Lott**). Importin has two subunits, importin α and importin β. Members of the importin-β family can bind and transport cargo by themselves, or can form heterodimers with importin-α. As part of a heterodimer, importin-β mediates interactions with the pore complex. The NLS-Importin α-Importin β trimer dissociates after binding to Ran GTP inside the nucleus, with the two importin proteins being recycled to the cytoplasm for further use (**249, Lowe**). Importins are vital regulatory proteins during the processes of gametogenesis and embryogenesis. (**250, Miyamoto**).

### 3.1.3.2.5- Others

Finally, Table I indicates two proteins that are considered as implicated in signal transduction: Annexin A1 and 14-3-3 protein epsilon.

**Annexin A1** belongs to the annexin family of $Ca^{2+}$-dependent phospholipid-binding proteins that are preferentially located on the cytosolic face of the plasma membrane. Annexin 1 has been involved in a broad range of molecular and cellular processes, including anti-inflammatory signalling, kinase activities in signal transduction, maintenance of cytoskeleton and extracellular matrix integrity, tissue growth, apoptosis, and differentiation (**251, Bizzarro 2012**). Annexin A1 protein has an phospholipase A2 inhibitory activity. Increasing the synthesis and function of annexin A1 by Glucocorticoids treats diseases caused by an overactive immune system, including allergies, asthma, autoimmune diseases, and sepsis (**252, Perretti**). Upon induction by anti-inflammatory drugs, annexin A1 inhibits the NFκB signal transduction pathway, which is exploited by cancerous cells to proliferate and avoid apoptosis (**253, Bist**). Exposure of MCF-7 breast cancer cells to high physiological levels of estrogen lead to an up-regulation of annexin A1 expression, suggesting that Annexin A1 may act as a tumor suppressor and modulates the proliferative functions of estrogens (**254, Ang**). In PCa cells, annexin A1 knockdown is able to inhibit epithelial to mesenchymal transition (EMT) and to reduce FAK and metalloproteases (MMP)-2/9 expression (**255, Bizzarro 2015**). In nasopharyngeal carcinoma (NPC), the biological functions of Annexin A1 can inhibit the *in vitro* invasive ability of NPC cells through S100A9 and Vimentin interaction (**256, Xiao 2017**).

**14-3-3 protein epsilon** is an adapter protein implicated in the regulation of a large spectrum of both general and specialized signaling pathways. It binds to a multitude of functionally diverse signaling proteins, including kinases, phosphatases and transmembrane receptors, usually by recognition of a phosphoserine or phosphothreonine motif. Binding generally results in the modulation of the activity of the binding partner. A significant reduction of 14-3-3ε protein expression was observed in gastric cancer samples compared to their matched non-neoplastic tissue (**257, Leal**). Neuron navigator 2 (NAV2) is required to induce neurite outgrowth in human neuroblastoma cells. Knockdown of 14-3-3ε leads to a decrease in neurite outgrowth, similar to the elongation defects observed when NAV2 is depleted or mutated. The discovery of an interaction between NAV2 and 14-3-3ε provided insight into the mechanism by which NAV2 and 14-3-3 participate in promoting cell migration and neuronal elongation (**258, Marzinke**).

In conclusion, the majority of proteins found as partners of PEBP1 are multifunctional proteins. They are implicated in numerous cellular processes and are related to membrane and cytoskeletal organization. The presence among them of glycolytic enzymes and protein biosynthesis factors supports the assumption that cytoskeleton changes are coordinated with energy and synthesis of newly translated proteins. There is a role for PEBP1 for modulating the interactions between them depending on the cellular status, signaling pathways and concentration of different substrates (particularly ATP) as well as natural or synthesized products coming from the cell external microenvironment.

## 3.2- S-nitrosylated proteins in Alzheimer's disease

PEBP1 was identified among 45 S-nitrosylted proteins in Alzheimer's disease (AD) (**259, Zahid**). Autopsied brain specimens were from well-characterized AD patients. Differential S-nitrosylation of proteins was studied in human brain hippocampus, *substantia nigra* and cortex. Apart PEBP1, eleven S-nitrosylated proteins are part of the 35 protein partners described above (Table I): these are tubulin beta chain, actin cytoplasmic 1, the six glycolytic enzymes (pyruvate kinases isozymes M1/M2, Glyceraldehyde-3-phosphate dehydrogenase, L-lactate dehydrogenase A and B chains, alpha-enolase and fructose-bisphosphate aldolase A), HScognate 71 kDa, VDAC1 and 14-3-3 epsilon (**259, Zahid**). S-nitrosylation is a result of the covalent binding of NO with cysteine residues of target proteins with the formation of nitrosothiols (SNOs); it is a reversible post-translational modification. SNOs extensively modify protein function and play a key role in the pathology of multiple neurodegenerative diseases (**260, Nakamura**). On the basis of all these findings, the authors postulated that NO-induced PTM may alter the normal structural/functional protein pattern as well as protein–protein interaction mechanisms that lead to AD-associated conditions (**259, Zahid**).

## 3.3- Interaction between PEBP1 and small ligands

Amazingly, PEBP1 was described to bind various natural or synthesized small ligands. The most part of these ligands binds the small cavity present near the surface of PEBP1. Thus, this small cavity is considered the binding site of PEBP1, it plays a role in binding small ligands but also in interacting with other proteins, particularly with protein kinases. **Figure 2** indicates the PEBP1 binding site and the edges of a furrow interacting with kinases of signaling pathways modulated by PEBP1 (**23, Serre; 24, Banfield; 261, Martin; 262, Park; 263, Shemon**).

**3.3.1- ligands of natural origin**: **organic anions, nucleotides, phosphotyrosine, phospholipids and model membranes**

As soon as the bovine brain PEBP1 was discovered, its association with membrane lipids was observed, and it was needed to delipidate the brain extract to purify PEBP1 (**1, Bernier**). Then, the purified PEBP1 was shown to bind organic anions and phosphatidylethanolamine (**5, Bernier**). The crystal structure of PEBP1 from bovine brain revealed a novel structural family of proteins. The PEBP1 tridimensional structure contains no internal hydrophobic pocket, but a small cavity close to the protein surface that displayed a high affinity for anions, such as phosphate and acetate, and also phosphorylethanolamine (the polar head group of phosphatidylethanolamine) (**23, Serre**). The ligand binding site for anions was confirmed by the crystal structure of human PEBP1 which was identical to the bovine PEBP1 structure and revealed a molecule of cacodylate set up in the cavity of PEBP1 (**24, Banfield**).

The affinity of PEBP1 toward model membranes such as large unilamellar vesicles (LUV), small unilamellar vesicles (SUV) and phospholipid monolayers was measured. It was shown that the whole PEBP1 was able to bind membranes (**264, Vallée**). The crystal structure of PEBP1 indicated that the N-terminal and C-terminal parts of PEBP1 were exposed at the protein surface and could be involved in the interaction with membranes. Therefore, two peptides were synthetized, one of them was a 12-amino-acid peptide corresponding to the N-terminal extremity (residues 1-12) of PEBP1, whereas the other one was a 19-amino-acid corresponding to the C-terminal helix (residues 168-186) of PEBP1. Both peptides bound model membranes and displayed the same behavior as whole PEBP1, indicating that they could participate in the binding of the whole protein to membranes. Overall, the results suggested that PEBP1 may directly interact with negatively charged membrane microdomains in living cells (**264, Vallée**).

Otherwise, the sequence determination of bovine PEBP1 (**2, Schoentgen**) allowed to build a molecular model through hydrophobic cluster analysis and molecular modelling. At this time on, the analysis of the model and comparison with other proteins present in the Protein Data Bank (PDB) revealed a putative ATP binding site (**265, Schoentgen**). Thus, the affinity of nucleotides (anionic components) toward PEBP1 was experimentally tested. The more effective nucleotides to bind PEBP1, presented in decreasing order, were: FMN > GTP > GDP > GMP > FAD > ATP > NADP > CTP > UTP > ADP (**6, Bucquoy**). Later on, the binding of FMN and GTP was confirmed by nuclear magnetic resonance (NMR) and mass spectrometry under conditions similar to what is found in cells regarding pH, salt concentration and temperature (**266, Tavel**). Another study described the binding of phosphotyrosine to the ligand site of PEBP1. The structure of PEBP1 bound to pTyr showed that the PEBP1 ligand binding pocket molded complementary to the shape of the pTyr side chain (**267, Simister**). As studies have previously indicated that phosphorylation of the $S^{338}SYY^{341}$ region of Raf-1 enhances binding to PEBP1 leading to suppression of MEK activation (**262, Park**), a mechanism was proposed for PEBP1/Raf-1 interaction in which PEBP1 regulates Raf-1 activity *via* direct steric inhibition of the $Tyr^{341}$ region (**267, Simister**). NMR titration studies further confirmed that the pocket region is the binding site for the tri-phosphorylated peptide $^{331}RPRGQRDpSpSYpYWEIEASEV^{349}$, corresponding to the minimal region 331-349 of Raf-1 required for rat PEBP1 binding. In addition, the results indicated that residues in the vicinity of the pocket rather than those within the pocket were required for interaction with Raf-1 (**266, Tavel**). These results raised the question whether PEBP1 could be classified as a *bona fide* pTyr binder but it was concluded that direct, physiological or structural, evidence for complex formation is required before PEBP1 can be classified as a *bona fide* pTyr binder (**268, Kaneko**). Finally, by Affinity Elution in Tandem Hydrophobic Interaction Chromatography, it was demonstrated that PEBP1 is an ATP-binding protein and that ATP attenuates the interaction between PEBP1 and Raf-1. In parallel, short-term ATP depletion in cultured HEK293 cells augments interaction between PEBP1 and Raf-1, resulting in decreased activation of the downstream ERK signaling. Therefore, the ATP-binding function renders the inhibition of Raf-1 by PEBP1 modulated by cellular ATP concentrations. These data shed light on how energy levels affect the propagation of cellular signaling (**269, Huang 2016**).

### 3.3.2- Binding of PEBP1 to synthetic inhibitors, drugs and prodrugs

Besides the binding of intracellular small ligands as nucleotides and phospholipids, PEBP1 was also found to bind small synthetic molecules. By identifying protein "interactors" of PF-3717842, a high-affinity phosphodiesterase-5 (PDE5) inhibitor, from rat testis tissue lysate, PDE5 and a few other PDEs were strongly and specifically enriched. However, in addition to these expected affinity-enriched proteins, PEBP2 (the PEBP isoform specific of testis) was found as a putative binder to the PDE5 inhibitor. By using recombinant forms of the related murine mPEBP2, mPEBP1 and human hPEBP1 it was confirmed that they all can bind strongly to immobilized as well as soluble PF-3717842. These results suggested that the synthetic PDE5 inhibitor PF-3717842 might form a platform to synthesize enhanced binders/inhibitors of the family of PEBP proteins (**8, Dadvar**).

Locostatin, another interesting product was fortunately discovered as interacting with PEBP1. Locostatin, a cell migration inhibitor, was demonstrated to bind four proteins in MDCK cell lysates: PEBP1, glutathione S-transferase omega 1-1, aldehyde dehydrogenase 1A1 and prolyl oligopeptidase (**270, Zhu**). The possibility for PEBP1 to be the main target of locostatin appeared very strong: PEBP1 is one of only four proteins specifically bound by locostatin and the only one of the four whose binding profile for locostatin analogs correlates with the effects of the analogs on cell locomotion (**271, Bement**). The protective effect of PEBP1 against drug inhibitor of migration suggested a new role for RKIP in potentially sequestering toxic compounds that may have deleterious effects on cells (**263, Shemon**). It was shown that locostatin disrupted interactions of PEBP1 with Raf-1 and also with GRK2. In contrast, locostatin did not disrupt binding of PEBP1 with IKKα and TAK1. These results implied that different proteins interact with different regions of PEBP1. After binding PEBP1, part of locostatin is slowly hydrolyzed, leaving a smaller PEBP1-butyrate adduct. The residue alkylated by locostatin was identified as His86, a conserved residue of the PEBP1 ligand-binding pocket (**272, Beshir**). The use of locostatin as an inhibitor of PEBP1 revealed the important roles of PEBP1 in the regulation of cell adhesion, as it positively controls cell-substratum adhesion while negatively controls cell-cell adhesion (**54, Mc Henry**). In the same way, locostatin revealed that PEBP1 induces extracellular matrix (ECM) in uterine leiomyoma and myometrial cells**.** Uterine Leiomyoma is a tumor with fibrotic characteristics and accumulates excessive amounts of ECM mainly composed of collagens, fibronectin, and versican. Locostatin treatment resulted in the activation of the MAPK signal pathway (ERK phosphorylation), reduced GSK3β expression and reduced ECM components. Moreover, the inhibition of PEBP1 by locostatin impaired cell proliferation and migration in both leiomyoma and myometrial cells (**273, Janjusevic**).

As mentioned above, it was observed that after binding PEBP1, part of locostatin is slowly hydrolyzed, leaving a smaller PEBP1-butyrate adduct (**272, Beshir**). Another product, namely Prasugrel, was also described to be hydrolysed by PEBP1. Prasugrel is a thienopyridine antiplatelet prodrug that undergoes rapid hydrolysis *in vivo* to a thiolactone metabolite by human

carboxylesterase-2 (hCE2) during gastrointestinal absorption. The thiolactone metabolite is further converted to a pharmacologically active metabolite by cytochrome P450 isoforms. PEBP1 was identified as another protein capable of hydrolyzing prasugrel to its thiolactone metabolite, and to play a significant role with hCE2 in prasugrel bioactivation in human intestine. The estimated contributions of these two hydrolyzing enzymes to the prasugrel hydrolysis were approximately 40% for PEBP1 and 60% for hCE2. Thus, PEBP1 was described for the first time as a hydrolase involved in drug metabolism (**274, Kazui**).

# 4- Cellular processes

PEBP1 is known to be implicated in several cellular processes, these are summarized in **Table II**. Apart from signaling pathways discussed in generalities paragraph, the other main processes modulated by PEBP1 are presented below.

## 4.1- Roles of PEBP1 in membrane remodeling and association with vesicles

Very soon, at the time of the identification of PEBP1, its localization along the plasma membrane of oligodendrocytes in brain (**275, Roussel**) and its implication in spermatogenesis (**276, Seddiqi**) led us to assume that PEBP1 may be implicated in membrane remodeling. Since then, PEBP1 was described to participate in several processes affecting directly cellular membrane features. Among them, we specifically discuss below sperm capacitation, receptors internalization, exosomes formation and neuronal synapse.

### 4.1.1- Sperm capacitation

The role of PEBP1 in the sperm capacitation was demonstrated by the study of a PEBP-1(-/-) mice. Testicular germ cells express high levels of PEBP1 mRNA during spermatogenesis, essentially from late pachytene spermatocytes to elongate spermatids. Sperm from PEBP1-deficient mice were precociously capacitated compared with their wild-type counterparts (**49, Moffit**). Data from mating experiments indicate decreased reproduction rates between crosses of PEBP1(-/-) male mice and either heterozygous or PEBP1(-/-) females. Furthermore, PEBP1 immunolocalization of epididymal sperm supported transfer of the protein from germ cell cytoplasm to the sperm *via* the cytoplasmic droplet during epididymal transport (**49, Moffit**). Interestingly, the homozygous mice developed dramatic olfaction deficits in the first year of life (**75, Theroux**).

### 4.1.2- Internalization of receptors and desensitization

GRK2 is well-known to phosphorylate and desensitize G protein-coupled receptors (GPCRs) (**277, Robinson**). It was shown that PEBP1 switched from Raf-1 to GRK2 when phosphorylated by protein kinase C (PKC) on serine 153. Thus, after stimulation of G-protein-coupled receptors (GPCRs), $pS^{153}$PEBP1 which associated with GRK-2, blocked its activity (**88, Lorenz**).

#### 4.1.2.1- Opioid receptors

Among GPCRs, opioid receptors were several times mentioned to be in close relationships with PEBP1, without PEBP1 being itself an opioid receptor. First, rat brain PEBP1 was eluted from a morphine-derived affinity column suggesting that although not an opioid receptor itself, PEBP1 may be associated with such a receptor (**4, Grandy**). Later on, it was shown that inside the secretory granules of bovine primary chromaffin cells, PEBP1 directly interacted with endogenous morphine glucuronides (M3G and M6G) but not with morphine. Thus, it was suggested that, during stress, M6G-PEBP1 complexes may be released into circulation to target peripheral opioid receptors (**278, Atmanene**). Even later, the role of PEBP1 in μ-opioid receptor-mediated ERK activation was investigated in Chinese hamster ovary/μ cells and SH-SY5Y cells, as well as in human embryonic kidney HEK293 cells. It appeared that PEBP1 did not have a significant role in μ-opioid receptor-mediated regulation of ERK. In the same study, only the activation of Gq-coupled adrenergic α1A receptors induced an upregulated phosphorylation of PEBP1, which was protein kinase C activity dependent (**279, Bian**).

Contrary to μ-opioid receptor (MOR), PEBP1 was proved to impact δ-opioid receptor (DOR) activity. While most clinical opioids target MOR, those that target the DOR class also demonstrated analgesic efficacy. However, DOR are analgesically incompetent in the absence of inflammation. It was reported that GRK2 naively associated with plasma membrane DOR in peripheral sensory neurons to inhibit analgesic agonist efficacy. This interaction prevented optimal Gβ subunit association with the receptor, thereby reducing DOR activity. Under the effect of the inflammatory mediator bradykinin, GRK2 moved away from DOR and onto PEBP1. Then, PKC-dependent PEBP1 phosphorylation induced GRK2 sequestration, restoring DOR functionality in sensory neurons (**90, Brackley**).

### 4.1.2.2- Beta-adrenergic receptors

$pS^{153}$PEBP1 inhibited agonist-induced internalization of β2-adrenergic receptors of HEK-293 cells and of cardiomyocytes in rat osteosarcoma ROS 17/2.8 cells. It was concluded that inhibition of GRK-2 by $pS^{153}$PEBP1 enhanced signaling and decreased receptor internalization (**88, Lorentz**). Recently, PEBP1 exhibited a combination of favorable effects for heart failure patients as the endogenous PEBP1 was found to activate adrenergic receptors (βAR) signaling of the heart. The first favorable effect was that, the gain/increase of cardiac contractile efficiency by the activation of Gs signaling/ β1AR led to functional recovery of the heart. Secondly, the protection of the heart under sympathetic stress from exaggerated β1AR downstream signaling included protection from apoptosis and proarrhythmic adverse effects via β2AR activation. In a more general way, it was observed that in the context of heart failure, PEBP1 comprises several favorable characteristic effects on calcium cycling, calcium sensitivity and G-protein recruitment to beta adrenergic receptors (**280, Lorentz, 2017).** In cardiomyocytes, the binding of GRK2 to the βγ subunits of activated G proteins (Gβγ), regulates the activation of beta-adrenergic receptor (βAR) signaling. PEBP1 binds GRK2 in the amino-terminal domain and the downregulation of PEBP1 inhibits beta-adrenergic signaling and contractile activity (**281, Sorriento 2016**).

### 4.1.2.3- Vasoactive intestinal peptide receptors

In the gastrointestinal tract, the neuropeptides vasoactive intestinal peptide (VIP) and pituitary adenylyl cyclase-activating peptide (PACAP) were colocalized in a subset of myenteric neurons that innervate the smooth muscle, and their release was functionally linked to smooth muscle relaxation (**282, Ulrich**). The biological actions of VIP and PACAP were mediated by a family of G protein-coupled receptors, which were designated as $VPAC_1$, $VPAC_2$, and $PAC_1$ receptors (**282, Ulrich**). In gastrointestinal smooth muscle cells, $VPAC_2$ receptor desensitization is exclusively mediated by GRK2. Muscle cells, stimulated with acetylcholine in the presence of the M2 receptor antagonist methoctramine (i.e., activation of M3 receptors) displayed inhibition of VPAC$_2$ receptor phosphorylation, internalization, and desensitization (**283, Huang 2007**). The inhibition was mediated by PKC, derived from Muscarinic receptor M3 activation. In these conditions, acetylcholine induced phosphorylation of PEBP1, increased PEBP1-GRK2 association, decreased PEBP1-Raf-1 association, and stimulated ERK1/2 activity (**283, Huang, 2007**).

### 4.1.2.4- Mitochondrial localization of GRK2

It is now validated that GRK2 regulates several intracellular signaling pathways not only through the phosphorylation of specific substrates but also through protein–protein interactions independently from its catalytic activity (**283, Sorriento 2016**). Moreover, it should be noted, that apart from heart failure, GRK2 was found, in fibroblasts, to increase ATP cellular content by enhancing mitochondrial biogenesis; also, it antagonized ATP loss after hypoxia/reperfusion and was detected to localize in the mitochondrial outer membrane (**284, Fusco**). In macrophages, in response to lipopolysaccharide, GRK2 accumulated in mitochondria, increasing biogenesis. The overexpression of the carboxy-terminal domain of GRK2 (βARK-ct), known to displace GRK2 from plasma membranes, induced earlier localization of GRK2 to mitochondria in response to lipopolysaccharide leading to increased mt-DNA transcription and reduced ROS production and cytokine expression (**285, Sorriento 2013**). Interestingly, the effect of PEBP1 on mitochondria via GRK2 is in accordance with its effect on Raf1. Indeed, during hepatitis B virus X-mediated hepatocarcinogenesis, the PEBP1 downregulation induced the translocation of Raf-1 into mitochondria by increase of the free Raf-1 level in the cytosol. Subsequently, the enhanced mitochondrial Raf-1 imparted the anti-apoptotic effect on cells (**286, Kim SY 2011**).

### 4.1.3- PEBP1 is associated with several types of vesicles

### 4.1.3.1- Exosomes

Numerous studies have described the association of PEBP1 with various types of vesicles. Among them exosomes were often reported to contain PEBP1. Exosomes are defined as vesicles in the range of 30–100 nm, they are released by several types of cells and have been found in numerous body fluids (**287, H Rashed**). Exosomes can contain microRNAs, mRNAs, DNA fragments, and proteins, which are shuttled from a donor cell to recipient cells. Notably, tumor cells have been shown to produce and secrete exosomes in greater numbers than normal cells. Recently, as it appeared that the tumor microenvironment influences cancer progression and metastasis, the role of exosomes in cancer was extensively studied (**288, Brinton; 289, Kahlert; 290, Zha**). PEBP1 was previously described as a released or a potentially secreted protein in several organs and cells. It was particularly found in testicular interstitial fluid (**291, Turner**), in epididyme (**292, Dacheux**), in chromaffin cell secretory granules (**293, Goumon**) and in non-small cell lung cancer (**294, Huang, 2006**). In addition, proteome profiling of exosomes have revealed the presence of PEBP1 in ovarian cancer ascites (**295, Shender**), in prostate cancer (**296, Kharaziha**) and in lipid rafts of prostasomes that are membranous vesicles released by the prostate gland epithelial cells into seminal fluid (**297, Dubois**). PEBP1 was also identified in soluble secretome of primary cancer cell line SW480 and lymph node metastatic colorectal cancer cells SW620 (**298, Ji**). Finally, independently of cancer, PEBP1 was also detected in extracellular vesicles from tissues of the central nervous system (**299, Gallart-Palau**).

### 4.1.3.2- Other cellular vesicles

A quantitative proteomics analysis revealed that PEBP1 belongs to a subset of the endoplasmic reticulum that contributes to the phagosome. Phagosomes, by killing and degrading pathogens for antigen presentation, are organelles implicated in key aspects of innate and adaptive immunity. It was well established that phagosomes consist of membranes from the plasma membrane, endosomes, lysosomes, and also from the endoplasmic reticulum (ER) membrane which contributes to about 20% of the early phagosome proteome. Only a subset of ER proteins (among which PEBP1) was recruited to the phagosome, suggesting that a specific subdomain of the ER might be involved in phagocytosis (**300, Campbell-Valois**).

PEBP1 was also detected in neuromelanin granules isolated from human brain *substantia nigra* (**301, Plum**). The *substantia nigra* is a basal ganglia structure located in the midbrain that plays an important role in reward and movement. Neuromelanin granules could form during aging by the fusion of many neuromelanin-laden endosomes and lysosomes in multivesicular bodies (**301, Plum**).

## 4.2- PEBP1 in neuronal synapse

PEBP1 was shown to have several neurophysiological roles in mammalian brain, including neural development and differentiation, circadian clock and synaptic plasticity with implications in certain pathologies such as Alzheimer's disease and brain cancer (**62, Ling**). Particularly, the implication of PEBP1 in neuronal plasticity was demonstrated by its role in cerebellar long-term synaptic depression (LTD) and in hippocampus long-term potentiation (LTP) in transgenic mice. LTD and LTP are two important mechanisms for learning and memory. Moreover, the level of PEBP1 expression was found to be regulated by neuronal activity and also by alcohol, nicotine and psychoactive drugs.

### 4.2.1- Regulation of PEBP1 expression in rat hippocampus

PEBP1 expression was found to be upregulated by excitation of hippocampal pyramidal neurons and to be dependent on muscarinic cholinergic and glutamatergic receptors. HCNP is derived from the N-terminal region of PEBP1 (11 N-terminal residues). It was originally isolated from the hippocampus of young rats and enhances acetylcholine synthesis in rat medial septal nucleus *in vitro*. The precursor of HCNP is PEBP1, thus the role and the expression of PEBP1 was specifically studied in hippocampus. The highest expression of PEBP1 mRNA is in hippocampal pyramidal neurons that are a type of multipolar neurons found in areas of the brain including the cerebral cortex, the hippocampus, and the amygdala. Pyramidal neurons are the primary excitation units of the mammalian prefontal cortex and the corticospinal tract. In an *in vitro* rat hippocampal slice, PEBP1 mRNA expression was regulated by neuronal activity (**304, Iwase**). Selective inhibition with pharmacological agents revealed that stimulation of ionotropic glutamate receptors, as well as activation of either muscarinic receptors or metabotropic glutamate receptors, may act cooperatively to modulate PEBP1 mRNA expression. Finally, the main results suggested that in the rat hippocampus: (1) membrane depolarization increases PEBP1 mRNA expression; (2) glutamatergic and cholinergic neurotransmitter pathways contribute to a reciprocal regulation of PEBP1 expression; (3) L-type $Ca^{2+}$ channels and NMDA receptors are both important for the regulation of PEBP1 expression; and (4) the activity-dependent and constitutive expressions of PEBP1 may be regulated by different routes, involving calcium influx via L-type Ca(2+) channels and NMDA receptors (**304, Iwase**).

### 4.2.2- Alcohol, nicotine and psychoactive drugs modify the expression of PEBP1.
#### *4.2.2.1- Alcohol, innate differences*

Two-dimensional gel electrophoresis (2-DE) was used to separate protein samples solubilized from the nucleus accumbens and hippocampus of alcohol-naïve, adult, male inbred alcohol-preferring (iP) and alcohol-nonpreferring (iNP) rats (**303, Witzmann**). Several protein spots were excised from the gel, destained, digested with trypsin, and analyzed by mass spectrometry. The results indicated that selective breeding for disparate alcohol drinking behaviors produced innate alterations in the expression of several proteins that could influence neuronal function within the nucleus accumbens and hippocampus. Particularly, PEBP1 was more abundant in hippocampus and nucleus accumbens of alcohol-non preferring rats than alcohol-prefering rats (1.2 fold in hippocampus and 1.4 fold in nucleus accumbens) (**303, Witzmann**).

### 4.2.2.2- Alcohol, long-term exposure

It was shown that long-term ethanol exposure impairs neuronal differentiation of human neuroblastoma cells. Chronic ethanol interfered with the development of a neuronal network consisting of cell clusters and neuritic bundles. Furthermore, neuronal and synaptic markers were reduced, indicating impaired neuronal differentiation. Brain-derived neutrophic factor (BDNF)-mediated activation of the ERK cascade was found to be continuously impaired by ethanol. However, BDNF also activated PKC signaling and PEBP1 phosphorylation, which finally led to ERK activation (**304, Hellmann**). In this study, PEBP1, acting as a signaling switch at the merge of the PKC cascade and the Raf/MEK/ERK cascade, was associated with neuronal differentiation and was found to be significantly reduced in ethanol treatment. Moreover, PKC expression itself was even more strongly reduced. Reduced PEBP1 and PKC levels and subsequently reduced positive feedback on ERK activation provided an explanation for the striking effects of long-term ethanol exposure on BDNF signal transduction and neuronal differentiation (**304, Hellmann**).

### 4.2.2.3- Nicotine exposure

Proteomic analysis of the hippocampus and cortex of mice treated for 6 months with nicotine was performed (**305, Matsuura**). The expression level of PEBP1 was not changed in the hippocampus but was increased by 1.2 fold in the cortex (**305, Matsuura**).

### 4.2.2.4- Psychoactive drugs

A study focused on determining changes in the proteome that occurred after long-term exposure of rats to two clinically effective antidepressant drugs, fluoxetine (a selective serotonin reuptake inhibitor) and venlafaxine (a dual serotonin/norepinephrine reuptake inhibitor). The hippocampus was chosen for analysis, taking into consideration the involvement of this important anatomic region in clinical depression (**306, Nestler**). Either venlafaxine or fluoxetine was administered systemically to adult rats for 2 weeks, and protein patterns from rat hippocampal cytosolic extracts were compared. Thirty-three proteins were modulated by both drug treatments compared to controls. Among them, the upregulation of PEBP1 was observed: 2.0 fold with venlafaxine and 1.9 with fluoxetine treatment. It was concluded that PEBP1 would contribute to the long-term maturation, extension of neurites, and integration of developing granular cells within the existing hippocampal circuitry (**306, Nestler**). Parallel studies to investigate further the effects of venlafaxine and fluoxetine on adult hippocampal neurogenesis *in vivo* revealed a significant drug-induced increase in the proliferation rate and long-term survivability of progenitor stem cells located in the subgranular zone. The upregulation of PEBP1 was 2.0 with venlafaxine and 1.0 with fluoxetine. These data suggested that the two drugs share wide-ranging proteome changes within the hippocampal formation, beyond serotonin/norepinephrine neurotransmission and that this may reflect long-term functional adaptations required for antidepressant activity (**131, Khawaja**).

### 4.2.2.5- Neurotoxics

Methylmercury (MeHg) is a neurotoxic and ubiquitous environmental contaminant for which the brain is considered to be the main target organ (National Research Council, 2000). In order to investigate molecular mechanisms underlying MeHg neurotoxicity as well as to identify protein biomarker candidates for environmental monitoring, changes in the brain proteome of Atlantic cod (Gadus morhua) following MeHg exposure were studied. The doses were given in two injections, half of the dose on the first day and the second half after 1 week, and the total exposure period lasted 2 weeks. The level of 71 proteins to be 20% or more significantly altered following MeHg exposure was observed, and 40 of these proteins were successfully identified. PEBP1 expression was 1.6 fold increased in brain of cods submitted to MeHg exposure. Several of the proteomic alterations seem to reflect well-established molecular targets and mechanisms of MeHg-induced neurotoxicity in mammals, such as mitochondrial dysfunction, oxidative stress, altered calcium homeostasis, tubulin/disruption of microtubules and calpain-cleavage. In addition, indications of glycolysis as a target for MeHg were strengthened by the demonstration of up-regulation of four glycolytic enzymes (fructose 1,6-diphosphate aldolase, glyceraldehyde-3-phosphate dehydrogenase, phosphoglycerate kinase, phosphoglycerate kinase). These results were considered indicative of the molecular mechanisms underlying MeHg neurotoxicity and defense responses (**307, Berg**).

Asiaticoside was found to counteract the neurotoxic effects of the natural pesticide rotenone. Asiaticoside (AS), a triterpenoid saponin isolated from the Indian medicinal herb *Centella asiatica* is known to exert a neuroprotective effect by attenuating the neurobehavioral, neurochemical and pathological changes in animal models. Rotenone (ROT), a commonly used natural pesticide extracted from the roots of tropical plants, such as *Derris elliptica*, is able to freely cross the blood brain barrier, plasma membrane and mitochondrial membranes thus inducing Parkinson disease-like symptoms (**308, Gopi**) . Expression of PEBP1 after ROT-infusion in rat model was significantly increased when compared to control group. However, increased PEBP1 expression was not observed on AS administration in ROT received rats. This implies that AS treatment may normalize hypercholinergic stimulation, and also maintain normal synaptic function by preserving PEBP1. It was concluded that ROT-infused generation of ROS and hence oxidation of cytosolic dopamine might be altered by AS not only by its antioxidant potential but also by its effect on proteins that need phosphoinositides for their function and localization, thus ensuring an improved cytoprotection, phosphoinositides assisted vesicular trafficking and synaptic integrity (**308, Gopi**).

### 4.2.3- Localisation of PEBP1 with CRMP2 at presynaptic terminals

It was demonstrated that the amount of collapsin response mediator protein-2 (CRMP-2) is increased in hippocampus of PEBP1 transgenic mice. PEBP1 co-localizes with CRMP-2 at presynaptic terminals and PEBP1 overexpression increases synaptophysin levels. These findings suggested that PEBP1, in association with CRMP-2, plays an important role in presynaptic function in the hippocampus (**56, Kato**). It was demonstrated that PEBP1 predominantly colocalized and associated with unphosphorylated and/or pSer-522-CRMP-2 at presynaptic terminals in the hippocampus. Interestingly, PEBP1 did not associate with pThr-509/514-CRMP-2, which is primarily localized at postsynaptic terminals. These results suggested that PEBP1, in association with unphosphorylated and/or pSer522-CRMP-2, plays an important role in presynaptic function in the mature hippocampus (**57, Mizuno**).

### 4.2.4- PEBP1 is implicated in LTD and LTP
#### *4.2.4.1- Long-term depression (LTD)*

LTD is an activity-dependent reduction in the efficacy of neuronal synapses lasting hours or longer following a long patterned stimulus. LTD is thought to result mainly from a decrease in postsynaptic receptor density, although a decrease in presynaptic neurotransmitter release may also play a role. Cerebellar LTD has been hypothesized to be important for motor learning. It was demonstrated that a positive feedback loop, including protein kinase C (PKC) and mitogen-activated protein kinase (MAPK), is required for the gradual expression of cerebellar long-term depression (LTD). PKC and MAPK are mutually activated in this loop. To test the hypothesis that PEBP1, a substrate of PKC, was implicated in this loop, Yamamoto et al. (**64, Yamamoto**) used a mutant form of PEBP1 that was not phosphorylated by PKC and thus constitutively inhibited Raf-1 and MEK. When this PEBP1 mutant was introduced into Purkinje cells of mouse cerebellar slices through patch-clamp electrodes, LTD was blocked, while wild-type PEBP1 had no effect on LTD. Physiological experiments demonstrated that PEBP1 worked downstream of PKC and upstream of MAPK during LTD induction. Furthermore, biochemical analyses demonstrated that endogenous PEBP1 dissociated from Raf-1 and MEK during LTD induction in a PKC-dependent manner, suggesting that PEBP1 inhibition of Raf-1 and MEK is removed upon LTD induction. It was concluded that PKC-dependent regulation of PEBP1 led to MAPK activation that is required for LTD (**64, Yamamoto**).

### 4.2.4.2- Long-term potentiation (LTP)

Long-term potentiation (LTP) is the opposing process to LTD; it is the long-lasting increase of synaptic strength. In conjunction, LTD and LTP are factors affecting neuronal synaptic plasticity. Glutamate is the primary excitatory neurotransmitter at almost all synapses in the central nervous system. Acetylcholine is an excitatory neurotransmitter involved in synaptic transmission at neuromuscular junctions, visceral motor systems and various sites within the central nervous system (**309, Picciotto**). Although all acetylcholine receptors respond to acetylcholine, they respond to other molecules as well. Nicotinic and muscarinic are two main kinds of "cholinergic" receptors. To examine the physiological function of PEBP1 on hippocampal neural activity, Ohi et al. (**310, Ohi**) investigated whether overexpression of PEBP1 strengthened the efficiency of neural activity in the hippocampus. The results suggested that muscarinic (M1) modulation of glutamatergic postsynaptic function may be involved in strengthening LTP transgenic mice overexpressing PEBP1. As it was reported that overexpression of PEBP1 in the transgenic mice increased the amount of cholineacetyltransferase in the septal nucleus, a possible explanation was that the remaining axon terminals from the septum may play some role in synthesizing large amounts of acetylcholine and reinforcing exocytosis in PEBP1 transgenic mice (**310, Ohi**).

### 4.2.5- Localization of PEBP1 in rat intestine nerve cells

Independently of neural cells in brain, HCNP and/or PEBP1 was also detected by immunoreactivity in nerve processes and nerve cell bodies in the rat small intestine (**311, Katada**). Labeled nerve processes were numerous in the circular muscle layer and around the submucosal blood vessels. The data provided evidence that HCNP and/or PEBP1 nerve cell bodies and nerve fibers are present in the submucosal and myenteric plexuses of the rat small intestine. By immunohistochemical staining, HCNP and/or PEBP1 was observed on the membranes of various subcellular organelles, including the rough endoplasmic reticulum, Golgi saccules, ovoid electron-lucent synaptic vesicles in axon terminals associated with submucosal and myenteric plexuses, and the outer membranes of a few mitochondria. The synaptic vesicles of HCNP and/or PEBP1 terminals were 60-85 nm in diameter. An immunohistochemical light microscopic study revealed that in both the submucosal and myenteric ganglia, almost all choline acetyltransferase (ChAT)-immunoreactive neurons were also immunoreactive for HCNP and/or PEBP1. These observations suggested that HCNP proper and/or PEBP1 is a membrane-associated protein with a widespread subcellular distribution, that PEBP1 may be biosynthesized within neurons localized in the rat enteric nervous system, and that HCNP proper and/or PEBP1 are probably stored in axon terminals (**311, Katada**).

## 4.3- PEBP1 prevents primary cilium formation

Using the Cep290 mutant mouse retinal degeneration (rd16), it was shown that Cep290-mediated photoreceptor degeneration is associated with aberrant accumulation of PEBP1, and that Cep290 physically interacts with PEBP1. For the first time, these data suggested that PEBP1 prevents cilia formation and is associated with Cep290-mediated photoreceptor degeneration (**75, Murga-Zamalloa**). PEBP1 co-localized with Cep290 at the transition zone of the photoreceptors, which is the region between the basal body and the axoneme. PEBP1 revealed to be present in the apical inner segment, transition zone, and to a lesser extent, at the basal body and centriole of the photoreceptors. The concentrated signal of PEBP1 at these sites and its ability to interact with Rab8A suggested its involvement in the regulation of docking and transport of membrane protein-containing vesicles in photoreceptors (**75, Murga-Zamalloa**). The ectopic accumulation of PEBP1 led to defective cilia formation in zebrafish and cultured cells, this effect being mediated by the interaction of PEBP1 with the ciliary GTPase Rab8A. PEBP1 preferentially associated with Rab8A in the presence of GDP substrate, suggesting that the PEBP1-Rab8A (GDP) complex may need to dissociate for the release of Rab8A-GDP and subsequent conversion to Rab8A-GTP for appropriate ciliary transport. These results suggested that PEBP1 modulates the localization of a distinct set of proteins required for cilia assembly and maintenance (see **Figure 3** for a schematic representation of primary cilium). Moreover, PEBP1 levels appeared to be critical for cilia formation, injection of mRNA encoding PEBP1 into wild-type zebrafish embryos resulting in defective ciliogenesis. Finally, the accumulation of PEBP1 seemed to be due to increased stability of the protein. The phosphorylation of PEBP1 at Ser153 did not affect its ability to reduce cilia formation (**75, Murga-Zamalloa**)**.** To complete these results, the study of double knockout mice *Cep290 $^{rd16}$: PEBP1 $^{ko/ko}$* showed that retinal degeneration in the *Cep290 $^{rd16}$* mice could be delayed by downregulating the expression of PEBP1 (**76, Subramanian**). The data also suggested a role of CEP290 in modulating PEBP1 levels for normal photoreceptor development and survival. Moreover, as it was previously shown that PEBP1 could be degraded *via* the ubiquitin-proteasome system (UPS) (**118, Qu 2013**) and given the involvement of cilia and basal body proteins in regulating the UPS (**312, Liu 2014**), it seemed conceivable that CEP290 targets PEBP1 to the UPS (**76, Subramanian**).

All these results are in accordance with other studies suggesting PEBP1 implication in primary cilium functioning. Particularly, cilia movement regulates expression of PEBP1 (**129, Sas**). It was shown that PEBP1 levels were significantly reduced in mice lacking cilia and that PEBP1 levels with cilia bending were diminished by a PKC inhibitor (**129, Sas**). Malfunction of primary cilium was associated with various pathological disorders, many of them leading to cystic disorders in kidney, liver and pancreas. In kidney, primary cilia were shown to be present on nearly all epithelial cells and extend from the apical surface into the tubular lumen, where they respond to luminal fluid flow. Cilia bending, with changes in apical flow, led to various intracellular signaling events including increases in intracellular calcium concentrations, thereby acting as mechanosensors (**313, Rydholm**) and also served an important role in the maintenance of cell differentiation (**314, Bacallao**).

## 4.4- PEBP1 regulates the mitotic spindle checkpoint

PEBP1 was described to associate with centrosomes and kinetochores and to regulate the spindle checkpoint in mammalian cells (**155, Eves**). The centrosome is an organelle that serves as the main microtubule organizing center (MTOC) of animal cells. It acts as a multifunctional platform for numerous signaling processes and is functionally implicated in processes such as cell cycle progression, cell polarity, migration, mitosis, and ciliogenesis (**315, Tollenaere**). In mitosis the nuclear membrane breaks down and the centrosome nucleated microtubules can interact with the chromosomes to build the mitotic spindle. During the process of cell division, the spindle checkpoint prevents separation of the duplicated chromosomes until each chromosome is properly attached to the spindle apparatus (**316, London; 317, Musacchio**). The kinetochores are large protein complexes assembled on the centromeric region of the chromosomes, they mediate spindle–microtubule attachment and control the movement of chromosomes during mitosis and meiosis (**318, Yamagishi**). PEBP1 depletion decreased the mitotic index, decreased timing from nuclear envelope breakdown to anaphase, and caused an override of mitotic checkpoints induced by spindle poisons such as Taxol (**155, Eves**). PEBP1 depletion reduced kinetochore localization and kinase activity of Aurora B which is a regulator of the spindle checkpoint (**155, Eves**). In complex with other "chromosomal passenger" proteins, Aurora B accumulates at inner centromeres during prometaphase and controls the interactions of microtubules with kinetochores. Loss of PEBP1 led to a defective spindle checkpoint through a mechanism involving enhanced Raf-1 activation and Aurora B kinase inhibition. Thus, it appeared that PEBP1 regulates normal timing of mitosis from nuclear envelope breakdown to anaphase, and accumulation of cells in metaphase. In addition, PEBP1 depletion increased chromosomal damage upon exposure of HeLa cells to Taxol, due probably to relaxation of the spindle checkpoint (**155, Eves**). **Figure 4** shows a schematic representation of mitotic spindle according to Prosser and Pelletier (**319, Prosser**).

Phosphorylated PEBP1 (pS$^{153}$PEBP1) was also localized at kinetochores and PEBP1 phosphorylation was specifically increased during mitosis, leading probably to the activation of the MAP kinase cascade (**155, Eves**). In contrast to elevated MAP kinase signaling during the G1, S or G2 phases of the cell cycle that activated checkpoints and induced arrest or senescence, loss of PEBP1 during M phase led to bypass of the spindle assembly checkpoint and the generation of chromosomal abnormalities. Thus, PEBP1 and the MAP kinase cascade seemed to play a role in ensuring the fidelity of chromosome segregation prior to cell division. The need for precise titration of the MAP kinase signal to ensure the integrity of the spindle assembly process was highlighted and it was proposed that PEBP1 status in tumors could influence the efficacy of treatments such as poisons that stimulate the Aurora B-dependent spindle assembly checkpoint (**156, Rosner**). Using HEK-293 PEBP1 depleted (termed HEK-499) it was observed that PEBP1 silencing accelerated DNA synthesis and G1/S transition entry. It was proposed that PEBP1 silencing deregulates the expression of key genes involved in DNA replication, G1/S phase transition, G2/M checkpoint and chromosomal stability, which in summation may accelerate S-phase transition and mitosis, leading to the observed higher proliferation rate in HEK-499 cells. PEBP1 depletion down-regulated G2/M checkpoint molecules like Aurora B, cyclin G1 and sertuin that slow the G2/M transition time. These changes in the kinetics of the cell cycle resulted in the higher proliferation rate of HEK-499 compared to control cells (**38, Al-Mulla 2011**).

## 4.5- Role of PEBP1 in apoptosis

### 4.5.1- Pro-apoptotic role of PEBP1

Since 2003, the decrease of PEBP1 in cancer cells was associated with metastasis occurring and, consequently, PEBP1 expression was considered as a possible marker of cancer clinical course (**61, Fu 2003**). Unfortunately, during cancer treatment, the acquisition of resistance to therapies such as radiation, chemo- and immunotherapeutic drugs appeared to be a major obstacle to eliminate tumor cells. Nevertheless, in tumorigenic human prostate and breast cancer cells, it was shown that low expression levels of PEBP1 were rapidly induced upon chemotherapeutic drug treatment with 9-nitrocamptothecin (9NC), sensitizing the cells to apoptosis (**120, Chatterjee**). Expression levels were low in prostate cancer cells and increase after DNA damage. The direct correlation of PEBP1 expression with the extent of apoptosis suggested that upregulation of PEBP1 sensitized cancer cells to apoptotic signals in response to DNA damage. Indeed, both DU145 and PC3 cells have a low level of PEBP1 expression, which could be significantly induced to high levels after DNA damage (**120, Chatterjee**). PEBP1 has been shown to abrogate the survival and anti-apoptotic properties of the Raf1/MEK/ERK and NF-kappa B signaling pathways *via* physical interaction with Raf-1 and MEK1/2 in the former and NIK or TAK in the latter. PEBP1 was then considered to play a pivotal role in cancer, regulating apoptosis induced by drugs or immune-mediated stimuli. However it appeared that chemosensitization or direct trigger of apoptosis by PEBP1 induction might be cell type specific or depending on the differentiation stage and the activation status of the signaling molecules in the cells (**130, Odabaei**).

In fact, a lot of studies on various cells and cancer types, have led to the description of several signaling pathways regulated by PEBP1 to trigger apoptosis. In most cases, the ERK and NFκB pathways were considered the targets of PEBP1 to regulate apoptosis, migration and invasion of cancer cells. In cultured SW1990 and AsPC-1 pancreatic tumor cell lines, overexpression of PEBP1 suppressed cell proliferation, promoted apoptosis and inhibited cell migration and invasion through downregulating Raf-1-MEK1/2-ERK1/2 signaling pathway. *In vivo*, PEBP1 could also suppress tumorigenesis by diminishing the volume of the tumors (**37, Dai**).

Artemisinin and its derivatives, such as dihydroartemisinin (DHA), were used for treatment of recurrent and metastatic cervical cancer. DHA have been shown to inhibit tumor growth with low toxicity to normal cells. The expression of PEBP1 was significantly upregulated and the expression of bcl-2 was significantly downregulated in DHA-treated HeLa and Caski cells compared with control cells. DHA treatment caused a significant inhibition of tumor growth and a significant increase in the apoptotic index in nude mice bearing Hela or Caski tumors. Thus, it was concluded that DHA inhibited cervical cancer growth *via* upregulation of PEBP1 and downregulation of bcl-2 (**136, Hu**).

Besides, the major genetic causes for the incidence and therapeutic refractoriness of neuroblastomas include amplification of the oncogene N-Myc and loss of tumor suppressor p53. The flavonoid didymin inhibited proliferation and induced apoptosis irrespective of p53 status in neuroblastomas. Mainly didymin inhibited N-Myc as confirmed at protein, mRNA, and transcriptional level. But, also, didymin upregulated PEBP1 levels and downregulated PI3K, pAkt, Akt and vimentin. Lastly, didymin induced G(2)/M arrest along with decreasing the levels of cyclin D1, CDK4, and cyclin B1 (**320, Singhal**). Similar results were obtained on human HepG2 liver carcinoma cells wherein didymin strongly inhibited the viability, clonogenicity and migration of HepG2 cells. Didymin attenuated the mitochondrial membrane potential, accompanied by the release of cytochrome c and inhibited the ERK/MAPK and PI3K/Akt pathways by increasing the level of PEBP1. Thus, the results indicated that didymin induced apoptosis of HepG2 cells through mitochondrial dysfunction and inactivation of the ERK/MAPK and PI3K/Akt pathways by up-regulating PEBP1 (**43, Wei**).

A particular role of NO on the pro-survival/anti-apoptotic loop NFκB/Snail/YY1/PEBP1/PTEN was identified in cancer (**321, Bonavida 2013**). It was shown that in tumor cells, the modification by NO of this dysregulated loop was responsible, in large, for the reversal of chemo and immune resistance and sensitization to apoptotic mechanisms by cytotoxic agents. Moreover, tumor cells treated with exogenous NO donors resulted in the inhibition of NFκB activity, inhibition of Snail, inhibition of transcription repression of YY1 and subsequent inhibition of epithelial-mesenchymal transition (EMT), but induction of PEBP1 and induction of PTEN. Each gene product modified by NO in the loop was involved in chemo-immunosensitization. Overall, NO-donors interference in the regulatory circuitry resulted in chemo-immunosensitization and inhibition of EMT (**322, Bonavida 2015**).

In human gastric cell lines, PEBP1 suppressed cell proliferation and invasion and enhanced apoptosis. In that case, the down-regulation of PEBP1 expression was attributed to miR-224 which negatively regulated the expression PEBP1 (**31, Liu**). In another study, PEBP1 was described to be implicated in sensitization to TRAIL apoptosis. LFB-R603 (a chimeric anti-CD20 monoclonal antibody) resensitized resistant Ramos cells to TRAIL-apoptosis through modification of the dysregulated NF-κB/Snail/PEBP1/PTEN circuitry. In this loop, both NFκB and Snail regulated resistance whereas both PEBP1 and PTEN regulated sensitivity, each of these gene products being involved in the regulation of tumor cells sensitivity to TRAIL. Cells overexpressing PEBP1 were resensitized to TRAIL apoptosis, thus, indicating the direct role of PEBP1 in the reversal of tumor resistance to TRAIL (**323, Baritaki**).

### 4.5.2- Anti-apoptotic role of PEBP1

Contrary to the above studies indicating a pro-apoptotic role of PEBP1, it was brought out that PEBP1 attenuates apoptosis in neural cells and in particular conditions as oxygen deprivation or under the effect of microwaves. PEBP1 was overexpressed or knocked down using lentivirus in PC12 cells (embryonic origin from the neural crest), which were then challenged by oxygen-glucose deprivation (OGD). PEBP1 overexpression significantly increased the viability of OGD cells, and attenuated apoptosis, cell cycle arrest, and reactive oxygen species generation. PEBP1 overexpression significantly inhibited NFκB pathway by inhibiting IKK, IκBα, and P65 phosphorylation and also inhibited the ERK pathway by inhibiting MEK, ERK, and CREB phosphorylation. These results on PC12 cells, were confirmed on animal models of focal cerebral ischemia (**92, Su**). In another study, it was observed that OGD downregulated PEBP1 in microglia cells. Overexpression of PEBP1 reduced focal cerebral ischemia injury and significantly attenuated microglia cells apoptosis. In addition, PEBP1 overexpression inhibited the upregulation of phosphorylation of NFκB induced by OGD and cerebral ischemia and decreased cell transwell migration. Overall, these results showed that PEBP1 protected against ischemic stroke through inhibition of microglial cells excessive activation, inhibited their motility, and promoted neuronal survival partly though IKKβ-IκBα-NFκB signaling axis (**324, Su 2017**).

The effects of electromagnetic fields on human health have been widely studied. Differentiated PC12 cells were exposed to continuous microwave radiation at 2.856 GHz for 5 min. The results showed that PEBP1 was downregulated after microwave. PEBP1 downregulation led up to excessive activation of the MEK/ERK/CREB pathway, decrease in the Bcl-2/Bax ratio, and activation of caspase-3, finally triggering the execution of apoptosis (**41, Zuo**).

## 4.6- Role of PEBP1 in autophagy

PEBP1 was identified to interact with the microtubule-associated protein 1 light chain 3 β (LC3) to regulate the initiation of autophagy (**44, Noh**). When deprived of nutrients or growth factors, cells trigger the self-eating process called autophagy in order to supply the nutrients necessary for cell survival. Upon signaling, double-membraned cup-shaped structures called phagophores engulf portions of cytoplasm. After fusion of the membranes closing the cup-shaped phagophore, the autophagosome is formed. These vesicles traffic along microtubules towards the microtubule-organizing center, where lysosomes are concentrated. This brings autophagosomes close to lysosomes, enabling fusion and degradation of the contents of the resulting autolysosomes by lysosomal hydrolases (**325, Rubinsztein**). LC3 is a cytosolic protein, it can associate with intracellular membranes through Phosphatidylethanolamine-dependent conjugation upon induction of autophagy and regulates initiation of autophagy through interaction with many autophagy-related proteins possessing an LC3-interacting region (LIR) motif. The LIR motif is composed of tryptophane and Leucine amino acids separated by 2 non-conserved amino acids (WXXL). The PE-conjugated LC3 proteins in the membranes (called LC3-II) play a critical role in regulating the formation of autophagosomes and sequestering cellular components into the phagophore lumen (**326, Kabeya**). Specifically, PEBP1 interacted with only LC3-I, the cytosolic unconjugated LC3 form, but not PE-conjugated LC3-II. A LIR motif was identified in the PEBP1 sequence (W[55]DGL[57]) which is situated in a loop of the PEBP1 tridimensional structure. Mutation of the conserved LIR motif (WXXL to AXXA) abolished the ability of PEBP1 to bind LC3. These data indicated that PEBP1 could associate with LC3 to regulate autophagy before inclusion of LC3-II proteins within autophagosomes (**44, Noh 2016**). Overexpression of PEBP1 stimulated the Akt/mTORC1 pathway, indicating that PEBP1 could suppress autophagy through activation of Akt/mTORC1 under conditions of nutrient deprivation. PEBP1 phosphorylation at Ser153 caused dissociation of LC3 from the PEBP1-LC3 complex for autophagy induction. PEBP1 was involved in holding LC3 in the PEBP1-LC3 complex that makes LC3 not accessible for PE-lipidation. Upon the activation of autophagy, kinases may phosphorylate PEBP1 to liberate LC3, leading to PE conjugation which is a critical step necessary for autophagosome biogenesis. Thus, overexpression of PEBP1 significantly prevented starvation-induced autophagy, while, knockdown of PEBP1 expression induced high levels of autophagy *via* stimulation of LC3-lipidation and Akt/mTORC1 activation (**44, Noh 2016**).

Another role of PEBP1 associated with LC3 was observed in preadipocytes. It was shown that LC3 overexpression in 3T3-L1 preadipocytes stimulated adipocyte differentiation *via* direct modulation of PEBP1-dependent ERK1 activity (**327, Hahm**).

In Parkinson disease, CDK5 was activated and was responsible for associated neuronal loss (**328, Cheung**). Surprisingly, this mechanism occurred *via* the degradation of PEBP1 by chaperone-mediated autophagy. Indeed, excessive CDK5 activation resulted in the phosphorylation of PEBP1 on T42, promoting the exposure of the motif [179]KLYEQ[183] situated in the C-terminal helix of PEBP1. Then, Hsc 70 recognized the KLYEQ motif and transported PEBP1 to lysosomes for destruction. In turn, the ERK cascade was activated, triggering neuron cells to S-phase reentry and causing neuronal death (**45, Wen**).

## 4.7- PEBP1 in cell motility

Numerous studies have been undertaken to decipher the mechanisms underlying migration and invasion in various types of cancer. The results revealed that PEBP1 regulates cell motility under various actions. In particular, PEBP1 was described to modulate several signaling pathways, to interact directly with key proteins involved in cellular migration, to regulate the expression of matrix metalloproteinases and also to regulate the expression of several miRNAs.

### 4.7.1- Signaling pathways

It was shown that PEBP1 modulates some of the main signaling pathways implicated in cellular motility, notably in cancer cells. Indeed, previous studies have shown that PEBP1 can interfere with the Raf-1-MEK1/2-ERK1/2 signaling pathway (**13, Yeung; 7, Kim**), inhibit the transcription factor nuclear factor kappa B (NFκB) (**329, Beshir**) and G protein-coupled receptor kinase signal transduction pathways (**330, Zeng**), all of which being closely related to tumorigenesis and cancer metastasis. In several cancer types, MEK/ERK and PI3K/Akt activities were indispensable for the cellular effects of a natural phosphoantigen upon γδ T-cell activation and proliferation, Th1 cytokine secretion and anti-tumor cytotoxicity (**331, Correia**). Other evidence exhibited that tamoxifen, an anti-estrogen reagent, inhibited tumor cell invasion and metastasis in mouse melanoma through suppression of both PKC/MEK/ERK and PKC/PI3K/Akt pathways (**332, Matsuoka**). In the AsPC-1 human pancreatic adenocarcinoma cell line, (-)-epigallocatechin 3-gallate, strongly regulated PEBP1/ERK/NF-κB and/or PEBP1/NF-κB/Snail to inhibit invasive metastasis (**7, Kim**). In colorectal cancer, the functional impact of proteins from different signaling pathways varies between tumor center and tumor front. Indeed, overexpression of pAkt, Bcl2, VEGF, APAF-1, pERK, EphB2, PEBP1, CDX2, E-cadherin, MST1 and pSMAD2 was more frequently observed in the tumor center, whereas matrix metalloproteinase 7 and Laminin5γ2 overexpression was associated with the invasive front. In other words, in colorectal cancer progression, VEGF overexpression seemed to play a role in the tumor center, whereas Laminin5γ2-overexpression combined with PEBP1 loss was associated with tumor invasion at the front (**333, Karamitopoulou).**

Functional studies on hepatocellular carcinoma cells HCCLM3 and HepG2 revealed the inhibitory role of PEBP1 in migration and invasion *in vitro* and tumor growth *in vivo***.** However, the engagement of PEBP1 in regulating Raf/ MEK1/2-ERK1/2 and NF-κB signaling pathways appeared to be cell type-specific. TNF-α-induced NFκB signaling transduction was negatively regulated by PEBP1 in HCCLM3 cells, but not in HepG2 cells. In addition, PEBP1 functioned through EGF-activated Raf/MEK1/2-ERK1/2 signaling pathway to control HCCLM3 cell invasiveness, but not in HepG2 cells. These contradictive findings suggested that the mechanisms of signaling regulation are complicated in different HCC cells, and PEBP1-mediated modulation may be cell type-specific and context-dependent (**334, Wu).**

GRK2 was also described to be involved in cell migration since data have revealed a complex GRK2 interactome that includes a variety of proteins related to cell motility (**335, Ribas**). Altered GRK2 expression or activity may differentially affect the functionality of its interaction partners, thus affecting cell motility in distinct ways, depending on the cell type and/or stimuli involved (**336, Penela**).

### 4.7.2- Direct interaction of PEBP1 with protein partners

It was described that PEBP1 influences cell motility by interacting with several molecular partners. In cancer, PEBP1 was primarily shown to inhibit cell motility and metastasis formation, but PEBP1 was also found to enhance cell motility in cells depending on GSK3β for motility (**36, Al-Mulla 2012b; 96, Al-Mulla 2011**).

### 4.7.2.1- PEBP1 inhibits cell invasion and metastasis
**STAT3**

The signal transducer and activator of transcription 3 (STAT3) is a latent cytoplasmic transcription factor; it is known to be activated by cytokines, growth factors, and oncogenic proteins. Once phosphorylated and activated, STAT3 enters the nucleus and regulates gene transcription (**337, Germain**). STAT3 is constitutively activated in a number of malignancies and promotes cell survival, EMT, invasion and metastasis (**338, Devarajan**). It has been reported that PEBP1 overexpression resulted in constitutive physical interaction with STAT3 and blocked c-Src-phosphorylating and activating STAT3 (**98, Yousuf**). To explore the role of PEBP1 in nasopharyngeal carcinoma (NPC) metastasis *in vivo*, NPC cell lines with stable PEBP1 expression changes were established. The effect of PEBP1 was tested in a xenograft metastasis model in which PEBP1 knockdown 6-10B cells, PEBP1 overexpression 5-8F cells, and their corresponding control cells were used to generate pulmonary metastases in nude mice. Then the effects of PEBP1 were tested on the phosphorylated level of ERK- 1/2, STAT3, NFκB (IKK-α/ β and IκB-α) and GSK3β. The results indicated that PEBP1 inhibited ERK, NFκB, and STAT3, and activated GSK3β in NPC cells. The general conclusion was that in NPC cells, PEBP1 acted essentially by inhibiting STAT3 activation. PEBP1 interacted with and then blocked STAT3 phosphorylation, inhibiting its activation. Consequently, PEBP1 downregulation promoted NPC invasion, metastasis and EMT-like molecular alterations by activating STAT3 signaling (**100, He**).

**Syntenin/MDA-9**

In contrast to PEBP1, melanoma differentiation associated gene-9 (MDA-9), also known as syntenin, is a positive regulator of melanoma progression and metastasis. It was documented that, upon extracellular matrix engagement, MDA-9 interacts with c-Src leading to c-Src phosphorylation and FAK auto-phosphorylation (**339, Bouckerche).** These biochemical changes result in formation of an active focal adhesion complex with resultant activation of the NFκB pathway and secretion of MMP-2, thereby causing augmented tumor cell motility and metastatic competence. It was found that MDA-9 transcriptionally downregulated PEBP1. Furthermore, MDA-9 and PEBP1 physically interacted. The interaction of PEBP1 created a steric interference for c-Src interaction with MDA-9 in a manner that correlated with a suppression of FAK and c-Src phosphorylation. Avoiding these crucial steps, necessary for MDA-9 to promote FAK/c-Src complex formation, prevented the initiation of signaling cascades leading to invasion and metastasis. Ectopic PEBP1 expression in melanoma cells counteracted MDA-9-mediated signaling, inhibiting cell invasion, anchorage-independent growth, and *in vivo* dissemination of tumor cells (**55, Das**).

**Notch1**

In H1299 cells (lung cancer cell line) and in HeLa (cervical cancer cell line), activation of Notch1 promotes tumor invasion/ migration and upregulation of epithelial-mesenchymal transition-related proteins. Upon activation of Notch signaling, Notch1 is cleaved in a stepwise manner mediated by the proteolytic actions of the γ-secretase complex, leading to the subsequent release of Notch intracellular domain (NICD), the activated form of Notch1. Therefore, inhibition of γ-secretase would prevent Notch1 activation, resulting in suppression of the Notch signaling pathway. Interestingly, PEBP1 can directly interact with full-length Notch1 but not with the only NICD. These data suggested that PEBP1 prevents activation of Notch1 by directly binding to the whole Notch1 (intracellular/transmembrane domain), consequently inhibiting the proteolytic activity of γ-secretase (**18, Noh 2016**).

### 4.7.2.2- Via GSK3, PEBP1 promotes cell invasion and metastasis

PEBP1 influences cell motility by interacting directly with several partners. A particular case is the glycogen synthase kinase 3β (GSK3β) because PEBP1 has been shown to inhibit Raf/MEK/ERK, NFκB and GRK2 but to activate the GSK3β signaling pathway (**53, Al-Mulla, 2012a**). It is well-known that in most cell types, PEBP1 overexpression inhibits migration of cells and conversely its silencing enhances it. On the contrary, in MDCK cells the effect of PEBP1 on cellular motility is reversed as PEBP1 expression induces cellular motility and *vice versa*. Now, these cells depend on GSK3β for their motility (**36, Al-Mulla 2012b**). GSK3β suppresses tumor progression by downregulating multiple oncogenic pathways including Wnt signaling and cyclin D1 activation. PEBP1 binds GSK3 proteins and maintains GSK3β protein levels and its active form. In colorectal cancer cells HEK-499 (silencing PEBP1) it was found that the phosphorylated and active form of p38 was significantly increased. PEBP1 silencing induced a rapid p38 activation and phosphorylation of GSK3β at Thr-390 that culminated in its degradation (**96, Al-Mulla 2011**). Therefore, loss of PEBP1 induces the inactivation and ubiquitination of GSK3β and loss of cellular motility in these cells that depend on GSK3β for cellular motion (**36, Al-Mulla 2012b**). In another study, conditioned medium from ionizing radiation-induced senescent MCF7 cells significantly increased cell proliferation, invasion, migration, and wound healing activity in MCF7 cells and HUVECs. Downregulation of PEBP1 suppressed MCF7 migration and prevented the potential effects of the secretome from ionizing radiation-induced senescent tumor cells on neighboring cell migration. It is to note that in this study, PEBP1 was found as a secretory proteins produced by senescent tumor cells treated by ionizing radiation (**35, Han**). Interestingly, GSK3β signaling was described as a key pathway in MCF7 cells, which notably regulates autophagy activation (**340, Gavilan**).

### 4.7.3- Regulation of matrix metalloproteinases expression

Independently of direct interaction with protein partners, PEBP1 acts also on cellular motility through the regulation of the expression levels of other proteins. As an example, it was shown that PEBP1 depletion in HEK-499 cells led to the overexpression or stabilization of beta-catenin, vimentin, tyrosine-protein kinase MET and serine/threonine protein kinase PAK1, SNAIL and SLUG (**38, Al-Mulla 2011b**).

Otherwise, migration and invasion of cells are tanglesome courses including cytoskeletal-reorganizations and degradations of extracellular matrix (ECM) (**341, Cheng 2010**). Normally, degradations of ECM are primary steps of cancer cell invasions and matrix metalloproteinases (MMPs) may be the mediations of these courses (**33, Lei**)**.** PEBP1 is especially known to inhibit cell invasion and metastasis through downregulation of several **matrix metalloproteinases** (MMPs) expression. In glioma cells U87, overexpression of PEBP1 diminished the MMP2, MMP9 expression and also the HMGA2 expression (**33, Lei**). In Cholangiocarcinoma cells that originate in the biliary epithelium, PEBP1 inhibited the invasive and metastatic ability of the cholangiocarcinoma cell line RBE, by downregulating MMP-9 and upregulating the tissue inhibitor of metalloproteinases (TIMP-4) mRNA expression (**342, Ma**)**.** In human pancreatic adenocarcinoma cells AsPC-1, (-)-epigallocatechin 3-gallate (EGCG) induced PEBP1 upregulation *via* the inhibition of histone deacetylase (HDAC) activity which increased histone H3 expression and inhibited Snail expression, NFκB nuclear translocation, MMP-2 and -9 activity and Matrigel invasion (**7, Kim**). PEBP1 inhibited the invasive and metastatic abilities of esophageal cancer cell line TE-1 by downregulating mRNA expression of LIN28 and MMP-14 (**104, Zhao 2013**). The NFκB pathway, but not the Raf pathway, appeared to positively control the invasion of breast cancer cells and a regulatory loop involving an opposing relationship between PEBP1 and the NFκB pathway was suggested to control the level of MMP expression and cell invasion (**329, Beshir 2010**). In a panel of breast cancer, colon cancer and melanoma cell lines, PEBP1 negatively regulated the invasion of the different cancer cells through extracellular matrix barriers by controlling the expression of various MMPs, particularly MMP-1 and MMP-2 (**329, Beshir**). In another study on breast cancer cells, it was considered that AP-2alpha, interleukin-4 (IL-4), E-cadherin, fibulin 1D, p16(INK4alpha), PTEN and PEBP1, are suppressors of cancer cell invasiveness, which are located upstream of MMPs in cell signaling pathways. In MDA-MB-231 breast cancer cells, MMP-1, MMP-2, MMP-9, MMP-11, MMP-13, MMP-14, MMP-16, and MMP-19 were enhancers of cancer cell invasiveness. The results indicated that each upstream cancer-progression determinant reaches these MMP expressions through different sets of signaling pathways. Particularly, PEBP1 could be involved in the control of the expression of MMP1, 3, 7, 11, 13 and 14 by modulating the Raf/MEK/ERK pathway that, in turn, regulates NF-κB, p53, cJun/cFos, PEA3 and CREB pathways (**343, Delassus 2008; 344, Delassus 2010**). In human prostate PC-3M cells, data indicated that PEBP1 does not affect cell proliferation in PC-3M cells, but inhibits both cell migration and cell invasion. In association with this inhibitory effect on motility, PEBP1 down-regulated MMP-2 and MMP-9, cathepsin B and urinary plasminogen activator (uPA) (**34, Xinzhou**). Finally, by using an unbiased PCR based screening and by analyzing DNA microarray expression datasets it was observed that the expression of multiple metalloproteases (MMPs) including MMP1, MMP3, MMP10 and MMP13 are negatively correlated with PEBP1 expression in breast cancer cell lines (MDA-MB231, 4T1 and 168 FARN) and clinical samples. It was shown that the expression signature of PEBP1 and MMPs is better at predicting high metastatic risk than the individual gene (**345, Datar**). Using a combination of loss- and gain-of-function approaches, the results showed that MMP13 is a cause of PEBP1-mediated inhibition of local cancer invasion. Moreover, PEBP1 negatively regulated MMP13 through the ERK2 signaling pathway and the repression of MMP13 by PEBP1 was transcription factor AP-1 independent (**345, Datar**). In conclusion, following all the results described above, PEBP1 appears to be able to regulate the expression of MMP1, 2, 3, 7, 9, 10, 11, 13, 14, 16 and 19. PEBP1 is probably involved in the control of the expression of all these MMPs by modulating different signaling pathways.

### 4.7.4- Regulation of miRNAs expression

Aside regulating the expression of proteins partners, PEBP1 controls also cell motility by regulating the expression of various miRNAs. Several studies have identified miRNAs operating to promote or suppress tumor invasion or metastasis via regulating metastasis-related genes. Using microRNA expression arrays to analyze small non-coding microRNAs that are regulated by PEBP1, 50 differentially expressed microRNAs were identified in breast cancer 1833 cells expressing S153E-PEBP1. These analyses showed that miR-200b was the most significant microRNA that positively correlated with PEBP1 expression in a cohort of breast cancer patient samples. It was found that PEBP1 induced expression of miR-200b that directly inhibited expression of lysyl oxidase (LOX), leading to decreased invasion. This was a novel signaling pathway targeted by PEBP1 that regulates remodeling of the extracellular matrix and tumor survival (**346, Sun**).

Using an orthotopic murine model, PEBP1 was shown to inhibit invasion by metastatic breast cancer cells and to repress cell intravasation and bone metastasis. The mechanism involved inhibition of MAPK, leading to decreased transcription of LIN28 via Myc. Suppression of LIN28 enhanced let-7 processing in breast cancer cells. Elevated let-7 expression inhibited HMGA2, a chromatin remodeling protein that activates pro-invasive and pro-metastatic genes, including Snail (**101, Dangi-Garimella**).

Another study of glioma tissues and cell lines U251, U57 and SHG44, showed that the expression of PEBP1 and miR-98 were significantly lower and HMGA2 was higher than that in normal brain tissues. MiR-98 belongs to the mature let-7 family of miRNAs and was initially found to be downregulated in leukemia cell lines (**347, Yu**). Overexpression of PEBP1 upregulated miR-98 expression and inhibited glioma cell invasion but had no effect on glioma cell proliferation. These results indicated that PEBP1 functions as a potent tumor invasion repression gene through regulating miR-98 expression. The PEBP1/miR-98 to HMGA2 axis provided a potential mechanism for glioma cell invasion but not proliferation (**348, Chen**).

In breast cancer tissues, the expression of PEBP1 and miR-185 was significantly lower than that in normal breast tissues. In breast cancer cell lines MCF7 and MDA-MB-231, over-expression of PEBP1 up-regulated miR-185 expression, inhibited breast cancer cell growth and invasion, and inhibited miR-185 target gene HMGA2. These findings described PEBP1/miR-185 to HMGA2 axis and provided one more potential mechanism for breast cancer cell growth and invasion (**30, Zou**).

## 4.8- Cell adhesion to substratum, cell-cell adhesion

Following initial solid tumor formation and neovascularization, metastasis is a complex process involving not only cell motility but also changes in cell adhesion, secretion of a range of proteases and the development of distant metastases (**349, Bogenrieder; 350, Kopfstein).** Exploring the role of PEBP1 in cell adhesion, it was found that overexpression of PEBP1 in Madin-Darby canine kidney (MDCK) epithelial cells increased adhesion on multiple substrata, while decreasing adhesion of the cells to one another. PEBP1 also increased the ability of cells to generate productive traction force. Knockdown of PEBP1 in both MDCK and MCF7 cells (human breast adenocarcinoma cells) resulted in reduced cell–substratum adhesion (**54, Mc Henry**). In contrast, PEBP1 appeared to play a negative role in the control of cell–cell adhesion, regulating the expression of the adherens junction protein E-cadherin and the localization of the tight junction protein ZO-1 in MDCK cells. Knockdown of PEBP1 in MCF7 cells led to a slight decrease in the rate of wound closure, an effect consistent with that of PEBP1 knockdown in MDCK cells. Thus, PEBP1 displayed important roles in the regulation of cell adhesion, positively controlling cell-substratum adhesion while negatively controlling cell-cell adhesion (**54, Mc Henry**).

Cadherins are a class of transmembrane proteins. They play main roles in cell adhesion, forming adherens junctions to bind cells within tissues together. They are dependent on Ca2+ ions to function. Cell-cell adhesion is mediated by extracellular cadherin domains, whereas the intracellular cytoplasmic tail associates with a large number of adaptor and signaling proteins. A study investigated the expression of PEBP1 and epithelial cadherin (E-cadherin) in lung squamous cell carcinoma tissue and its correlation with the clinical pathology of lung squamous cell carcinoma (**351, Zhu**). The rates of positive PEBP1 and E-cadherin mRNA expression were significantly lower in lung squamous cell carcinoma than in the surrounding normal tissues. The positive expression rates were significantly lower in those with lymph node metastasis than in those without. The rates of positive PEBP1 and E-cadherin mRNA expression were significantly lower in patients at advanced (III, IV) stages than in patients at early (I, II) stages. It was concluded that low expression levels of PEBP1 and E-cadherin may be associated with initiation, invasion and/or metastasis, as well as with the inhibition of lung squamous cell carcinoma differentiation (**351, Zhu**). Similar studies were performed in esophageal squamous cell carcinoma (ESCC) to detect the expressions of PEBP1, E-cadherin and NF-κB p65. The expressions of PEBP1 and E-cadherin in ESCC tissues were obviously lower than those in the paracancerous tissues, the expressions in ESCC tissues from cases with lymph node metastasis were lower than those from cases without lymph node metastasis. Downregulation or depletion of PEBP1 was related to the onset and progression of ESCC, and facilitated the invasion and metastasis of ESCC by downregulating E-cadherin and upregulating NF-κB p65 (**352, Ping**).

Focal adhesion kinase (FAK) can influence the cytoskeleton, structures of cell adhesion sites and membrane protrusions to regulate cell movement. FAK signaling has been associated with the disassembly of integrin-based adhesion sites (**353, Webb**)**.** In melanoma cells it was documented that upon adhesion to fibronectin, MDA-9 activated c-Src through phosphorylation at Tyr418 leading to FAK phosphorylation and initiated a signaling cascade that resulted in enhanced cell invasion. In metastatic melanoma cells, restored PEBP1 physically interacts with MDA-9 and potentially interferes with c-Src phosphorylation and subsequent FAK phosphorylation (**55, Das**).

In conclusion, PEBP1 is mainly implicated in processes affecting the plasma membrane features such as receptors internalization, neuronal synapse, adhesion, motility, primary cilium formation, and neuronal synapse. PEBP1 appears to be particularly involved in membrane sensing, participating in signal transduction. PEBP1 expression is directly bound to what happens at the plasma membrane. Autophagy is a process that involves also partly plasma membrane as well as other cellular membranes and that needs vesicles formation and trafficking. Apoptosis is a programmed cell death that is strongly depending of signaling pathways and that leads to important biochemical events, changes in cell morphology and finally to cell death. Interestingly, the role of PEBP1 in mitotic spindle checkpoint reveals its implication in the control of attachment between microtubules and kinetochores. Finally, all the processes modifying membrane changes and cellular shape are underlain by cytoskeleton organization.

# 5- Relationships between the cellular processes modulated by PEBP1

PEBP1 appears to be concerned by numerous cellular phenomena that implicate connection between membrane and cytoskeleton, this is the case for signal transduction, endocytosis, exocytosis, division, differentiation, adhesion and motility. It is interesting to note that several of the processes modulated by PEBP1 need to be coordinated and some of them are most of the time mutually exclusive. Indeed, throughout their life cycle, the cells have to decide between division and differentiation, survival and death or even adhesion and motility. PEBP1 appears to follow and modulate the signaling pathways that govern the various options leading to cellular fate. But, in addition, PEBP1 appears to play a role in the coordination of cellular processes by binding to the main proteins and complexes that govern each of them. In order to point out the strong level of interplay and coordination existing between numerous cellular processes, we present here several examples of well-known relationships between several cellular mechanisms that are modulated by PEBP1.

## 5.1- Endosomes, multipurposes and crosstalks

### *5.1.1- Endosomes are platforms for signaling complex assemblies, interactions between different signaling pathways and spatiotemporal regulation of transcription factor activation*

Endosomes efficiently recycle the large array of cargo that is taken up during endocytosis back to the cell surface or deliver it to other endocytic compartments. Therefore, endosomes control many important physiological processes, including nutrient absorption, signal transduction from membrane receptors, immune surveillance and antigen presentation (**354, Gould**). Endosomes function to terminate signaling processes (for example, through the internalization of receptors) and function in signal propagation by facilitating the recruitment and integration of signaling cascades on the surface of endocytic vesicles (**355, Emery**). An interesting study suggested that canonical GPCR signaling occurs from endosomes as well as the plasma membrane (**356, Irannejad**). The authors describe the application of conformation-specific single-domain antibodies (nanobodies) to directly probe activation of the β2-adrenoceptor, a prototypical GPCR, and its cognate G protein Gs in living mammalian cells. The results showed that the adrenergic agonist isoprenaline promotes receptor and G protein activation in the plasma membrane as expected, but also in the early endosome membrane, and that internalized receptors contribute to the overall cellular cyclic AMP response within several minutes after agonist application (**356, Irannejad**).

Endosomes can regulate the localization of signaling complexes either by spatially restricting signaling activity to particular loci in the cell or by acting as vesicular carriers to transport signaling proteins to cellular locations that are unreachable by diffusion (**357, Howe**). Endosomes can also isolate signaling components and prevent unwanted signaling interactions. This strategy is likely to occur in the regulation of glycogen synthase 3 beta (GSK3-β), a kinase that has numerous phosphorylation targets in distinct pathways, including Wnt, Hedgehog, epidermal growth factor (EGF)/mitogen-activated protein kinase (MAPK), and transforming growth factor beta (TGF-β) signaling. Thus, the endosome membrane facing the cytoplasm forms a physical platform for signaling complex assemblies where endosomal scaffolds can facilitate signaling reactions between the recruited components (**358, Palfy**).

Scaffold proteins can not only facilitate signal transduction within a pathway, but are also known to regulate interactions between different signaling pathways at the plasma membrane or in the cytoplasm. On Wnt-induced signaling, GSK3-β, dishevelled (DVL), and AXIN have been shown to localize to the cell membrane as well as to endosomes following internalization of the β-catenin destruction complex (**359, Cliffe**). AXIN also can integrate TGF-β and Wnt signals. The endosomal scaffold protein APPL can facilitate crosstalk between Akt and GSK3-β kinases, which are key components of insulin and Wnt pathways, respectively (**360, Schenck**) and the endosomal scaffold EEA1 can mediate crosstalk between EGF/MAPK and insulin pathways by connecting p38 and Akt (**361, Nazarewicz**). Four scaffold proteins were found (GRB1, CCNE1, SARA, AXIN1) that were potentially capable of connecting proteins from five different pathways (JAK/STAT, Notch, Wnt, TGF-β, and the RTK pathway, which contains insulin and EGF/MAPK cascades) (**358, Palfy**). Endosomes traverse the cell as they move inwards from the cell membrane toward the nucleus and their outer membrane is a unique surface to initiate artificial connections between signaling components. Therefore, because of the numerous deactivating mechanisms (e.g., phosphatase activity on protein kinases) and the long distances that signals travel within the cell, especially in highly polarized cells such as neurons with long axons, endosomes that are actively moved across the cytoplasm could play an essential role in ensuring that weak crosstalk signals exert their effect (**362, Ascano**). Thus, an important function of endosomes may be the spatiotemporal regulation of transcription factor activation (**358, Palfy**). It is conceivable that PEBP1 may participate in this spatiotemporal regulation of transcription factor activation by blocking the kinases of the signaling pathways that are transported by endosomes from plasma membrane to their final place of action.

### 5.1.2- Endosomes during cytokinesis

During cytokinesis, after replication of the genetic material, a furrow forms between the two cells and constricts using an actomyosin-based contractile ring, leaving the daughter cells attached by a thin bridge. The resolution of this bridge, by a process called abscission, separates the two cells (**363, Barr**). Recent studies have implicated both secretory and endosomal trafficking processes in these events. In late telophase, both secretory and endocytic vesicles move rapidly into the midbody and accumulate in the intercellular bridge immediately adjacent to the midbody ring. The fusion of these vesicles, both with each other and with the surrounding plasma membrane (a process known as compound fusion), precedes abscission (**364, Gromley**).

Endocytosis also has a central role in abscission itself, as Rab11-positive endosomes have been shown to traffic from centrosomes into the furrow from both daughter cells (**365, Fielding**). Phosphatidylinositol-4,5-bisphosphate (PtdIns(4,5)P2) accumulates at the cleavage furrow (**366, Emoto**) whereas phosphatidylinositol-3,4,5-trisphosphate (PtdIns(3,4,5)P3) is enriched at the poles of the dividing cells. The two enzymes that control the distribution of PtdIns(3,4,5)P3 (e.g. PTEN and class I PI3K) are enriched in the furrow and the poles of the dividing cells, respectively, and cells that lack these enzymes fail to complete cytokinesis (**367, Janetopoulos**). Such data might imply that endosomal vesicles could provide a unique platform for the assembly of the abscission machinery.

### 5.1.3- Endosomes in cell polarity and cell migration

Moreover endosomes are implicated in cell polarity. Membrane trafficking is essential for the generation and maintenance of polarity and it was shown that clathrin, a key protein involved in endocytosis also regulates basolateral polarity in mammalian cells (**368, Deborde**). Many cells polarize in response to a specific signal, and this is characterized by the reorganization of the cytoskeleton and the redistribution of certain organelles. This is clearly of importance in epithelial cells, where apical and basolateral surfaces need to be established, as well as during directed cell migration, where front-rear polarization enables the asymmetric morphology of cells migrating toward an extracellular signal. Endocytosis is an important regulator of cell polarization and cell migration since it regulates the trafficking of adhesion and polarity proteins (**369, Lobert**). E-cadherin is a type I transmembrane protein and an important component of Adherens junctions. Its recycling is thought to contribute to maintaining cell-cell junction integrity and is dependent on Rab8 and Rab13 (**370, Yamamura**).

Recent work has suggested that endosomes might fulfil additional functional roles in cell migration. New studies have identified a so-called Wnt-mediated receptor–actin–myosin polarity (WRAMP) structure, which accumulates asymmetrically at the cell periphery in response to WNT5A. W-RAMP formation involves the recruitment of actin, myosin IIB and the Wnt ligand receptor frizzled 3, is dependent on Rab4 and Rab8, requires endosome trafficking, and is associated with multivesicular bodies (**371, Witze**). The formation of lamellipodia, waves or circular dorsal ruffles is important in cell migration. It is controlled by small GTPases such as Rac and Rabs, and is known to involve endocytic proteins. Indeed, Dynamin, a critical endocytic protein, is required for the formation of circular ruffles, and the early endosomal protein Rab5, a master regulator of endocytosis, is needed for actin dynamics leading to circular ruffling. In some settings, Rab5 can also induce the formation of lamellipodia. Rac-containing vesicles are transported from endosomes to cell surface ruffles in an ARF6-dependent step (**372, Palamidessi**). Altogether, the studies discussed above reinforce the hypothesis that, by using endosomal membranes as a platform, cells could assemble specific molecular machines in space and time (**354, Gould**). By accompanying GPCR internalization, signaling pathways and interacting with specific proteins such as Rab8, PEBP1 may likely use the platforms formed by endosomal membranes to help in assembly/disassembly of specific molecular machines in precise localization and time during cellular processes.

## 5.2- Relationships between primary cilium and centrosome

It was shown that Cep290-mediated photoreceptor degeneration was associated with aberrant accumulation of its interacting partner PEBP1. Ectopic accumulation of PEBP1 led to defective cilia formation in zebrafish and cultured cells, an effect mediated by its interaction with the ciliary GTPase Rab8a (**75, Murga-Zamalloa**). CEP290 tethers flagellar transition zone microtubules to the membrane and regulates flagellar protein contents (**373, Craige**). CEP290 binds Rab8 under ciliated and non-ciliated conditions, and its regulation of ciliogenesis is dependent on its interaction with Rab8. Dysfunction or absence of CEP290 perturbs the ciliary influx and localization of specific photo-transducers and odorant signaling molecules such as Golf and Gɣ in the cilia of olfactory receptor neurons (**374, Chang**). CEP290 is also a centriolar protein, it localizes to the distal ends of centrioles and to the cloud of pericentriolar material that surrounds the centrosome (**374, Chang**). When a living cell initiates its program of ciliogenesis, the centrosome migrates to the apical surface of the cell and, at only the distal end of the mother centriole, CEP290 is activated, resulting in the growth of the ciliary axoneme and the formation of the primary cilium (**375, Tsang**).

Otherwise, PEBP1 was also described to regulate the mitotic spindle checkpoint by inhibiting Aurora B (**155, Eves**). To investigate a possible similar or common role of PEBP1 both in primary cilium and centrosome, we discuss below the separate role of primary cilium and centrosome as well as the known relationships between them.

### 5.2.1- Role of primary cilium

The primary cilium is a single antenna-like projection of the apical membrane, it is found in nearly every human cell type. This organelle plays important roles in numerous physiological phenomena, including tissue morphogenesis, signal transduction, determination of left-right asymmetry during development, and adult neurogenesis (**376, Huang**). Primary cilium has important sensory roles affecting multiple cellular processes, with a large proportion of G-protein coupled receptors, ion channels, and downstream effector molecules sequestered and confined to its membrane and lumen. To perform specialized sensory tasks, cilia are enriched with various receptors, for example, odorant receptors and rhodopsin to establish olfaction and vision (**377, Singla**). Furthermore, the primary cilium plays an active role in multiple signaling pathways including Wnt, PDGF and Hh, in which receptors traffic into/out of the ciliary axoneme during signal transduction (**378, Conduit**). It is to note here, that as the primary cilia, the motile cilia not only cause fluid flow but also sense signals from the flow to determine their orientation (**379, Marshall**). It has been shown that defects in primary cilium formation and function are responsible for a variety of human diseases and developmental disorders, termed ciliopathies (**380, Drivas**). The major organs affected by ciliopathy syndromes are the brain, kidney and eye. The cilium is formed as a membrane-bound microtubule extension, known as the ciliary axoneme, protruding from the distal end of the mother centriole at the apical surface of the cell (**381, Qin**). The barrier between the cilium and the cytosol is maintained at the transition zone of the cilium, the region of the organelle just proximal to the axoneme, in plane with the cortex of the cell.

### 5.2.2- The centrosome

The centrosome is an important cellular organelle which nucleates microtubules (MTs) to form the cytoskeleton during interphase and the mitotic spindle during mitosis. Cep290 was found specifically localized in the mitotic centrosome and involved in spindle position. Depletion of Cep290 caused a reduction of the astral spindle, leading to misorientation of the mitotic spindle. CEP290 interacts with the centriolar satellites that are small, spherical granules with a diameter of approximately 70–100 nm that cluster in the vicinity of the centrosome. They served as vehicles for protein trafficking towards the centrosome and the physical coupling between the satellites and microtubule-associated motor proteins can be mediated by several proteins such as BBS4, Par6a, and CEP290 (**382, Kim 2008**).

### 5.2.3- The primary cilium membrane is associated to centrosome

The primary cilium is nucleated at its base by the basal body, consisting of the eldest centriole in the cell (named the mother centriole) with associated appendage proteins that dock it to the plasma membrane (**383, Seeley; 384, Kim 2013; 385, Broekhuis**). In a series of experiments the ciliary membrane was followed in dividing embryonic neocortical stem cells and cultured cells (**386, Paridaen**). Interestingly, in these cells, ciliary membrane attached to the mother centriole was endocytosed at mitosis onset, persisted through mitosis at one spindle pole, and was asymmetrically inherited by one daughter cell, which retained stem cell character. This daughter re-established a primary cilium harboring an activated signal transducer earlier than the non-inheriting daughter. These results are important as they highlight two major facts. First, upon entry into mitosis, the ciliary membrane was internalized together with one of the centrosomes, indicating that, contrary to previous concepts, primary cilia are not completely disassembled prior to mitosis. Secondly, the data implied that centrosome-associated ciliary membrane acts as a determinant for the temporal-spatial control of ciliogenesis. Moreover, it was found that, at the G2-M phase transition, the shortening cilium was internalized with the basal body. Because the basal body/mother centriole maintains its attachment to the remaining ciliary membrane, it serves a dual function as a basal body as well as a part of one mitotic spindle pole (**386, Paridaen**).

The primary ciliary pocket is defined as the membrane domain starting from the transition fiber to the region where ciliary sheath emerges into extracellular environment. In photoreceptors, the ciliary pocket is part of the connecting cilium/transition zone (CC/TZ) where CEP290 and PEBP1 were colocalized (**75, Murga-Zamalloa**). The ciliary pocket was detected in various types of cells, particularly in retinal epithelial cell line (RPE1), epithelial kidney cells (IMCD3), murine embryonic fibroblasts (MEF) and murine oocytes. Interestingly, in these cells, ciliary pocket was shown to play a key role in membrane trafficking and endocytosis (**387, Molla-Herman**). In addition, the ciliary pocket connected to the actin cytoskeleton and probably served as an interface between the primary cilium and the actin cytoskeleton (**388, Ke**).

The abnormal expression of Aurora Kinase family is closely associated with the development of tumor, because of the crucial role of Aurora Kinase family plays in the mitosis and cell cycle. Aurora A and B are tightly regulated during the cell cycle and are overexpressed in many carcinomas, such as hepatocellular, ovarian, esophageal, bladder, breast, prostate, pancreas and lung cancers (**389, Wakahara**). It was observed that the cancer cells with elevated activated Aurora A rarely possess a primary cilium (**390, Yuan**). As a consequence, the loss of primary cilia may render multiple regulatory signals incapable of tumor growth suppression (**388, Ke**).

The localization of PEBP1 both at the primary cilium and at the mitotic spindle is in accordance with the fact that upon entry in mitosis, the ciliary membrane is internalized together with one of the centrosomes. Moreover, GRK2 was localized to centrosomes and described to play a central role in mitogen-promoted centrosome separation (**391, So**). PEBP1 appears to be associated with a set of key proteins, such as CEP290, Rab8, and GRK2 which govern in a coordinated manner main cellular events such as receptors internalization, primary cilium and mitotic spindle. This assumption is reinforced by the role of ciliary pocket in membrane trafficking and endocytosis and its connection to the actin cytoskeleton.

## 5.3- Primary cilium is involved in endocytosis, exocytosis, cell migration and immune synapse

### 5.3.1- Primary cilium regulates environmental factors and controls extracellular matrix

Primary cilia are known to regulate different ciliary signaling systems, such as those operated by G protein-coupled receptors (**392, Schou**) and receptor tyrosine kinases (**395, Christensen**) but also by transcient receptor potential (TRP) ion channels (**394, Phua**) or receptors for extracellular matrix proteins (**395, Seeger-Nukpezah**). Indeed, the transcient receptor potential (TRP) channels are known to accumulate inside the primary cilium and conduct calcium ions (Ca2+). Each subfamily member, namely TRPP2 TRPP3, TRPC1 and TRPV4, is gated by multiple environmental factors, including chemical (receptor ligands, intracellular second messengers such as Ca2+), mechanical (fluid shear stress, hypo-osmotic swelling), or physical (temperature, voltage) stimuli (**394, Phua**). Other interesting findings have revealed extensive signaling dialog between cilia and extracellular matrix (ECM), with defects in cilia associated with fibrosis in multiple contexts. It was demonstrated that primary cilia are essential for chondrocyte mechano-transduction and the control of ECM secretion in response to physiological compressive strain. In particular, the chondrocyte primary cilium is required for ATP-induced $Ca^{2+}$ signaling. In these cells, under normal growth, cilia-localized polycystin 1 mediates elevated extracellular ATP induced indirectly by cellular interactions with collagen during matrix compression. Thus, in chondrocytes, primary cilia are essential for mechanotransduction and the control of ECM secretion in response to physiological compressive strain (**396, Wann**). Finally, it appeared that primary cilium plays an important role in cellular homeostasis, based on its ability to integrate chemical cues and flow and compression forces. Disruptions in cilia deregulate cell growth and polarity, and produce an extracellular environment leading frequently to fibrosis. In turn, disruptions in the extracellular environment alter the signals received by cilia, again influencing cell growth properties (**395, Seeger-Nukpezak**).

### 5.3.2- Endocytosis and exocytosis at the ciliary pocket

As we have seen in a previous paragraph, the region between the plasma membrane and the ciliary membrane is structurally organized to produce a ciliary pocket in many cell types. This invagination comprises an interphase for the actin cytoskeleton and may function as a site for exocytosis and endocytosis of ciliary proteins, which regulates the spatiotemporal trafficking of receptors into and out of the cilium to control its sensory function (**397, Ghossoub**). Indeed, the periciliary membrane, comprises a site for clathrin-dependent endocytosis, which regulates signaling events associated with receptor internalization and subsequent recycling or degradation in the late endosomes/lysosomes compartment (**398, Pedersen**).

During primary cilium formation, the Rab GTPase membrane trafficking regulators Rab8a, -17, and -23, and their cognate GTPase-activating are involved in primary cilia formation. Additionally, Rab8a specifically interacts with cenexin/ODF2, a basal body and microtubule binding protein required for cilium biogenesis and is the sole Rab enriched at primary cilia. (**399, Yoshimura**). At the ciliary base, active Rab11 promotes a cascade of signaling events, including the activation of Rab8 and its localization along the ciliary axoneme, where it controls cilium elongation (**400, Westlake**). These findings provide a basis for understanding how specific membrane trafficking pathways cooperate with the microtubule cytoskeleton to give rise to the primary cilia.

### 5.3.3- Primary cilium and cell migration

Several works showed that primary cilia coordinate a series of signaling pathways critical to fibroblast cell migration during development and in wound healing. In particular, platelet-derived growth factor receptor alpha (PDGFRα) is compartmentalized to the primary cilium to activate signaling pathways that regulate reorganization of the cytoskeleton required for lamellipodium formation and directional migration in the presence of a specific ligand gradient (**401, Christensen**). It was noted that the primary cilium points in the direction of cell migration (**402, Albrecht-Buehler**). Especially, PDGFRa is transported to and imported into the growing cilium where it dimerizes, becomes phosphorylated, and signals via the Akt and MEK1/2–ERK1/2 pathways to control directional cell migration by influencing the transport and positioning of an Na+/H+ exchange protein to the lamellipodium (**403, Clement; 404, Schneider**).

### 5.3.4- Primary cilium and immune synapse

The intraflagellar transport (IFT) system was identified responsible for the assembly of the primary cilium, in the non-ciliated T-cell, where it controls immune synapse assembly by promoting polarized T-cell receptor recycling**.** It was shown that immune synapse and the primary cilium, which are both characterized by a specialized membrane domain highly enriched in receptors and signaling mediators, share architectural similarities and are homologous structures (**405, Finetti 2013**). Despite their obvious morphological differences, the immune synapse and the primary cilium both represent specialized membrane domains dedicated to sense the microenvironment and orchestrate the cellular response to extracellular cues. Moreover, similar to primary cilia, the immune synapse is enriched in lipid rafts, which are an important determinant in assisting the segregation of specific proteins to this location (**406, Finetti 2011**)**.** Vesicular trafficking ensures the transport to the immune synapse of both receptors and membrane-associated signaling mediators from intracellular pools that are associated with recycling endosomes. In parallel, the finding that IFT/BBS proteins interact with Rab11, which is associated with the pericentrosomal recycling compartment, and Rab8 (**407, Nachury**) which is implicated in protein recycling to the apical membrane in polarized cells, strongly suggests that ciliary proteins may be targeted to the periciliary membrane by recycling endosomes. Moreover, IFT20 is centrally implicated in the assembly of the immune synapse, which strongly supports the notion that the immune synapse and the primary cilium are homologous structures sharing an ancestral trafficking machinery for receptor targeting to the respective membrane domains. Finally, among the regulators of signaling recruited to both the immune synapse and the primary cilium are specific ubiquitin-conjugating enzymes that are centrally implicated both in the regulation of signaling and in the breakdown of these structures (**405, Finetti 2013**).

The implication of PEBP1 in primary cilium is coherent with its own role in modulating signaling pathways from various receptors, including chemical cues, flow and compression forces. Particularly, the ciliary pocket which is part of the connecting cilium/transition zone (CC/TZ) where CEP290 and PEBP1 were colocalized, is a site for exocytosis and endocytosis, which regulates the spatiotemporal trafficking of receptors into and out of the cilium to control its sensory function. It is to note that the ciliary pocket is also an interphase for the actin cytoskeleton. Finally, the relationships between primary cilium, centrosome, extracellular matrix, cell migration and immune synapse are in complete accordance with the multiple functions of PEBP1.

## 5.4- Crosstalk between apoptosis and autophagy

PEBP1 was described to control cell death signaling by regulating the pathways Raf/MEK/ERK (**13, Yeung 1999**), TAK/NIK/IKK/NF-κB (**91, Yeung 2001**), PI3K/Akt/mTOR (**16, Lin**) and MEK/ERK/CREB (**41, Zuo**). Moreover, PEBP1 is a member of the regulation loop NF-κB/Snail/YY1/PEBP1/PTEN implicated in the NO-mediated sensitization of resistant tumor cells (**408, Bonavida**). In cancer, cells overexpressing PEBP1 were resensitized to TRAIL apoptosis (**323, Baritaki**). The expression levels of PEBP1 were observed to be associated with variation in expression of Bcl-2 (**136, Hu**). In cellular senescence, PEBP1 induction is fully dependent on p53 and in turn, PEBP1 is responsible for p53-mediated ERK suppression and senescence (**409, Lee**). Interestingly, it turns out that Bcl-2 and p53 are two important proteins in both apoptosis and autophagy. Finally, PEBP1 was found to bind LC3, a major protein regulating autophagy (**44, Noh**). Altogether, these data indicated that PEBP1 is involved in pro-apoptotic or anti-apoptotic processes and that it may participate in bifurcation between intrinsic and extrinsic apoptosis and in autophagy. *De facto*, it is well-known that apoptosis and autophagy are two complicated processes which furthermore are intricately linked. To clarify this complex issue, we describe below the broad outlines for each apoptosis and autophagy and we discuss the known crosstalk between them.

### 5.4.1- Apoptosis, intrinsic and extrinsic pathways

Apoptosis is a form of programmed cell death, or "cellular suicide". Apoptosis plays critical roles in development, immune response, and tissue homeostasis; its deregulation leads to accumulation of "unwanted" cells and contributes to cancer development (**410, Su; 411, El Khattouti**). The pathways, which are involved in the regulation of apoptosis, are extremely complicated and appear to be both agent-dependent and tissues-specific (**412, Ng**). Apoptosis can be initiated through two major pathways, namely the intrinsic and the extrinsic pathways. In the *intrinsic pathway* the cell kills itself in response to cellular stress. The intrinsic pathway is triggered by a wide range of cellular stress signals, including DNA damage, endoplasmic reticulum stress, oxidative stress and growth factor withdrawal (**413, Tait**). A key step in the initiation of intrinsic apoptosis is mitochondrial outer membrane permeabilization (MOMP), which enables the release of pro-apoptotic factors, from the mitochondrial intermembrane space to the cytosol. The translocation of these intermembrane pro-apoptotic proteins, including cytochrome c and Smac/Diablo, is modulated by the members of Bcl-2 family (**413, Tait**). In the *extrinsic pathway* the cell kills itself because of signals from other cells. This pathway is initiated by transmembrane receptors such as Fas receptor, TNF-α receptor and death receptors DR3, DR4 and DR5. TNF-related apoptosis-inducing ligand (TRAIL) is a cytokine that is secreted by most normal tissue cell, it binds to the death receptors DR4 and DR5. TRAIL also binds the receptors DcR1 and DcR2, activating NF-κB and leading to transcription of genes known to antagonize the death signaling pathway and/or to promote inflammation (**414, Chicheportiche; 415, Luo**). Furthermore, it should be noted that Bcl-2 family proteins are critical in regulating both the intrinsic and extrinsic apoptotic pathways (**416, Zhao**).

### 5.4.2- Autophagy is a cytoplasmic quality-control mechanism

Autophagy is the natural, regulated, destructive mechanism of the cell that disassembles unnecessary or dysfunctional components leading to their orderly degradation and recycling. Autophagy is an ancient and highly conserved intracellular degradation process that is induced under various forms of cellular stress. Autophagy is involved in vesicular formation, transport, tethering, and fusion. In most cases, autophagy is a process of programmed cell survival and blocks the induction of apoptosis (**416, Zhao**). As a cytoplasmic quality-control mechanism, autophagy is vital for cellular homeostasis, and the dysfunction of autophagy may lead to many diseases such as neurodegenerative diseases, cardiomyopathy, tumorigenesis, and pathogenic infection (**151, Ao**). Autophagy starts with the formation of an isolation membrane called phagophore, usually at the contact sites between mitochondria and the endoplasmic reticulum. Upon autophagy induction, a portion of cytoplasm, including organelles, is enclosed by the phagophore to form the autophagosome. The outer membrane of the autophagosome subsequently fuses with the lysosome, and the internal material is degraded in the autolysosome, after which the degradation products are recycled by the cell (**410, Su**).

Here, we must remember that LC3, a cytosolic protein, can associate with intracellular membranes through PE-dependent conjugation to regulate initiation of autophagy. The PE-conjugated LC3 protein plays a critical role in regulating the formation of autophagosomes and sequestering cellular components into the phagophore lumen. PEBP1 interacted with the cytosolic unconjugated LC3 form, but not PE-conjugated LC3, indicating that PEBP1 could regulate autophagy before inclusion of PE-conjugated LC3 within autophagosomes (**44, Noh**).

Besides its role in death and cell survival, autophagy is also involved in regulating cell cycle arrest. The mechanism responsible for modulating autophagy-induced cell cycle arrest is thought to be regulated by inhibition of the mTOR pathway. What is more, autophagy may also help to maintain optimal, high ATP levels that may facilitate the apoptotic process (**417, Maiuri**).

### 5.4.3- Bcl-2, p53 and caspases are molecular crosstalks between apoptosis and autophagy
#### 5.4.3.1- Bcl-2 role in apoptosis and autophagy

Apoptosis and autophagy share some common upstream signaling components including Beclin 1, Bcl-2, p53, and mTOR, all known to interfere with PEBP1 activity. They may act independently in parallel pathways or may influence each other *via* interconnected signaling proteins (**416, Zhao**). Bcl-2 is the founding member of a family of regulator proteins that regulate cell death, by either inducing or inhibiting apoptosis. In most cancer types Bcl-2 is overexpressed while PEBP1 is downregulated. An exception was found in multiple myeloma, in which the overexpression of Bcl-2 was observed along with overexpression of PEBP1 but in this case, PEBP1 was in its phosphorylated form (**418, Shvartsur**). Bcl-2 family members are divided into three subgroups: 1- Bax and Bak, which permeabilize the mitochondrial outer membrane by forming proteinaceous pores; 2- the anti-apoptotic family members Bcl-2, Bcl-xL, Mcl-1, Bcl-W, and Bcl-2a1 which oppose mitochondrial outer membrane permeabilization (MOMP); 3- Bim, Bid, Puma, Noxa, Bad, Bik, Bmf, and Hrk, which induce apoptosis by facilitating Bax and/or Bak activation (**419, Dai**).

Anti-apoptotic Bcl-2 family proteins are important regulators of intrinsic apoptosis by controlling the mitochondrial outer membrane permeabilization (MOMP) (**420, Youle**) and inhibiting cytochrome c release from the mitochondria (**421, Kilbride**). Bcl-2 family proteins serve mainly to inhibit pro-apoptotic proteins, such as Bax and Bak. Beclin-1 is a component of the class III PI3K/Vps34 complex and is necessary for forming the autophagy vesicle. Beclin 1, acting as a critical autophagy initiator, interacts with Bcl-2, thereby regulating both apoptosis and autophagy. Thus, the interplay between Bcl-2 and Beclin-1 is essential to regulate the crosstalk between autophagy and apoptosis (**422, Rubinstein**).

The balance between apoptosis and autophagy is complicated and depends of the cellular status. For instance, under the condition of sufficient nutrition, Beclin-1 and Bax/Bak bind to Bcl-2 to inhibit autophagy or to Bcl-xL to inhibit apoptosis. Under conditions of starvation, C-Jun N-terminal protein kinase 1 (JNK1) is activated and phosphorylates Bcl-2 leading to Bcl-2-Beclin 1 complex destruction, which induces autophagy (**423, Wei**). However, in the situation of extreme starvation, JNK1 promotes hyper-phosphorylation of Bcl-2 leading to Bcl-2 separating from Bax. Then, it induces apoptosis via caspase-3-dependent pathways (**424, Li**).

It is important to note that compartmentalization of Bcl-2 to different cellular organelles allows independent control of autophagy and apoptosis. At the endoplasmic reticulum, Bcl-2 inhibits autophagy by binding to Beclin 1. The binding of Bcl-2 to Beclin 1 is regulated by several mechanisms, including competitive binding of Bik to Bcl-2, phosphorylation of Bcl-2 by JNK1, and phosphorylation of Beclin 1 by DAPK. At the mitochondrion, Bcl-2 inhibits apoptosis by interacting with Bax and Bak. Bad, Noxa, Bid and others activate apoptosis by inhibiting Bcl-2. The specific proteins Bid, Bim and Puma are also able to induce apoptosis by directly activating Bax and Bak (**422, Rubinstein**). Thus, the role of Beclin-1 in regulating autophagy is mediated by the localization of Bcl-2 at the endoplasmic reticulum. In turn, the regulation of apoptosis can be mediated through the localization of Bcl-2 to mitochondria. Accordingly, the cellular outcome between autophagy and apoptosis appears to depend on the entity of the mitochondria rather than the molecular mechanism induced (**411, El-Khattouti).**

### 5.4.3.2- p53 role in apoptosis and autophagy

Besides the Bcl2 protein family, p53 is also implicated in both apoptosis and autophagy. Upon stress, a cytoplasmic pool of p53 rapidly translocates to the mitochondrial surface, where it physically interacts with both anti- and pro-apoptotic Bcl-2 family members to inhibit or activate their respective functions, leading to MOMP and apoptosis (**425, Nikoletopoulou**). p53 is a master regulator of apoptosis in both the extrinsic and the intrinsic pathways. In the extrinsic pathway, nuclear p53 increases the expression of death receptors such as the Fas receptor and the TRAIL receptor DR4/5 (**426, Kuribayashi**), while cytoplasmic p53 activates caspase 8 and caspase 3. In the intrinsic apoptotic pathway, nuclear p53 activates the expression of the proapoptotic proteins such as Pidd, Puma, Noxa, Bax, and Bid; leading to cytochrome c release, and activation of caspase-9 and caspase-8. Meanwhile, cytoplasmic p53 translocates to the mitochondria and forms a complex with Bcl-2/Bcl-xL to liberate the proapoptotic proteins Bax and Bak (**427, Moll**). In contrast to apoptosis upregulation by p53, cytoplasmic and nuclear p53 have contradictory roles in regulating autophagy. Cytoplasmic p53 inhibits autophagy through the activation of mTOR signaling *via* the inactivation of AMP kinase, while nuclear p53 activates autophagy by transcriptional activation of DRAM (damage-regulated autophagy modulator) which promotes the formation of autophagolysosomes (**410, Su**).

Moreover, the class I PI3K/Akt/mTOR signaling is a well-known pathway involved in the regulation of both apoptosis and autophagy and may contribute to simultaneous or sequential induction of both processes (**416, Zhao; 428, Levine**).

### 5.4.3.3- Role of Caspases in apoptosis and autophagy

Caspases play also an important role in the interconnection between apoptosis and autophagy. For example, when Beclin 1 is bound to, and inhibited by Bcl-2 or Bcl-xL, this interaction can be disrupted by phosphorylation of Bcl-2 and Beclin 1, or ubiquitination of Beclin 1. But more interestingly, caspase-mediated cleavage of Beclin 1 promotes crosstalk between apoptosis and autophagy (**429, Kang**). While autophagy mediates the degradation of caspases, caspases can also target the cleavage of autophagy-related proteins (**411, El-Khattouti**).

## 5.4.4- Apoptosis and autophagy in cancer

In sum, autophagy and apoptosis are both cellular degradation pathways essential for organismal homeostasis. Therefore, both autophagy and apoptosis have been implicated in protecting organisms against a variety of diseases, especially cancer (**430, Liu**). Initially, autophagy was identified as a tumor suppressor pathway, as in normal cells it facilitates the degradation of oncogenic molecules. However, unlike apoptosis, the role of autophagy in cancer appears to be very diverse, dictated primarily by cellular contexts (**410, Su**). Apoptosis and autophagy often occur in the same cell, mostly in a sequence in which autophagy precedes apoptosis. For example, some anticancer agents induce autophagy preceding apoptosis through mitochondrial damage and/or endoplasmic reticulum stress (**431, Park**). The concept that inhibition of apoptosis results in autophagic/necrotic cell death is now extended, and it is clear that apoptosis and autophagy can act as partners to induce cell death in a synergistic or cooperative fashion (**432, Kumar**).

In conclusion, apoptosis and autophagy are two intertwined processes regulated by common proteins and pathway. In this context, it is not surprising that PEBP1 participates in the two events by interacting with several shared main regulators as Beclin 1, Bcl-2, p53 and PI3K/Akt/mTOR pathway.

## 5.5- Autophagy, focal adhesion and cell migration

PEBP1 was found to interact with LC3 to inhibit its binding to PE, preventing starvation-induced autophagy. These data indicated that PEBP1 could associate with LC3 to regulate autophagy before inclusion of LC3-II proteins within autophagosomes (**44, Noh 2016**). Upon the activation of autophagy, kinases may phosphorylate PEBP1 to liberate LC3, leading to PE conjugation which is a critical step necessary for autophagosome biogenesis. Thus, overexpression of PEBP1 significantly prevented starvation-induced autophagy, while, knockdown of PEBP1 expression induced high levels of autophagy *via* stimulation of LC3-lipidation and Akt/mTORC1 activation (**44, Noh 2016).**

A large number of proteins regulate various stages of autophagy, in particular, actin nucleation play an important role in the regulation of autophagosome formation and maturation. (**433, Coutts**). It was also shown that autophagy inhibition reduces tumor cell migration and invasion *in vitro* and attenuates metastasis *in vivo* (**434, Sharifi**). In contrast to starvation-induced autophagy, which involves nonspecific bulk uptake of cellular constituents, basal autophagy is highly selective in order to prevent the unwanted removal of cellular material. Selective autophagy should be modulated in time and space for proper coordination of focal adhesion assembly and disassembly (**435, Kenific**).

### 5.5.1- Autophagy and focal adhesion

Cell-matrix focal adhesions (FAs) are transmembrane protein complexes that traverse cytoskeletal infrastructures all the way to the extracellular matrix, producing traction at the leading edge of the cell, thus allowing for motility. Using live-cell imaging, it was seen that autophagy promotes optimal migratory rate and facilitates the dynamic assembly and disassembly of cell-matrix FAs which is essential for efficient motility. Additionally, autophagosomes associate with FAs primarily during disassembly, suggesting autophagy locally facilitates the destabilization of cell-matrix contact sites (**435, Kenific**). Furthermore, the selective autophagy cargo receptor neighbor of BRCA1 (NBR1) was identified as a key mediator of autophagy-dependent FA remodeling. It appeared that NBR1 enhances FA disassembly and reduces FA lifetime during migration (**435, Kenific**).

Otherwise, numerous abnormally large FAs accumulate in autophagy-deficient tumor cells, reflecting a role for autophagy in FA disassembly through targeted degradation of paxillin. It was demonstrated that paxillin interacts with processed LC3 through a conserved LIR motif in the amino-terminal end of paxillin and that this interaction is regulated by oncogenic Src activity (**434, Sharifi**). It appeared that interaction between the LC3-interacting region of NBR1 and LC3 proteins promotes the targeting of autophagosomes to focal adhesions which leads to sequestration of focal adhesion proteins (**436, Kenific**). The resulting focal adhesion disassembly drives turnover of cell–matrix adhesions and promotes cell migration (**437, Xu**). Clearly, autophagy contributes to the destabilization and turnover of cell–matrix contacts by locally capturing focal adhesion proteins. Tension induces autophagy, as shown by an increase in autophagosome-incorporated LC3 proteins coinciding with increased tension, suggesting that localized tension at focal adhesions could be one mechanism through which autophagy is spatially and temporally regulated (**438, Ulbricht**).

## 5.5.2- Focal adhesions direct cell migration

It was proposed that, at the front-most part of the cell, chemotactic GPCRs activated by a gradient of ligand could inhibit autophagy to favor the efficient formation of adhesions, while autophagy would remain active at distance from the site of GPCR activation/signaling in order to enable focal adhesion disassembly. Specifically, degradation of focal adhesion components, through selective autophagy, was shown to participate in the turnover of adhesions during cancer cell migration (**439, Coly**). Autophagic degradation of key proteins participating in actin remodeling may also constitute an efficient way of clearing these proteins from the cell rear and concentrating them at the cell front, in order to initiate the expansion of a single lamellipodium in the direction of the chemotactic stimulus (**439, Coly**).

Moreover, focal adhesions are highly dynamic at the leading edge of migrating cells, undergoing continuous cycles of assembly and disassembly. GSK3β could have a role in spatially restricting selective autophagy to leading-edge focal adhesions. GSK3β is locally inactivated at leading-edge lamella where focal adhesions turnover (**440, Barth**) and phosphorylation of NBR1 by GSK3β inhibits its ability to turnover ubiquitylated proteins (**441, Nicot**). This regulation suggests that NBR1 is preferentially activated at leading-edge focal adhesions where GSK3β is not active.

Focal adhesions direct migration by mechanically transmitting forces generated by the actin cytoskeleton for movement such that the formation of focal adhesion sites at the front of the cell establishes traction for forward propulsion (**442, Gardel**). Filamin A is an actin cross-linker that stabilizes the actin cytoskeleton. It also functions to link integrins to actin and is recruited to focal adhesions in a tension dependent manner to reinforce tensional stresses on focal adhesions (**443, Wolfenson**). It was also demonstrated that targeting of microtubules to focal adhesions promotes their disassembly and it was proposed that microtubules might have an important role in promoting recruitment of autophagy proteins, including LC3 and NBR1, to focal adhesions.

In conclusion, it appears that interaction between the LC3-interacting region of NBR1 and LC3 proteins promotes the targeting of autophagosomes to focal adhesions which leads to sequestration of focal adhesion proteins. The fact that PEBP1 interacts with LC3 and on the other hand may physically interact with MDA-9 in a manner that correlates with a suppression of FAK and c-Src phosphorylation is strongly indicative that PEBP1 may take part in the regulation of focal adhesion disassembly, a main event to allow cell motility.

# 5.6- Connections between endosomes, exosomes and autophagy

## 5.6.1- Autophagy and endosomes
### 5.6.1.1- The origin of autophagosomes membranes
The formation and maturation of autophagosomes is a complex process, entailing vesicular trafficking and fusion events, and substantial reorganization of existing compartments, or subdomains of compartments, in particular, the endoplasmic reticulum but including the plasma membrane, Golgi, and mitochondria. In canonical autophagy, which originates in the endoplasmic reticulum, remodeling of the endoplasmic reticulum-derived autophagosome must occur to allow for their fusion with the endosome and lysosome (**444, Tooze**). Endocytosed material first enters endocytic vesicles, which then consolidate into early endosomes. After sorting, the material destined to be degraded is delivered to late endosomes, with which lysosomes fuse in a transient manner to pick up the material for final digestion and absorption. Lysosomes also drive autophagy and eventually fuse with the autophagic vacuole (**445, Kiselyov**). Late endosomes are also known as multivesicular bodies, because late endosomes have a multivesicular morphology. Taken together, the early endosomes, late endosomes, endolysosomes, and lysosomes provide a dynamic and adaptable continuum. The pathway is elusive because the vesicles are scattered, and undergoing continuous maturation, transformation, fusion, and fission (**446, Huotari**).

### 5.6.1.2- Early endosomes and autophagy
In HeLa and HEK293A cells it was shown that fusion of autophagic vacuoles with functional early endosomes is required for autophagy. Inhibition of early endosome function by loss of the coatomer COPI subunits (β', β, or α) resulted in accumulation of autophagosomes. The data demonstrated that LC3–positive autophagic vacuoles accumulated in COPI-depleted cells due to an inhibition of early endosome function, resulting in the formation of autophagic vacuoles that cannot mature. These results provided direct evidence that early endosome fusion with autophagic vacuoles is a prerequisite for autophagic vacuole maturation, and occurs before fusion with late endosomes and multi-vesicular bodies (**447, Razi**).

### 5.6.1.3- Late endosomes and autophagy
Proteins and complexes of the endocytic pathway involve Rab proteins, clathrin coats and ESCRTs (endosomal sorting complex required for transport) (**448, Papandreou**). The role of ESCRTs was demonstrated to interplay between late endosomes and autophagosomes formation. The ESCRT complexes enable a unique mode of membrane remodeling that results in membranes bending/budding away from the cytoplasm. Moreover, the ESCRT machinery plays a vital role in a number of cellular processes including multivesicular body (MVB) biogenesis, cellular abscission, and viral budding (**449, Schmidt; 450, Babst).** Defects in the endosomal-lysosomal pathway have been implicated in a number of neurodegenerative disorders. A key step in the endocytic regulation of transmembrane proteins occurs in a subset of late-endosomal compartments known as multivesicular bodies (MVBs), whose formation is controlled by ESCRT. Particularly, ESCRT-III dysfunction was associated with the autophagy pathway, that frontotemporal dementia linked to chromosome 3 (FTD3) (**451, Lee**). More recently, an interaction between the ATG12-ATG3 conjugate and the ESCRT-associated protein PDCD6IP/Alix was identified to promote basal autophagy and endolysosomal trafficking. In addition, the results highlighted the importance of late endosomes for basal autophagic flux and revealed distinct roles for the core autophagy proteins ATG12 and ATG3 in controlling late endosome function. Moreover, ATG12-ATG3 was required for diverse functions including late endosome distribution, exosome secretion, and viral budding (**452, Murrow**).

### 5.6.1.4- Recycling endosomes and autophagy

A recent result was that the two autophagic proteins ATG9 and ATG16L1 both traffic from the plasma membrane to autophagic precursor structures following two different routes. ATG9 localizes on the plasma membrane in clathrin-coated structures and was internalized following a classical endocytic pathway through early and then recycling endosomes. By contrast, ATG16L1 was also internalized by clathrin- mediated endocytosis but via different clathrin-coated pits, and appeared to follow a different route to the recycling endosomes, bypassing the early endosomal compartment. Then, The R-SNARE VAMP3 mediates the coalescence of the 2 different pools of vesicles (containing ATG16L1 or ATG9) in recycling endosomes (**453, Puri**).

The small GTPase Rab11 regulates endosomal traffic and function at the level of recycling endosomes. In *D. melanogaster*, loss of Rab11 led to accumulation of autophagosomes and late endosomes. Under fed conditions, Rab11 localized to recycling endosomes, and Hook anchored late endosomes to microtubules, thereby suppressing endosome maturation and fusion events. Autophagy induction by starvation resulted in the translocation of Rab11 from recycling endosomes to autophagosomes. Meanwhile, Rab11 removed Hook from late endosomes, allowing subsequent fusion of late endosomes with autophagosomes (**454, Szatmari**).

In conclusion, collectively, all the results described above demonstrate a role for both early endosomes, recycling endosomes and late endosomes in autophagy.

### 5.6.2- Autophagy and exosomes

### 5.6.2.1- Secretory autophagy

During autophagy, all cellular macromolecules including lipids, polysaccharides, and proteins are degraded in lysosomes by virtue of their hydrolase. Autophagy summarizes several pathways, by which macromolecules can access the lysosomal lumen from the cytosol. Macro-, micro- and chaperone-mediated autophagy are the main pathways (**455, Münz**). While micro- and chaperone-mediated autophagy perform this import directly across lysosomal or late endosomal membranes, macroautophagy generates new vesicles around its substrate. These double-membrane surrounded autophagosomes then fuse with lysosomes for degradation of the inner autophagosomal membrane and its cargo. However, these autophagosomes do not automatically fuse with lysosomes and they can also be diverted to fuse with the cell membrane for non-canonical exocytosis (**455, Münz**). Thus**,** the autophagic machinery is also involved in a process termed secretory autophagy, a route for unconventional secretion of cytoplasmic substrates (**456, Hessvik**). In conventional secretion, proteins possessing a signal peptide enter the lumen of the endoplasmic reticulum followed by the Golgi apparatus for secretion at the plasma membrane. In unconventional secretion, cytoplasmic proteins that lack a signal sequence are not delivered to the lumen of the endoplasmic reticulum but are transported into the extracellular milieu *via* diverse unconventional secretory pathways, including secretory autophagy (**457, Ponpuak**). It was demonstrated that inhibition of the phosphoinositide kinase PIKfyve increased exosome secretion and induced secretory autophagy. This might be caused by impaired fusion of lysosomes with both MVBs and autophagosomes, and increased fusion of MVBs with autophagosomes. The generated enlarged MVB-like structures are called amphisomes (**456, Hessvik**).

### 5.6.2.2- Proteins and lipids in secretory autophagy

Secretory autophagy is associated specifically with the autophagy pathway defined by ATG factors that govern the biogenesis of autophagic membranes. The ATG factors directing canonical autophagy include ULKs (unc-51-like kinases), Beclins and LC3s (**457, Ponpuak**). Although the molecular machinery of autophagy, namely Atg proteins, was originally identified by its ability to allow cells to survive starvation via lysosomal degradation to recycle cellular components, it has recently become apparent that it is used by cells to secrete cytoplasmic constituents. Furthermore, viruses have learned to use this Atg supported exocytosis to exit cells, acquire envelopes in the cytosol and select lipids into their surrounding membranes that might allow for increased robustness of their virions and altered infection behavior (**455, Münz**).

Rab GTPases regulate many steps of membrane trafficking, including vesicle budding, transport of vesicles along actin and tubulin, as well as membrane fusion. Knockdown of Rab2b, Rab5a, Rab9a, Rab27a and Rab27b decreased exosome secretion (**458, Hessvik 2017**). In mammalian cells, secretory autophagy of IL-1b depends upon the autophagy factor Atg5, and the small GTPase Rab8a, which regulates vectorial sorting to plasma membrane (**459, Dupont**). Rab8a has also been reported to regulate autophagic secretion of other cargo, including α-synuclein (**460, Ejlerskov**), a major constituent of Lewy bodies, which are the hallmark protein aggregates associated with Parkinson's disease. Whereas Rab8a**,** a regulator of polarized sorting to the plasma membrane, was shown to be needed for secretory autophagy (**459, Dupont**), its closely related isoform Rab8b may be more important for maturation of the degradative autophagosome (**461, Pilli**). Furthermore, astrocytes normally secrete an insulin-degrading enzyme (IDE) extracellularly in the brain, one of the major proteases of the amyloid-β peptide which prevents its accumulation, a major marker of Alzheimer's disease. IDE is secreted into the cerebrospinal fluid by Rab8a and GORASP (Golgi reassembly stacking protein) (**448, Papandreou**).

Secretory autophagy of α-synuclein (a cytosolic cargo) was enhanced by inhibiting fusion of autophagosomes with lysosome using bafilomycin A1 (**460, Ejlerskov**). In addition to α-synuclein aggregate expulsion, secretory autophagy has also been implicated in extrusion of amyloid beta peptide aggregates associated with Alzheimer's disease (**462, Nilsson**). Additional leaderless cytosolic proteins have been reported in the autophagy-dependent secretome, including galectins (galectin-3), cytoskeletal proteins (tubulin), and others such as annexin-I (**463, Öhman**).

Finally, lipids may also be important for sorting of specific proteins into exosomes because exosomal membranes may contain lipid rafts, membrane subdomains enriched in cholesterol and glycosphingolipids that play important roles in signaling and sorting (**458, Hessvik 2017**).

### 5.6.2.3- Omegasomes, autophagosomes, amphisomes, exosomes; overlaps between exosome pathway and autophagy

At the endoplasmic reticulum exit sites, in mammalian cells, omegasomes facilitate the formation of degradative autophagosomes, of which the elongation and closure require LC3. Upon closure, degradative autophagosomes display motility toward the minus end of the microtubules where they fuse with lysosomes resulting in the degradation of the engulfed contents by lysosomal hydrolases (**457, Ponpuak**). Conversely in secretory autophagy, omegasomes located near endoplasmic reticulum exit sites aid the formation of secretory autophagosomes. Secretory autophagosomes show vectorial motility toward the plus end of the microtubules and eventually fuse with the plasma membrane releasing their contents into the extracellular environment.

Autophagosome maturation includes autophagosome elongation, trafficking and autophagosome/lysosome fusion (**448, Papandreou**). Autophagosomes fuse with lysosomes, but might also give rise to exosome secretion. Cytoplasmic substrates are recruited to forming autophagic membranes, the isolation membrane, *via* proteins that contain LC3-interacting regions (LIRs). After completion of autophagosome formation, LC3 is recycled from the outer autophagosomal membrane prior to fusion with lysosomes (**455, Münz**). Lysosomal hydrolysis degrades autophagosome cargo and the inner autophagosomal membrane. However, the inner autophagosome membrane and its content can also be secreted and might give rise to exosomes (**455, Münz**).

Exosomes are nanovesicles (30–150 nm in diameter) released from cells by fusion of multivesicular bodies (MVBs) with the plasma membrane. Exosomes were suggested to play a role in cell-to-cell communication (**464, Mathivanan; 465, Record**) and have been implicated in numerous physiological and pathological functions. Exosomes contain proteins, lipids and other metabolites, mRNAs, small RNAs such as miRNAs and probably DNA fragments, which can be delivered to recipient cells (**456, Hessvik 2016**).

Thus, several overlaps between the MVB-exosome pathway and autophagy exist (**466, Baixauli**). Particularly, in mammalian cells, MVBs are affected by autophagic machinery since autophagosomes and MVBs interact to form hybrid structures referred to as amphisomes (**467, Gordon; 468, Klionsky**). Finally, the notion that autophagy, endosomes and secretion are three distinct pathways which share components should be reconsidered as they are intertwined in a highly intricate manner and there are no clear-cut borders between these processes (**448, Papandreou**).

### 5.6.2.4- Roles of exosome secretion

The secreted exosomes might be targeted to and degraded by phagocytes, but they might also have other destinations. Since secretory autophagy is induced in cells with lysosomal dysfunction this might be, an alternative way of eliminating waste products (**458, Hessvik 2017**). However, exosomes secreted as waste are likely to affect neighboring cells and possibly induce pathological conditions. Another possibility is that cells might communicate to neighboring cells about intracellular stress by increasing exosome release (**458, Hessvik 2017**). Finally, exosome secretion may function in close relation with autophagy pathway to preserve protein and RNA homeostasis, and to mediate the spreading of signals to surrounding cells in order to coordinate organismal systemic responses (**466, Baixauli**).

Exosome secretion was described to play a role in several pathologies. It is established that loss of basal autophagy causes *neurodegeneration* and that exosomes are involved in the spread of toxic proteins in neurodegenerative diseases such as Alzheimer, Huntington, Parkinson, and prion diseases (**466, Baixauli**). In heart, reperfusion after an ischaemic insult might cause infarct extension. Mesenchymal stem cell (MSC)-derived exosomes could attenuate myocardial remodeling in animal models of myocardial ischemia reperfusion injury (MIRI) (**469, Liu**). *In vitro*, in H9C2 cells exposed to H2O2 for 12 h, treatment with exosomes enhanced autophagy via the AMPK/mTOR and Akt/mTOR pathways. Likewise, *in vivo* exosome injections in rats that underwent ischemia/reperfusion injury indicated that MSC-derived exosomes could reduce MIRI by inducing cardiomyocyte autophagy via AMPK/ mTOR and Akt/mTOR pathways (**469, Liu**).

The concept that autophagy, endosomes and secretion are intertwined in a highly intricate manner and that there are no clear-cut borders between them is important. It suggests that PEBP1 may play a global role in modulating and regulating molecular interactions in this set of events.

## 5.7- Primary cilium and autophagy

By analyzing multiple ATG, IFT and centriolar satellite components, it was found that autophagy initiates ciliogenesis on serum starvation but limits ciliary growth under nutrient-rich conditions. Autophagy and ciliogenesis are regulating each other and nutrition deprivation is a shared signal (**470, Pampliega; 471, Tang**). Autophagy plays both a positive and a negative role in the formation and length of the primary cilium (**470, Pampliega; 471, Tang**), whereas cilium can serve as a staging ground for autophagic machinery components (**471, Tang**). Autophagy helps initiate cilium formation under starvation conditions by reducing levels of OFD1, a protein that accumulates at the centriolar satellites where it inhibits ciliogenesis (**471, Tang**). Basal autophagy limits uncontrolled cilium growth under nutrient-rich conditions by diminishing intraflagellar transport protein IFT20 levels.

Thus, autophagy provides alternative trafficking routes for integral membrane proteins to reach the plasma membrane, but it also controls biogenesis of complex domains and organelles at the plasma membrane (**470, Pampliega**).

## 5.8- Connection between cytoskeleton and autophagy

Actin filament assembly and disassembly provide the mechanical forces for a wide range of cellular activities that involve membrane deformation, such as cell motility, phagocytosis, endocytosis and cytokinesis (**472, Anitei**). In fact, the membrane deformations implicated in numerous cellular processes are underspinned by cytoskeleton elements: in particular actin filaments of cell cortex and microtubules. Actin cytoskeleton dynamics play vital roles in most forms of intracellular trafficking by promoting the biogenesis and transport of vesicular cargoes. Thereby, branched actin polymerization is necessary for the biogenesis of autophagosomes from the endoplasmic reticulum membrane (**473, Kast**).

In higher eukaryotes, following their detachment, mature autophagosomes are transported using both actin- and microtubule based mechanisms (**474, Mauvezin**). In selective autophagy, the fusion of autophagosomes with lysosomes depends on the recruitment of cortactin to fusion sites by the ubiquitin- binding deacetylase HDAC6 (**475, Lee**). Cortactin is a monomeric protein located in the cytoplasm of cells that can be activated by external stimuli to promote polymerization and rearrangement of the actin cytoskeleton, it is important in promoting lamellipodia formation, invadopodia formation, cell migration and endocytosis (**476, Kirkbride**).

A tight spatiotemporal coordination of actin cytoskeleton and membrane dynamics is a distinctive feature of many cellular processes, including cell migration, morphogenesis and endocytosis. In these processes, Bin/Amphiphysin/Rvs (BAR) domain proteins have emerged as essential regulators, linking signaling pathways to the actin cytoskeleton and membrane remodeling. BAR domain proteins feature a curved membrane binding surface as well as other domains involved in protein–protein interactions, protein–membrane interactions and signaling that contribute to the membrane-binding activity and the recruitment of signaling and cytoskeletal proteins (**473, Kast**).

The annexin family of Ca2+-regulated phospholipid-binding proteins constitutes another group of adaptors implicated in membrane budding and fusion and in the recruitment of binding partners to specific membranes (**477, Gerke**). Of the 12 annexins expressed in humans, several interact directly with actin, including annexins A1, A2, A5 and A6, and may therefore physically connect membranes to actin filaments (**478, Hayes**). Of these, annexins A2 and A5 have been implicated in autophagy. Annexin A5 accumulates on lysosomal membranes during starvation, and overexpression and silencing experiments indicate that it induces autophagosome–lysosome fusion, thus increasing lysosomal protein degradation (**479, Ghislat**). Both BAR and annexin families that play a role in autophagy share the ability to recruit binding partners, including signaling proteins and actin (**473, Kast**).

Strikingly, each cellular process implicating PEBP1 is governed by molecular mechanisms that are shared with other processes. Of particular interest is that autophagy, endosomes and secretion are intertwined in a highly intricate manner suggesting a global role for PEBP1 in modulating this set of events. In a same way, apoptosis and autophagy are strongly networked, being regulated by several shared proteins (especially Beclin1, Bcl-2, p53) and PI3K/Akt/mTOR pathway. Primary cilium formation and mitosis are also interdependent processes as upon entry in mitosis the ciliary membrane is internalized together with one of the centrosomes. Moreover, CEP290 and Rab8a which were found to localize with PEBP1 in the transition zone of primary cilium, are also components of centrosomes. Indeed, CEP290 is a centrosomal protein and Rab8-dependent ciliary assembly is initiated by the relocalization of Rabin8 to Rab11-positive vesicles that are transported to the centrosome. All these data suggest that PEBP1 is associated with molecular machines to regulate and coordinate the assembly/disassembly of these organelles. In addition, endosomes, primary cilium and centrosomes may constitute platforms where PEBP1 can interact with partners and accompany or deliver the signaling pathways where needed.

Ultimately, the main cellular processes that underlie the primary PEBP1 features are probably primary cilium, autophagy and mitotic spindle checkpoint. Effectively, primary cilium is a sensory organelle capable of sensing chemical, physical and mechanical signals such as tension, pressure, flow, movement and temperature. In addition, the ciliary pocket which is part of the connecting cilium/transition zone where PEBP1 was localized, is a site for exocytosis and endocytosis, which regulates the spatiotemporal trafficking of receptors into and out of the cilium to control its sensory function. Autophagy also reflects a main feature of PEBP1 because it is a very important process to allow cell motility by governing focal adhesions. Yet PEBP1 plays an important role in controlling cell motility with dramatic effects in cancer when PEBP1 is downregulated. At last, mitotic spindle checkpoint is significant of PEBP1 implication in cell division. During mitosis PEBP1 controls Aurora B activity in order to trigger microtubules elongation when kinetochores are properly attached to the spindle apparatus.

Interestingly, many of the processes modulated by PEBP1 concerned membrane changes (invagination, protrusion, engulfment, vesicle formation or fusion) and all of them need fine-tuned cytoskeleton organization.

## 6- Some other PEBP isoforms: human PEBP4, TFL1 and TF1 in plants

PEBP isoforms have been characterized from numerous organisms such as mimivirus, bacteria, parasites, yeast, plants and mammals. In each studied species several PEBP isoforms were described. In human, apart from PEBP1, PEBP2 isoform was only colocalized in late spermatocytes and spermatids and on the midpiece of epididymal sperm. In addition to PEBP1, PEBP2 was demonstrated to participate in the MAP kinase signaling pathway, with a role in spermatogenesis and posttesticular sperm maturation (**11, Hickox**). Thus, the role of PEBP2 in membrane maturation and capacitation of sperm was clearly established (**49, Moffit**). PEBP4 appeared also to play a role in plasma membrane changes, but surprisingly, some of its final effects are opposite to those of PEBP1. This is particularly true in cancer where PEBP4 favors tumor and metastasis development. We discuss below some features of PEBP4.

## hPEBP4

The molecular cloning and characterization of PEBP4 was derived from human bone marrow stromal cells (BMSCs). The cDNA potentially encoded a 227-residue protein. RT-PCR analysis revealed that hPEBP4 mRNA is expressed in most human tissues and in a variety of tumor cells and freshly isolated cells. The expression of PEBP4 in tumor cells was enhanced upon TNF-alpha treatment. Whereas hPEBP4 normally co-localized with lysosomes, TNF-alpha stimulation triggered its transfer to the cell membrane, where it bound to Raf-1 and MEK1 (**12, Wang**). L929 cells overexpressing hPEBP4 were resistant to both TNF-alpha-induced ERK1/2, MEK1, and JNK activation and TNF-alpha-mediated apoptosis. The effects of TNF-alpha on hPEBP4 mRNA expression in A549 and LoVo cells, which do not normally express hPEBP4 provided evidence that hPEBP4 might be induced to translocate from lysosomes to the plasma membrane following cellular exposure to TNF alpha and that this translocation requires the binding site of PEBP4. Given that MCF-7 breast cancer cells expressed hPEBP4 at a high level, small interfering RNA was used to silence the expression of hPEBP4. It was demonstrated that down-regulation of hPEBP4 expression sensitizes MCF-7 breast cancer cells to TNF alpha-induced apoptosis. Finally, hPEBP4 was described to promote cellular resistance to TNF-induced apoptosis by inhibiting activation of the Raf-1/MEK/ERK pathway, JNK, and phosphatidylethanolamine (PE) externalization. hPEBP4 was thought to be targeted to the cell membrane and binds to PE situated on the inner leaflet of the plasma membrane, preventing PE externalization, and maintaining membrane phospholipid asymmetry (**12, Wang**).

The study of primary human skeletal muscle myoblasts (HSMM, non-cancerous cells) suggested that PEBP4 can form ternary complexes with Raf-1 and MEK, and might function as a scaffold protein for Raf-1/MEK interactions. This sharply distinguishes it from PEBP1, which disrupts the interaction between Raf-1 and MEK (**13, Yeung et al, 1999**), acting as a competitive inhibitor of MEK binding (**78, Yeung et al, 2000**). Importantly, a ternary endogenous Raf-1/MEK/PEBP4 complex also exists in primary human skeletal muscle cells (SkMC) and MCF7 cells. In contrast to PEBP1, which disrupts the Raf-1/MEK interaction, PEBP4 forms ternary complexes with Raf-1 and MEK, and can scaffold this interaction. PEBP4 expression was induced during the differentiation of primary human myoblasts. Consistent with the properties of a scaffold, PEBP4 enhanced the Raf-1/MEK interaction and the activation of MEK at low expression levels, whereas it inhibited these parameters at higher expression levels. Downregulation of PEBP4 by short hairpin RNA in human myoblasts increased MEK signaling and inhibited differentiation; by contrast, PEBP4 overexpression enhances differentiation (**480, Garcia**).

A sequence alignment of several isoforms of mouse and human PEBPs predicted that PEBP4 contains a signal peptide. To test if PEBP4 was secreted, constructs were made with Myc epitope at the amino (N) terminus or carboxyl (C) terminus to mask the signal sequence or keep it free, respectively. The data revealed that both mouse and human PEBP4 were secreted when the epitope was tagged at their C-terminus. Surprisingly, secretion was dependent upon the C-terminal conserved domain in addition to the N-terminal signal sequence. When the epitope was placed to the N-terminus, the recombinant protein failed to secrete and instead, was retained in the cytoplasm. Transfection of PEBP4 shRNA into HEK293T cells did not appear to affect ERK activation, suggesting that PEBP4 does not participate in the regulation of this pathway. In contrast, PEBP4 siRNA suppressed phosphorylation of Akt at S473 into HEK293T (**481, He**).

A homolog of human PEBP4 was purified from swine seminal plasma, it was named sPEBP4. The sequenced sPEBP4 displayed 201 aminoacids in length while the full-length cDNA encoded a protein of 222 residues, revealing a signal peptide of 21 residues in the N-terminal terminus of the protein. Immunohistochemical staining and western blotting analysis showed that sPEBP4 was secreted from epithelial cells in the epididymis to the seminal plasma. Sperms suspended in phosphate-buffered saline began to swim after the addition of purified sPEBP4, but not when swine serum albumin was added, indicating that sPEBP4 promotes sperm motility (**482, An**).

### PEBP4 in cancer

PEBP4 was described to be upregulated in various cancers. In colorectal carcinoma, the positive expression rate in the cancer tissues from patients with positive lymph node and distant metastasis was significantly higher than that from the patients negative for lymph node and distant metastasis. The over-expression of PEBP4 protein was related to the tumorigenesis, development, metastasis, and invasion of colorectal cancer (**483, Liu**). The same effects were observed in non-small cell lung cancer (**484, Wu 2012**), in lung squamous cell carcinoma (**485, Yu 2011**) and in gliomas tissues (**486, Huang**).

Numerous studies have linked the activation of Akt to the progression of cancer. This was the case in samples of pancreatic ductal adenocarcinoma (PDAC) in which the mRNA and protein levels of PEBP4 were elevated. It was found that overexpression of PEBP4 elevated the phosphorylation of Akt in the serine 473. The interaction between Akt and PEBP4 was demonstrated by immunoprecipitation in BXPC3 cells. Since PEBP4 interacted with Akt and promoted the phosphorylation of serine 473 in Akt, it was suggested that PEBP4 might promote the progression of PDAC through activating Akt signaling (**487, Zhang**). PEBP4 was found to cause tumor cell proliferation, migration and invasion by activating PI3K/Akt/mTOR signaling pathway in human breast cancer (**488, Wang**) and in gastric cancer (**489, Wu**). In another study, PEBP4 mRNA and protein expression were also markedly increased in the human prostate cancer tissues and cell lines. Knockdown of PEBP4 significantly inhibited hypoxia-induced migration/invasion and EMT program. Furthermore, knockdown of PEBP4 prevented the expression of p-Akt and p-mTOR induced by hypoxia in prostate cancer cells (**490, Li**). Finally, the overexpression of PEBP4 increases the phosphorylation levels of Akt and mTOR in lung cancer cells HCC827 suggesting that the PI3K/Akt/mTOR signaling axis may be a key molecular pathway *via* which PEBP4 promotes the proliferation and invasion of non-small cell lung cancer cells (**491, Yu 2015**). PEBP4 was highly expressed in various lung cancer cells (HCC827, A549, NCI-H661, NCI-H292, and 95-D), but its expression was low in a normal human bronchial epithelial (HBE) cell line. Cell viability, cell proliferation, and invasion of HCC827 cells in the PEBP4 knockdown group were significantly lower than that in the negative control and blank control groups. In HCC827 cells, the expression levels of cyclin D1, Bcl-2, MMP-2, and MMP-9 in the PEBP4 knockdown group were significantly lower and the expression of p53 protein was significantly higher than that in the negative and blank control groups (**492, Yu 2013**).

It was shown that the expression of PEBP4 could be on the control of miRNAs. Increased miR-15b expression was detected in tumor tissues sampled from lung adenocarcinoma patients treated with cisplatin-based chemotherapy and was proved to be correlated with low expression of PEBP4, decreased sensitivity to cisplatin and poor prognosis. These results suggested that upregulation of miR-15b could suppress PEBP4 expression and in turn contribute to chemoresistance of lung adenocarcinoma cells to cisplatin (**493, Zhao**). Another work demonstrated that overexpression of PEBP4 reduced the sensitivity of lung cancer cell line A549 to cisplatin-induced cytotoxicity and that PEBP4 expression was significantly decreased after transfection with miR-34a. These results suggested that PEBP4 is a target of miR-34a (**494, Yu 2014**).

The PEBP4 features are sharply different from that of PEBP1. Particularly, PEBP4 does not inhibit the Raf/MEK/ERK pathway but conversely activates the PI3K/Akt/mTOR pathway. Furthermore PEBP4 is normally localized with lysosomes and is transferred to cell membrane by TNF-alpha activation. These results suggested that the binding properties of PEBP4 toward molecular partners are significantly dissimilar than that of PEBP1.

## **PEBP family members in plants**

It is to note that several PEBP isoforms were described in plants and numerous experiments were performed in various species due to the important role of PEBP isoforms in vegetative growth and flowering. Two main families of PEBP isoforms are known to regulate flowering and vegetative growth, they are called Flowering Locus T (FT) and Terminal Flower 1 (TFL). Just to give some examples concerning the FT family, in sugar beet (Beta vulgaris), FT1 and FT2 have antagonistic functions in the control of flowering (**495, Pin**), and in poplar, FT1 regulates reproductive onset in response to winter temperatures, whereas vegetative growth and the inhibition of bud set are promoted by FT2 in response to warm temperatures and long days in the growing season (**496, Hsu**). In Norway spruce, gene expression and population genetic studies have suggested a role for TFL2 in the control of growth cessation and bud set as well as in local adaptation resulting in variation for timing of bud set. The temporal and spatial expression of FTL1 and FTL2 largely complemented each other, which suggested that they act in concert to control perennial growth in Norway spruce (**497, Karlgren**).

The floral inhibitor Terminal Flower1 (TFL1) is a homolog to FT. TFL1 acts as a repressor of flowering and extends the vegetative growth state while maintaining the indeterminate state of inflorescences (**498, Shannon; 499, Bradley**). TFL1 is expressed in the nucleus and cytoplasm but interacts with Flowering Locus D solely in the nucleus and represses genes activated by FT (**500, Hanano**).

These some examples are indicative of the role of FT and TFL1 as sensors of plant cell membrane changes during temperature variations and length of daylight dependent on season. FT and TFL1 are implicated in signal transduction, allowing the plants to adapt to their environment.

# **7- Discussion**

## **7.1- PEBP1 is a regulator and a switch in cell signaling**

PEBP1 was demonstrated to regulate several signaling pathways such as Raf/MEK/ERK (**13, Yeung 1999**), GRK2 (**88, Lorentz**), NF-κB (**91, Yeung 2001**), PI3K/Akt/mTOR (**16, Lin**), GSK3 (**96, Al-Mulla 2011**), p38 MAPK (**86, Vandamme**), Wnt (**14, Shin**), STAT3 (**98, Yousuf**), LIN28/Let-7/HMGA2 and BACH1 (**101, Dangi-Garimella; 102, Yun**), Notch1 (**18, Noh**). In most cases PEBP1 was demonstrated to regulate the phosphorylation cascades by inhibiting the kinases implicated in the pathway. Inhibition is due to physical interaction of PEBP1 with the kinases, preventing them to interact with their cascade targets. Only one exception was noted with GSK3β which is activated by PEBP1 (**96, Al-Mulla 2011**). The regulation of signaling pathways is dependent of the cell. This fact could be explained by the ability of PEBP1 to regulate the pathways that are actually activated in a given cell at a given time (**340, Gavilan**).

In addition to its role in regulation of signaling pathways, PEBP1 reveals to be potentially a switch between different pathways. Indeed, phosphorylation of PEBP1 at Ser153 by PKC switches the inhibitory effect of PEBP1 to GRK2 while releasing the Raf/MEK/ERK pathway (**89 Corbit**). In this case, PEBP1 functions as a switch between the Raf/MEK/ERK and GRK2 pathways. The same kind of mechanism is suggested for other processes. For instance, in autophagy, PEBP1 interacts with LC3 preventing the association of LC3 with phosphatidylethanolamine. The phosphorylation of PEBP1 at Ser153 induces the dissociation of PEBP1 from LC3, initiating autophagy (**44, Noh**). Another example is the phosphorylation of PEBP1 at Thr42 during mitosis that may trigger the mitotic spindle checkpoint by Aurora B activation (**140, Dephoure**). Loss of PEBP1 during mitosis phase led to bypass of the spindle assembly checkpoint and the generation of chromosomal abnormalities (**156, Rosner**).

## 7.2- Properties of protein partners and small ligands. Role of PEBP1 in molecular interactions and stability

Most of the proteins that bind PEBP1 are multifunctional, they are implicated in main cellular processes or machineries and bring into play several targets. Among them are CEP290 (**75, Murga-Zamalloa**), Rab8 (**75, Murga-Zamalloa**), LC3 (**44, Noh**), GSK3 (**96, Al-Mulla 2011**), syntenin (**55, Das**). In addition, the interactome of overexpressed PEBP1 in gastric cancer cell (**157, Gu**) revealed cytoskeletal proteins, biosynthesis proteins, glycolytic enzymes, molecular chaperones and mitochondrial proteins such as ATP synthase and VDAC1 (**see Table 1**). Among the cytoskeletal proteins are tubulin beta that polymerizes with tubulin alpha into microtubules and actin that form microfilaments. Actin filament assembly and disassembly are crucial to create mechanical forces for motility, phagocytosis, endocytosis and cytokinesis (**472, Anitei**). In addition, several proteins that bind actin such as plectin, F-actin-capping protein and vinculin were identified by the interactome analysis. Finally, proteins implicated in intermediate filaments such as keratins and vimentin were also characterized. Thus, the main components of cytoskeleton, essential to constitute actin filaments, microtubules and intermediate filaments were found as first or second partners of PEBP1.

Interestingly, the most part of the proteins partners of PEBP1 found by interactome analysis possess a binding site for nucleotides. In particular actin is known to be an ATPase and tubulin a GTPase, the 90-kDa heat shock protein (Hsp90) is a molecular chaperone that assists in an ATP-dependent manner the folding of numerous targets such as nuclear hormone receptors and protein kinases. CEP290 is a protein with an ATP/GTP binding site motif A and a region with similarity to the chromosome segregation ATPases SMC (structural maintenance of chromosomes). All the proteins implicated in protein biosynthesis display ATP or GTP binding sites, glycolytic enzymes hydrolyze different nucleotides and VDAC1 displays a binding site for NAD. Nucleotide concentrations may influence the binding of PEBP1 to these proteins. PEBP1 was demonstrated to be an ATP-binding protein and ATP attenuated the interaction between PEBP1 and Raf-1 resulting in increased activation of the downstream ERK signaling (**269, Huang**). Therefore, the ATP-binding function of PEBP1 renders its inhibition on Raf-1 modulated by cellular ATP concentrations, revealing how energy levels may affect the propagation of cellular signaling.

Apart from nucleotides and phospholipids, various small ligands were described to bind PEBP1 suggesting that in living cell, competition can be installed between small ligands that in turn could modulate interaction between PEBP1 and its protein targets. Surprisingly, PF-3717842, a soluble inhibitor of PDE5 was found to bind PEBP1 (**8, Davdar**). Locostatin, a cell migration inhibitor, was demonstrated to bind PEBP1 (**501, Rudnitskaya**). Locostatin disrupted interactions of PEBP1 with Raf-1 and also with GRK2. In contrast, locostatin did not disrupt binding of PEBP1 with IKKα and TAK1. After binding PEBP1, part of locostatin was slowly hydrolyzed, leaving a smaller PEBP1-butyrate adduct (**273, Beshir**). More astonishingly, PEBP1 was able to hydrolysate another ligand, namely the prodrug Prasugrel. This synthetic product is a thienopyridine antiplatelet prodrug that undergoes rapid hydrolysis *in vivo* to a thiolactone metabolite by human carboxylesterase-2 (hCE2) during gastrointestinal absorption. The estimated contributions to prasugrel hydrolysis were approximately 40% for PEBP1 and 60% for hCE2 (**275, Kazui**). Thus PEBP1 appears to destabilize locostatin and prasugrel. In contrary, PEBP1 was described to stabilize proteins such as Kelch-like ECH-associated protein1 (Keap1). PEBP1 enhanced KEAP1 stability in colorectal cancer tissues and HT29 CRC cell line. PEBP1 silencing in immortalized HEK-293 cells correlated significantly with Keap 1 protein degradation and subsequent NRF2 addiction in these cells (**53, Al-Mulla 2012 a**). Keap1 associates with actin filaments in the cytoplasm. Disruption of the actin cytoskeleton promotes nuclear entry of Nrf2. The actin cytoskeleton therefore provides scaffolding that is essential for the function of Keap1, which is the sensor for oxidative and electrophilic stress (**502, Kang**). In lung adenocarcinoma cells, Keap1 expression regulates rearrangements of F-actin-containing microfilament bundles, suggesting that F-actin-containing microfilament bundle rearrangements may be involved in Keap1-mediated cell invasiveness (**503, Chien**). Moreover, PEBP1 was also found among the major proteins responsible for the maintenance of human tear film stability (**504, Perumal**).

### 7.3- PEBP1 with membranes and vesicles

One can notice that several of the main cellular functions of PEBP1 match with plasma membrane changes and reorganization. Indeed, when phosphorylated by PKC, PEBP1 is known to regulate receptor internalization by inhibiting GRK2 (**88, Lorentz**). PEBP1 is also involved in autophagy (**44, Noh**), in sperm capacitation (**49, Moffit**), cilium formation (**75, Murga-Zamalloa**), adhesion (**54, Mc Henry**), motility (**36, Al-Mulla 2012 b**). PEBP1 was also found to be associated with different types of vesicles such as lipid rafts (**505, Goumon**) and exosomes (**35, Han**).

In brain hippocampus, a role for PEBP1 was also described in synapse and more particularly in presynapse (**56, Kato**). Especially, it was suggested that PEBP1, in association with unphosphorylated and/or pSer522-CRMP-2, plays an important role in presynaptic function in the mature hippocampus (**57, Mizuno**). The role of PEBP1 in presynaptic function was also described in neuroblastoma where didymin inhibits MYCN through Raf-dependent and -independent mechanisms by inhibiting clathrin-dependent endocytosis through PEBP1, GRK, PKC, and Let-7 micro-RNA (**103, Singhal**). Clathrin which is a partner of PEBP1 in the interactome analysis (**Table 1**) plays a role in synaptic vesicle endocytosis. Indeed, it is known that in neuromuscular junctions actin polymerization is required for fast endocytosis (**506, Watanabe 2013**). Thus, when actin polymerisation was blocked in cultured neurons, the failure of ultrafast endocytosis leads to clathrin-mediated endocytosis on the plasma membrane. In other words, in the absence of ultrafast endocytosis, synaptic vesicles are retrieved directly from the plasma membrane by clathrin-mediated endocytosis (**507, Watanabe, 2014**).

The possibility that PEBP1 plays also a role at the mitochondrial membrane was suggested by the fact that mitochondrial ATP synthase was identified in the interactome of overexpressed PEBP1 in gastric cancer cell (Table1). Furthermore, GRK2 antagonized ATP loss after hypoxia/reperfusion and was detected to localize in the mitochondrial outer membrane. Raf-1 was also encountered at the mitochondrial membrane (**284, Fusco**). During hepatitis B virus X-mediated hepatocarcinogenesis, the PEBP1 downregulation induced the translocation of Raf-1 into mitochondria by increase of the free Raf-1 level in the cytosol (**286, Kim**).

Vesicles are common features between endocytosis, autophagy and exocytosis. The early endosomes, late endosomes, autophagosomes and exosomes are in fact a continuum of vesicles and their membranes originate from different cell organelles such as plasma membrane but also from reticulum endoplasmic and mitochondria (**446, Huotari**). Endosomes play a role in signal propagation by facilitating the recruitment and integration of signaling cascades on the surface of endocytic vesicles (**355, Emery**) and it was demonstrated that canonical GPCR signaling occurs from plasma membrane and from endosomes (**356, Irannejad**). In fact, endosomes are platforms to transport signaling pathways, but they are also a physical platform for signaling complex assemblies where endosomal scaffolds can facilitate crosstalk between the recruited kinases (**358, Palfy**). Considering that PEBP1 controls signal transduction by inhibiting kinases, it is tempting to propose that PEBP1 uses endosomes to preserve signal intact during transport. It seems likely that the inhibition of kinases is needed to synchronize the fast kinases activities with the time necessary to transfer the signal across the whole cell during vesicle trafficking. It is also essential to synchronize the downstream physical membrane and cytoskeleton reorganization in response to proper signaling. We propose that PEBP1 may play a role in signal conservation during vesicle trafficking and synchronize the pathway signaling with cellular processes. In addition, endosomes could also be platforms where PEBP1 may coordinate various signaling pathways between them.

### 7.4- Common features and molecular interactions in cellular processes.

On the molecular point of view, PEBP1 appears to prevent several cellular mechanisms by dissociating interaction between partners.

### 7.4.1- PEBP1 dissociates the pathway cascades.

PEBP1 inhibits signaling pathways by physical interaction with the enzymes of the pathways. This was well demonstrated for Raf/MEK/ERK (**78, Yeung 2000**), TAK/NIK/NF-κB (**91, Yeung 2001; 93, Tang**). Direct interaction was also found with GRK2 (**88, Lorenz**), GSK3 (**96, Al-Mulla, 2011**), STAT3 (**98, Yousuf**) and Notch1 (**18, Noh**).

### 7.4.2- PEBP1 prevents the primary cilium formation.

It was shown that Cep290-mediated photoreceptor degeneration is associated with aberrant accumulation of PEBP1, and that Cep290 physically interacts with PEBP1. The concentrated signal of PEBP1 at the apical inner segment, transition zone, and to a lesser extent, at the basal body and centriole of the photoreceptors, together with its ability to interact with Rab8A suggested its involvement in the regulation of docking and transport of membrane protein-containing vesicles in photoreceptors (**75, Murga- Zamalloa**). The ectopic accumulation of PEBP1 led to defective cilia formation in zebrafish and cultured cells, this effect being mediated by the interaction of PEBP1 with the ciliary GTPase Rab8a.  PEBP1 preferentially associated with Rab8a in the presence of GDP substrate, suggesting that the PEBP1-Rab8a (GDP) complex may need to dissociate for the release of Rab8a-GDP and subsequent conversion to Rab8a-GTP for appropriate ciliary transport (**75, Murga-Zamalloa**). Primary cilia are microtubule-based organelles that are generated from mother centrioles (also called basal body) by an evolutionarily conserved process called intraflagellar transport (IFT). IFT and other ciliary proteins facilitate cilia formation and maintenance by cooperating with small GTPases such as Rab8a (**75, Murga-Zamalloa**). Activation of Rab8 is linked to the formation of filopodia, lamellipodia, protrusions, ruffles, and primary cilia, whereas inhibition of Rab8 affects negatively their appearance. All these structures are containing actin (**508, Peranen**). Rab8 was also associated with focal adhesions, promoting their disassembly in a microtubule-dependent manner. This Rab8 effect involved calpain, MMP14 and Rho GTPases. Moreover, the role of Rab8 was also demonstrated in the cell migration process. The data revealed that Rab8 drives cell motility by mechanisms both dependent and independent of Rho GTPases, thereby regulating the establishment of cell polarity, turnover of focal adhesions and actin cytoskeleton rearrangements, thus determining the directionality of cell migration (**152, Bravo-Cordero**). A specific role of Rab8 in signaling and endocytosis (**509, Esseltine**) and in protrusion (**510, Vidal-Quadras**) was also described in details.

### 7.4.3- **PEBP1 delays mitosis**.

By allowing the control of mitotic spindle checkpoint PEBP1 delays mitosis. Loss of PEBP1 accelerates G1/S and leads to the bypass of mitotic checkpoint (**155, Eves**). At the kinetochore, Aurora B phosphorylates, and thereby inactivates, the microtubule-destabilizing proteins mitotic centromere-associated kinesin and stathmin, which promotes microtubule growth around kinetochores (**319, Prosser**). By inhibiting Aurora B, PEBP1 prevents microtubule growth around kinetochores that is consistent with the report that PEBP1 depletion causes an override of mitotic checkpoints induced by spindle poisons such as Taxol (**155, Eves**).

### 7.4.4- **PEBP1 blocks autophagy initiation**.

By sequestering LC3, PEBP1 prevents its association with phosphatidylethanolamine (**44, Noh**). LC3 was first identified as one of three light chains associated with purified microtubule‑associated protein (MAP)1A and MAP1B and is considered to be involved in microtubule assembly/disassembly (**142, Tang**). However, the function of LC3 is known to extend above and beyond its role in autophagosome formation. Especially, the cytoplasmic nonlipidated form, interacts with the *Chlamydia trachomatis* inclusion as a microtubule-associated protein rather than an autophagosome-associated component. Inclusion-bound LC3 was shown to be connected with the microtubular network and depolymerization of the microtubular architecture disrupted the association of LC3/MAP1s with the inclusion (**511, Al-Younes**).

Thus, in signaling pathways, primary cilium, mitosis and autophagy, PEBP1 delays the mechanisms to allow their initiation at the correct time and space. PEBP1 acts by targeting key proteins such as GRK2, Aurora B, LC3 and Rab8. In all these processes, PEBP1 appears to protect the inactive form of its targets preventing their activation. The association/dissociation of PEBP1 from its target is controlled by posttranslational modifications of PEBP1, especially phosphorylation.

## 7.5- Direct partners of PEBP1 interact with cytoskeletal elements

Interestingly, most of the described partners of PEBP1 are known to interact with cytoskeletal elements. Here are some examples.

### 7.5.1- **ERK.**

The signaling ERK pathway is known to be implicated in cytoskeleton organization and cell motility but it may act in association with PI3K pathway. In human breast tumor cells MDA-MB-231 LM2, the Ras-ERK pathway was found responsible for the disruption of stress fibers and for the expression of genes associated with the lung metastasis signature. However, inhibition of the Ras-ERK pathway did not result in inhibition of cell motility but was accompanied by activation of the PI3K pathway. Inhibition of both ERK and PI3K pathways was required to inhibit motility of MDA-MLM2 cells (**512, Choi**).

### 7.5.2- **NF-κB inducing kinase (NIK).**

A constitutive association between filamin A and the NF-κB inducing kinase (NIK), in both Jurkat and human primary T cells was identified. CD28 (Cluster of Differentiation 28) is one of the proteins expressed on T cells that provide co-stimulatory signals required for T cell activation and survival. Both proline and tyrosine residues within the C-terminal proline-rich motif of CD28 are involved in the recruitment of filamin A/NIK complexes to the membrane as well as in the activation of NIK and IKKα (**513, Muscolini**).

### 7.5.3- GRK2

Since a long time GRK2 is considered to bind and phosphorylate tubulin (**514, Haga**). More recently, GRK2 appears to be a key integrative node in cell migration control. GRK2 was identified as a modulator of diverse molecular processes involved in motility, such as gradient sensing, cell polarity or cytoskeletal reorganization. GRK2 dynamically associates with and phosphorylates HDAC6 to stimulate its α-tubulin deacetylase activity at specific cellular localizations such as the leading edge of migrating cells, thus promoting local tubulin deacetylation and enhanced motility (**515, Lafarga**). Thus, GRK2 can play an effector role in the organization of actin and microtubule networks and in adhesion dynamics, by means of substrates and transient interacting partners, such as the GIT1 scaffold or the cytoplasmic HDAC6. The overall effect of altering GRK2 levels or activity on chemotaxis would depend on how such different roles are integrated in a given cell type and physiological context (**516, Penela**).

### 7.5.4- GSK3β

Several studies described the relationships between GSK3 and cytoskeleton. As an example, mechanistically, GSK3β signaling promoted preferential hematopoietic stem and progenitor cell migration by regulating actin rearrangement and microtubule turnover, including CXCL12-induced actin polarization and polymerization (**517, Lapid**).

### 7.5.5- STAT3

Cytoplasmic STAT3 was demonstrated to interact with both tubulin and microtubules. The localization of STAT3 on microtubules and its activation are independent of HDAC6 activity. Moreover, cytoplasmic STAT3 is physically associated with microtubules, whereas its activation and nuclear translocation are independent of microtubule dynamics, implicating that the association of STAT3 with microtubules might be involved in the regulation of non-canonical functions of STAT3 in the cytoplasm (**518, Yan**).

### 7.5.6- Notch

The Notch pathway has been recognized as one of the main contributors in regulating neural development and has been proposed as a key mediator in neuroplasticity by modulating neuronal cytoskeleton (**519, Bonini**). Several pathways are implicated in the microtubule assembly/disassembly process, among them Notch acts as a microtubule dynamics regulator. It was demonstrated that activation of Notch signaling results in increased microtubule stability and changes in axonal morphology and branching. By contrast, Notch inhibition leads to an increase in cytoskeleton plasticity with intense neurite remodeling (**520, Bonini**).

### 7.5.7- Drugs and prodrugs

The drugs fortunately found to be captured by PEBP1 (i.e. PDE5 inhibitors, locostatin, prasugrel) may interact also with cytoskeletal elements. Chronic treatment with the PDE5 inhibitor Tadalafil used in diabetic patients was demonstrated to modulate proteins involved in cytoskeletal rearrangement and redox signaling of the heart, which may explain the beneficial effects of PDE-5 inhibition in diabetes (**521, Koka**). The protective effect of PEBP1 against locostatin, a drug inhibition of migration, suggested a new role for PEBP1 in potentially sequestering toxic compounds that may have deleterious effects on cells (**263, Shemon**). The prodrug Prasugrel hydrolyzed by PEBP1 is a P2Y12 receptor antagonist. In cultured Vascular

Smooth Muscle Cells (VSMCs), the activation of P2Y$_{12}$ receptor inhibited cAMP/protein kinase A signaling pathway, which induced cofilin dephosphorylation and filamentous actin disassembly, thereby enhancing VSMCs motility and migration (**522, Niu**).

All these data suggest that PEBP1 may be implicated in protection and organization of cytoskeleton.

## 7.6- Cytoskeleton in cancer

Interestingly, metastatic disease, or the movement of cancer cells from one site to another, is a complex process requiring dramatic remodeling of the cell cytoskeleton. The various components of the cytoskeleton, actin (microfilaments), microtubules (MTs) and intermediate filaments, are highly integrated and their functions are well orchestrated in normal cells. In contrast, mutations and abnormal expression of cytoskeletal and cytoskeletal-associated proteins play an important role in the ability of cancer cells to resist chemotherapy and metastasize. Studies on the role of actin and its interacting partners have highlighted key signaling pathways, such as the Rho GTPases, and downstream effector proteins that, through the cytoskeleton, mediate tumour cell migration, invasion and metastasis. An emerging role for MTs in tumour cell metastasis is being unravelled and there is increasing interest in the crosstalk between key MT interacting proteins and the actin cytoskeleton (**523, Fife**).

It was illustrated that the proportion of microtubules and actin microfilaments is different in grade I and grade IV colon cancer cells in a manner that microtubules attain an effectual role in progressive reorganization of cytoskeleton in transition from nonaggressive to malignant phenotypes in cancer cells. Furthermore, it was concluded that larger instantaneous Young's modulus value (which defines the relationship between stress and strain in a material) for high grade cells is related to the existence of extensively build-up actin networks at the cell cortex (**524, Pachenari**). The ability of a eukaryotic cell to resist deformation, to transport intracellular cargo and to change shape during movement depends on the cytoskeleton, an interconnected network of filamentous polymers and regulatory proteins. Recent work has demonstrated that both internal and external physical forces can act through the cytoskeleton to affect local mechanical properties and cellular behavior. An important insight emerging from this work is that long-lived cytoskeletal structures may act as epigenetic determinants of cell shape, function and fate (**186, Fletcher**).

## 7.7- Actin filaments and microtubules are tension sensors

While the actin filament (F-actin) has been viewed as dynamic in terms of polymerization and depolymerization, more recent results suggested that F-actin itself may function as a highly dynamic tension sensor (**525, Galkin**). Tensile forces generated by stress fibers drive signal transduction events at focal adhesions and it was reported that stress fibers act as a platform for tension-induced activation of biochemical signals. ERK is activated on stress fibers in a myosin II-dependent manner. By quantifying myosin II- or mechanical stretch-mediated tensile forces in individual stress fibers, ERK activation on stress fibers correlated positively with tensile forces acting on the fibers, indicating stress fibers as a tension sensor in ERK activation. Myosin II-dependent ERK activation is also observed on actomyosin bundles connecting E-cadherin clusters, thus suggesting that actomyosin bundles work as a platform for tension-dependent ERK activation. (**526, Hirata**). A single actin filaments was placed under tension using optical tweezers. When a fiber was tensed, it was severed after the application of cofilin with a significantly larger delay in comparison with control filaments suspended in solution. The binding rate of cofilin to an actin bundle decreased when the bundle was tensed, resulting in a decrease in its effective severing activity (**527, Hayakawa**).

In yeast kinetochore, the Ndc80 complex (Ndc80, Nuf2, Spc24 and Spc25) is a highly conserved protein essential for end-on anchorage to spindle microtubule plus ends and for force generation coupled to plus-end polymerization and depolymerization. Spc24/Spc25 at one end of the Ndc80 complex binds the kinetochore, the N-terminal tail and calponin-homology domains of Ndc80 bind microtubules, and an internal domain binds microtubule-associated proteins such as the Dam1 complex. A mechanical model of force generation at metaphase was proposed where the position of the kinetochore relative to the microtubule plus end reflects the relative strengths of microtubule depolymerization, centromere stretch and microtubule-binding interactions with the Ndc80 and Dam1 complexes (**528, Suzuki**).

### 7.8- Main properties and functions of PEBP1

In summary of the main molecular specifications of PEBP1, we can say:
- PEBP1 is essentially implicated in cell processes implying membrane changes and cytoskeleton reorganization (e.g. cell motility, autophagy, cytokinesis, primary cilium formation).
- By direct interaction, PEBP1 inhibits several proteins containing nucleotide sites, particularly kinases of main signaling pathways, GRK2, Aurora B and the GTPase Rab8.
- PEBP1 is a switch that inhibits or releases specific targets depending on its own posttranslational modifications. By blocking signaling pathways kinases, GRK2, Aurora B and LC3, PEBP1 exerts a spatiotemporal control of signaling pathways, receptors internalization, mitotic spindle checkpoint and phagocytosis initiation, respectively.
- PEBP1 binds small ligands such as nucleotides, locostatin, polyphenols and drugs. It may act as a hydrolase.
- Depending on the small ligands present in the cell, PEBP1 is a modulator of interactions between protein kinases of main signaling pathways.
- PEBP1 may act as a cytoskeleton protector by stabilizing proteins (Keap1), or destabilizing toxic compounds that may have deleterious effects on cells (locostatin).
- As an interaction controller, PEBP1 may synchronize the fast activity of enzymes (e.g. kinases and Rab8 GTPase) with the time necessary for the downstream physical organization of membrane and cytoskeleton that is needed to perform cellular processes.

All the cellular processes in which PEBP1 is implicated are in very close relationships, they constitute a variety of cellular responses to external or internal signals. A large majority of these processes leads to membrane changes and all of them are underlain by cytoskeleton rearrangement. PEBP1 participates in sensing the plasma membrane status by acting on receptors internalization and by controlling the formation of primary cilium that is a sensory organelle. The continuity between the sensing of physical and chemical membrane status (captured notably by actin) and the cytoskeletal reorganization is assumed by several signaling pathways. It is likely that the circuits regulating the cellular response to signals may be multiple and more or less long, depending on cell status and signaling pathways in action.

Most of the proteins interacting with PEBP1 are themselves involved in great cellular functions such as proliferation, differentiation, division, survival and motility by interacting with several targets, thus confirming the existence of consistent molecular complexes regulating cell processes. Furthermore, the fact that in cancer cells, overexpressed PEBP1 interacts with proteins governing cytoskeletal network, glycolysis and protein biosynthesis is in concordance with the need of energy and protein synthesis for cancer cells. It is likely that these interactions may also exist in non-pathological conditions, suggesting a coordinated link of cytoskeletal organization with energy and protein biosynthesis needed to perform cytoskeleton changes.

It is interesting to note that PEBP4 was described to act as a chaperone and did not display the regulator effect of PEBP1. Consequently, its global effect is the opposite of that of PEBP1, favoring motility and migration of cells (**529, Zhang**). On the contrary, TFL1 and FT1 in plants confirm a role of PEBP family to sensing cell membrane and controlling signaling pathways, consistent with their implication in adaptation of cells to their environment by regulating circadian rhythm and temperature depending thermo-sensitive flowering regulation (**530, Kim**).

## In conclusion

In addition to regulate protein kinases, PEBP1 possibly regulates also guanine exchange factors as GPCRs and GTPases such as Rab8. PEBP1 is involved in integrative system(s) which lead and control various phenomena such as membrane sensing, cellular shape and vesicular traffic, all depending on cytoskeleton organization. PEBP1 is a modulator of molecular interactions which adjusts signaling pathways and works as a switch between them in order to block or initiate specific cellular mechanisms, allowing the cell to constantly adapt to its microenvironment. Mechanically, PEBP1 acts as a node between signaling pathways and cytoskeleton organization by targeting specific proteins. Then, it is convenient that PEBP1 may interact directly or indirectly with components of cytoskeleton as actin, tubulin and vimentin.

PEBP1 is a modulator of molecular interactions, it may stabilize or destabilize some protein partners and ligands. It inhibits several signaling pathway kinases and key proteins having hand in cytoskeleton organization such as GRK2, Aurora B and LC3. Cytoskeletal organization is a complex and integrative mechanism, it underlies numerous cellular mechanisms and is implicated in main cellular diseases such as cancer, Alzheimer's, ciliopathies and diabetic nephropathy.

The levels of PEBP1 concentration seem to be essential for its action as well as the presence and concentration of other partners and different ligands in the cell, particularly ATP. The level of PEBP1 expression is controlled by complicated loops and is influenced by various small ligands such as ATP and polyphenols. Its rapid degradation is conducted by chaperone-mediated autophagy. In cancer cells, methylation of the promotor region in the gene encoding PEBP1 was demonstrated explaining the downregulation of PEBP1 in these cells.

Thus PEBP1 seems to adapt interactions between partners according to the cell status and to the presence of different ligands into the cell. This action may allow the correct response of the cell according to its external environment and also to its internal composition. In addition, the sharp regulation of molecular interactions likely permit to adapt reciprocal conversion of chemical forces represented by molecular interactions to physical forces generated by membrane deformation and cytoskeleton tension.

As saying that the fine-tuned interactions between PEBP1 and its partners are far from to be determined. In the future, a particular effort should be made to decipher what leads to the complex and kinetic interactions between all these molecular actors. Probably, in addition to the basic affinity between them, it will be useful to consider the intracellular medium, the location of the molecular partners, their reciprocal concentration and input order of appearance as well as the dynamics of the events in which they are implicated.

## Acknowledgments

This review article was fueled by Canceropôle du grand Ouest meetings and discussions. We thank Isabelle Callebaut (IMPMC) for helpful advises and fruitful comments. We thank also William Sacks (IMPMC) for reading the manuscript and valuable corrections of the text.

527- Actin filaments function as a tension sensor by tension-dependent binding of cofilin to the filament. Hayakawa K, Tatsumi H, Sokabe M. J Cell Biol. 2011 Nov 28;195(5):721-7. doi: 10.1083/jcb.201102039.

528- How the kinetochore couples microtubule force and centromere stretch to move chromosomes. Suzuki A, Badger BL, Haase J, Ohashi T, Erickson HP, Salmon ED, Bloom K. Nat Cell Biol. 2016 Apr;18(4):382-92. doi: 10.1038/ncb3323.

529- Promotion of cellular migration and apoptosis resistance by a mouse eye-specific phosphatidylethanolamine-binding protein. Zhang Y, Wang X, Xiang Z, Li H, Qiu J, Sun Q, Wan T, Li N, Cao X, Wang J. Int J Mol Med. 2007 Jan;19(1):55-63

530- Generation and analysis of a complete mutant set for the Arabidopsis FT/TFL1 family shows specific effects on thermo-sensitive flowering regulation. Kim W, Park TI, Yoo SJ, Jun AR, Ahn JH. J Exp Bot. 2013 Apr;64(6):1715-29. doi: 10.1093/jxb/ert036.

**Table I – Close partners of PEBP1**

| Uniprot ID | Protein name | Protein function |
|---|---|---|
| | | |
| | I-  Cytoskeleton-related proteins (28%) | |
| | *Microtubules* | |
| P07437 | Tubulin beta chain (**SNO**) | *Major constituent of microtubules* |
| | *Actin filaments and actin binding proteins* | |
| P60709 | Actin cytoplasmic 1 (**SNO**) | *Involved  in various types of cell motility* |
| Q15149 | Plectin, isoform 1 | *Actin binding* |
| P52907 | F-actin-capping protein subunit alpha 1 | *Actin binding* |
| P18206 | Vinculin, isoform 1 | *Actin-filament binding protein* |
| | *Intermediate filaments* | |
| P05783 | Keratin type I cytoskeletal 18 | *Cytoskeleton of intermediate filaments* |
| P05787* | Keratin type II cytoskeletal 8* | *Cytoskeleton of intermediate filaments* |
| P08670 | Vimentin | *Cell motility, structural constituent of cytoskeleton* |
| | ***Organization of the cytoskeleton, multifunctional proteins*** | |
| P35579 | Myosin-9 | *Cytokinesis, secretion, affects actin cytoskeleton* |
| P46940 | **Ras GTPase-activating-like protein IQGAP1** | *Actin cytoskeleton, cell adhesion, cell cycle, ERK pathway modulation, transcription* |
| | **II Protein biosynthesis (22.8 %)** | |
| P68104 | **Elongation factor 1-alpha 1** | *GTPase, delivery of aminoacyl-tRNAs to the ribosome* |
| P41252 | **Isoleucyl-tRNA synthase cytoplasmic** | *Isoleucine-tRNA ligase, ATP binding* |
| P26640 | Valyl-tRNA synthase | *Valine-tRNA ligase, ATP binding* |
| P26641 | Elongation factor 1-gamma | *Delivery of aminoacyl tRNAs to the ribosome* |
| P13639 | Elongation factor 2 | *GTP-dependent translocation of the ribosome* |
| P60842 | Eukaryotic initiation factor 4A-1 | *Binding of mRNA to 40S ribosomal subunits, ATP-dependent RNA helicase* |
| P63244 | Guanine-nucleotide-binding protein subunit beta-2-like1 | *Major component of translating ribosomes* |
| P23396 | 40S ribosomal protein S3 | *Ribosomal domain where translation is initiated* |
| | **III Glycolytic enzymes (17%)** | |
| P14618 | **Pyruvate kinase isozymes M1/M2 (SNO)** | *Final step of glycolysis, production of pyruvate and ATP* |
| P04406 | Glyceraldhehyde-3-phosphate dehydrogenase (**SNO**) | *G3P dehydrogenase, nitrosylase activity* |
| P07195 | L- lactate dehydrogenase B chain (**SNO**) | *L-lactate dehydrogenase activity* |
| P00338 | L- lactate dehydrogenase A chain (**SNO**) | *L-lactate dehydrogenase activity* |
| P06733 | Alpha-enolase (**SNO**) | *Multifunctional enzyme* |
| P04075 | Fructose-bisphosphate aldolase A (**SNO**) | *Production of DHAP and G3P, Actin binding* |

| | | IV- Molecular chaperones (14.3%) | |
|---|---|---|
| P08238* | Heat shock protein HSP-90 beta* | *Stress related* |
| P11142 | Heat shock cognate 71 kDa (**SNO**) | *Stress related* |
| P07900 | Heat shock protein HSP-90 alpha | *Stress related* |
| P38646 | Stress-70 mitochondrial | *Related with cell proliferation and aging* |
| P10809 | 60kDa heat shock protein, mitochondrial | *Stress related* |
| | **V- Mitochondrial proteins (5.7%)** | |
| P06576 | **ATP synthase subunit alpha mitochondrial** | *Mitochondrial membrane ATP synthase* |
| P21796 | **Voltage-dependent anion-selective channel protein 1 VDAC1 (SNO)** | *Voltage-gated anion channel activity* |
| | **VI- Transport of proteins and vesicles (5.7%)** | |
| Q14974 | Importin subunit beta 1 | *Protein domain specific binding (NLS), transport into the nucleus* |
| Q00610 | **Clathrin heavy chain 1** | *Polyhedral coat of coated pits and vesicles* |
| | **VII- Others (5.7%)** | |
| P04083 | Annexin A1 | *Ca2+-dependent phospholipid-binding proteins, phospholipase A2 inhibitory activity* |
| P62258* | 14-3-3 protein epsilon* (**SNO**) | *Binding of signaling proteins, recognition motif* |

**Table 1- The 35 proteins close partners of PEBP1 identified by MiMI analysis (159, Gu H 2013).** They are gathered in seven families. The percentage following the name of each family is representative of the number of proteins in the family against the total of 35 proteins. The 7 proteins identified as close partners of PEBP1 with the three protein network diagrams are indicated in bold characters and underlined. The three proteins identified for the first time as partners of PEBP1 are indicated by a star. (**SNO**), S-nitrosylated proteins in Alzheimer's disease (**261, Zahid**).

## Table II - Main cellular processes implying PEBP1

| Cellular processes | PEBP1 activity |
|---|---|
| **Modulation of signaling pathways** | *Inhibition of Raf/MEK/ERK, GRK2, NFkB, PI3K/Akt, p38MAPK, Wnt, STAT3, LIN28/Let-7 miRNA* <br> *Activation of GSK3β* |
| **Internalization of receptors and desensitization** | *GPCRs: opioid receptors, beta-adrenergic receptors, VPACs* |
| **Neural development** | *Differentiation into neurons and oligodendrocytes* |
| **Neuronal synapse** | *Neuronal plasticity: long-term synaptic depression, long-term potentiation* |
| **Hippocampal presynaptic terminals** | *Co-localization with CRMP-2* |
| **Rat intestine nerve cells** | *Stored in axon terminals* |
| **Vesicles formation and trafficking** | *Exosomes, autophagosomes, endosomes* |
| **Primary cilium** | *Prevention of primary cilium formation* |
| **Mitotic spindle checkpoint** | *Inhibition of Aurora B* |
| **Apoptosis** | *- Pro-apoptotic via Raf1/MEK/ERK and NFkB pathways* <br> *- Anti-apoptotic under oxygen-glucose deprivation, through IKKβ/IkBα/NFkB pathway* |
| **Autophagy** | *Inhibition of autophagy initiation; interaction with LC3-I* |
| **Cell motility** | *Inhibition of cell invasion and metastasis, except via GSK3* <br> *Regulation of MMPs and miRNAs expression* |
| **Cell adhesion** | ***Increased adhesion to substratum, decreased cell-cell adhesion*** |

**Table II - Main cellular processes implying PEBP1.** PEBP1 was described to participate in several cellular processes, most of them implicate plasma membrane and cell shape changes. The role of PEBP1 in each process is mentioned.

**Figure 1**

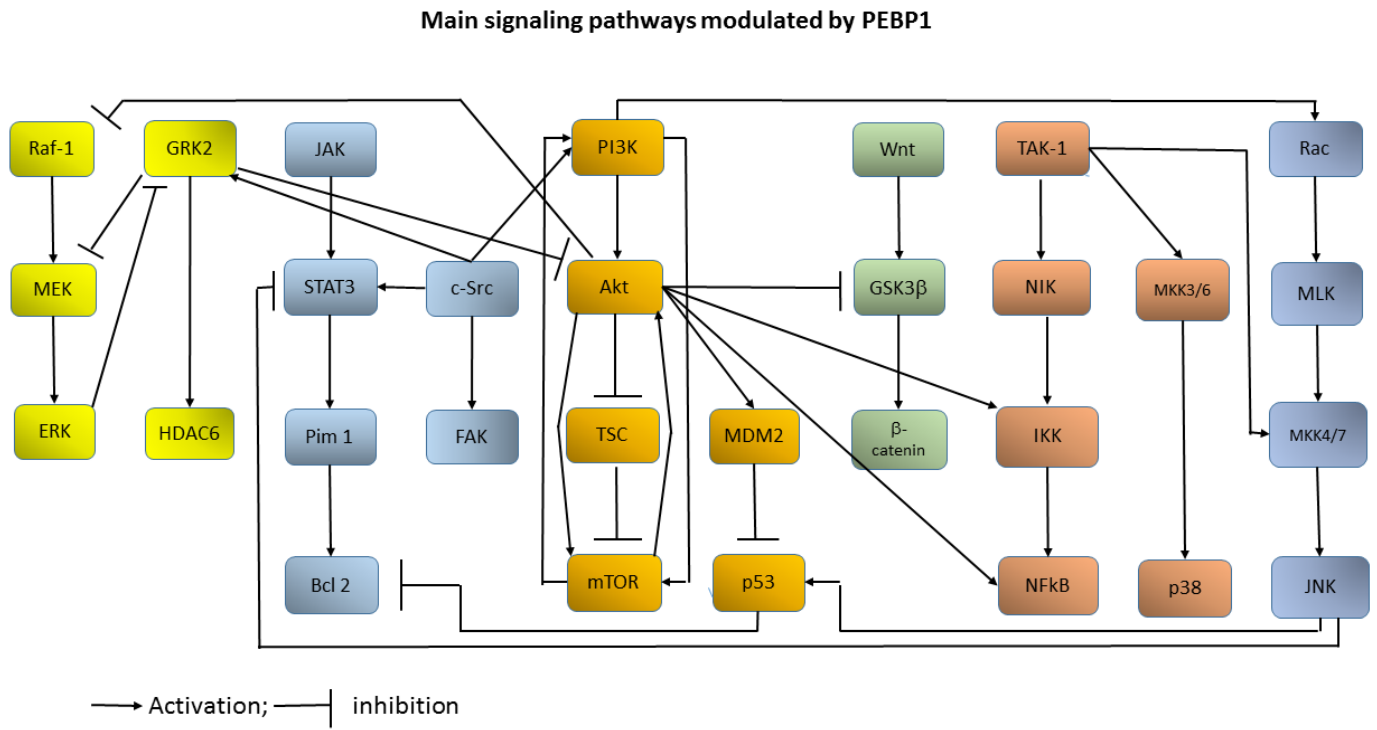

**Figure 1. Main signaling pathways modulated by PEBP1.** Direct binding of PEBP1 with several pathway kinases were demonstrated. Moreover, the pathways are known to be interconnected and to crosstalk between them. → Stimulation**,** ⊣ inhibition.

**Figure 2**

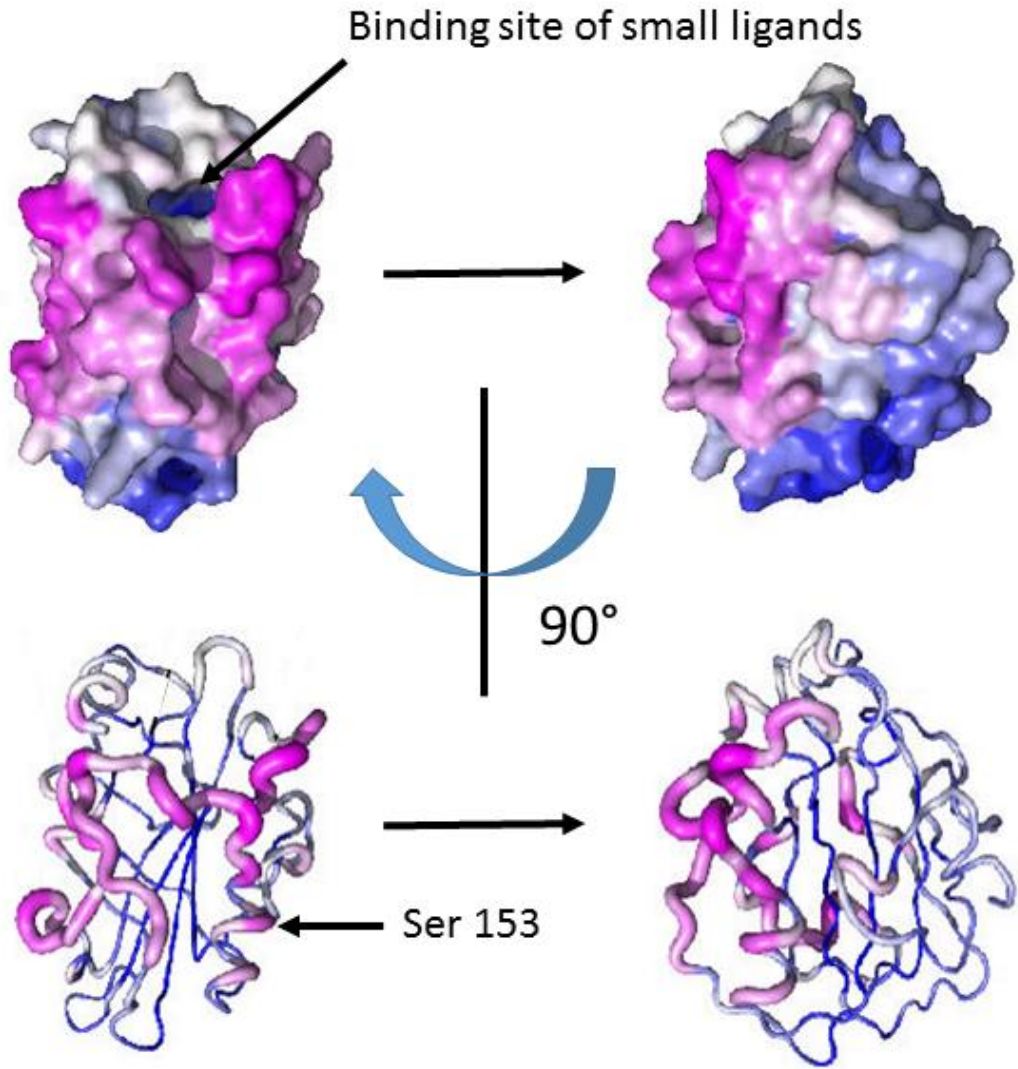

**Figure 2. PEBP1 interactions with kinases and small ligands**. Upper part, the surface model of PEBP1, lower part, the ribbon model. A pocket is indicated, it is the binding site of PEBP1 for small ligands (particularly nucleotides, cacodylate, phosphate, acetate, phospholipids and PDE5 inhibitor). The protein kinases modulated by PEBP1 were described to interact with the binding site and also with two regions situated on the edges of a shallow furrow beneath the binding site, these two regions are colored in magenta. When phosphorylated, Ser153 shifts PEBP1 binding from Raf-1 to GRK2. The 3D structure was published by Serre et al. (**23, Serre**) and the predicted interaction sites based on this structure was published in (**263, Martin).**

**Figure 3**

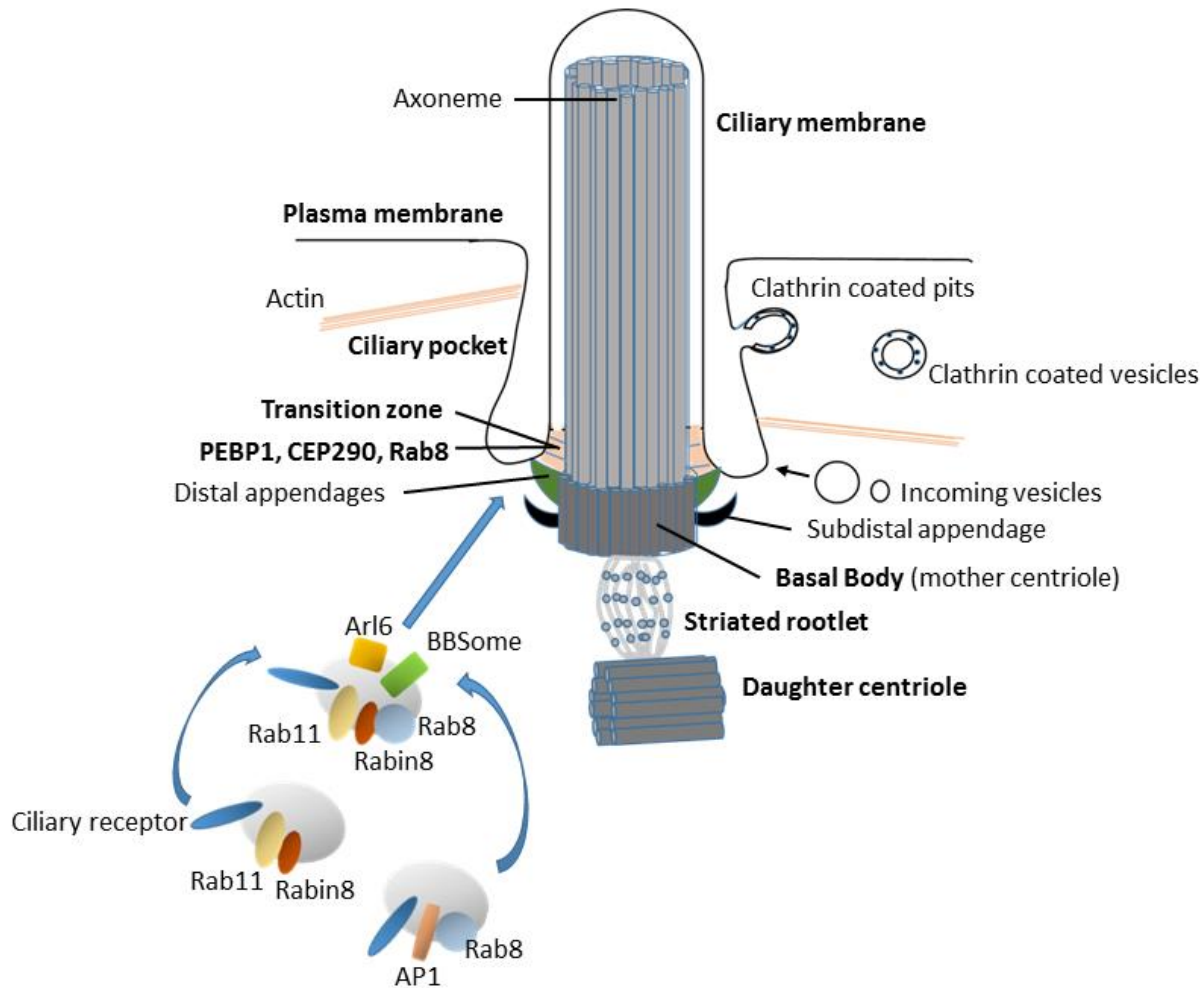

**Figure 3. Schema of primary cilia structure.** The ciliary skeleton is composed of axoneme and basal body. The transition zone is indicated by Y-shaped bridges extending from microtubule doublets to ciliary membrane and distal appendages, it corresponds to the transition between microtubule triplets (basal body and distal appendages) to microtubule doublets (axoneme). PEBP1, CEP290 and Rab8 were described to interact at the transition zone (**77, Murga-Zamalloa**). Primary ciliary pocket is defined to be membrane domain starting from distal appendages to the region where ciliary sheath emerges to extracellular environment. Primary cilia pocket plays a role in endocytosis. Incoming vesicles (from Golgi complex) dock at membrane junction at distal appendages and keep balances between ciliary membrane addition and removal (**390, Ke YN**). One pathway for protein targeting to the ciliary membrane is led by the BBSome and its interactor Arl6, which recruits Rab8 and its guanine nucleotide exchange factor Rabin8 at the base of the cilium. Rabin8 is transported to this location in Rab11$^+$ vesicles. Some membrane receptors destined for the ciliary membrane are transported from the trans-Golgi network apparatus to Rab8$^+$ vesicles with the assistance of the adapter AP-1 (**407, Fenetti**). In this figure, the general drawing of the primary cilium is based on Fig. 1 from (**390, Ke YN** ) and the transport of Rab proteins is inspired by fig.1 from (**407, Finetti).**

**Figure 4**

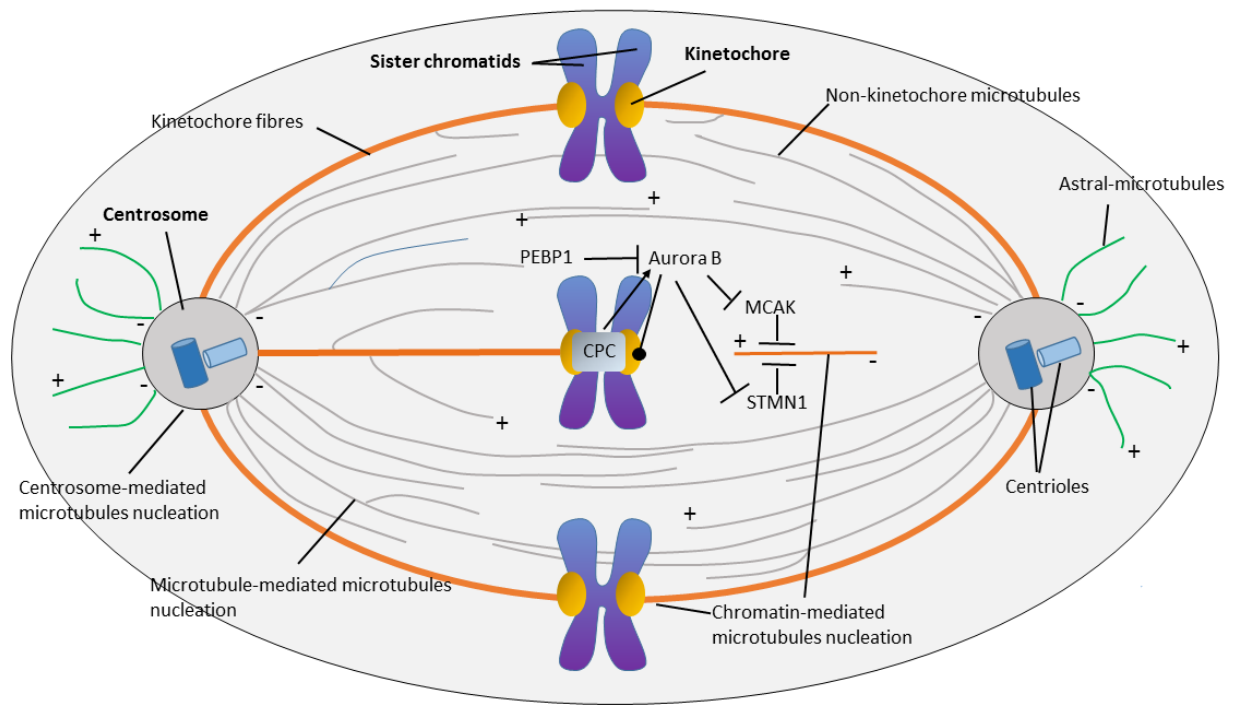

**Figure 4. Schema of mitotic spindle.** In this figure, the overview of the mitotic spindle is drawn according to Fig.1 in (**321, Prosser**).The chromosomes attach to the microtubules via kinetochores on the centromere of each sister chromatid. Three nucleation pathways drive the formation of microtubules to form the mitotic spindle: nucleation pathway from the centrosome, from the chromatin and from microtubules. The resulting spindle comprises three different types of microtubules: the kinetochore microtubules which attach the chromosomes to the spindle poles and move the sister chromatids apart, non-kinetochore microtubules which originate from the opposing poles, help to separate them and provide stability to the spindle, and astral microtubules which radiate out from the poles, contact the cell cortex, and play a part in spindle positioning. Once 20–30 Kinetochore microtubules have attached to the kinetochore, they become stabilized into bundles called kinetochore fibers.

The centre of the figure shows the mechanism by which microtubules grow from kinetochore. At the inner centromere (the part of chromosome that links sister chromatids) Haspin phosphorylates histone H3, which leads to the recruitment of the chromosome passenger complex (CPC) and *trans*-autophosphorylation activation of Aurora B. Then Aurora B phosphorylates, and thereby inactivates, the microtubule-destabilizing proteins mitotic centromere-associated kinesin (MCAK) and stathmin (STMN1) which promotes microtubule growth around kinetochores. PEBP1 is an inhibitor of Aurora B, it prevents microtubule growth around kinetochore (**321, Prosser**).